\documentclass[11pt, english]{article}%
\usepackage{amsfonts}
\usepackage{sw20elba}
\usepackage[T1]{fontenc}
\usepackage[latin9]{inputenc}
\usepackage[margin=1in,bottom=1in,top=1in]{geometry}
\usepackage{lscape}
\usepackage{setspace}
\usepackage{color}
\usepackage{amsmath}
\usepackage{amsthm}
\usepackage{bm}
\usepackage{amssymb}
\usepackage{stmaryrd}
\usepackage{graphicx}
\usepackage{float}
\usepackage{setspace}
\usepackage{rotating}
\usepackage{subcaption}
\usepackage{ragged2e}
\usepackage{esint}
\usepackage{ulem}
\usepackage{babel}
\usepackage{longtable}
\usepackage{booktabs}
\usepackage{caption,booktabs,array}
\usepackage[round]{natbib}
\usepackage[hypertexnames=false]{hyperref}
\usepackage{colortbl}
\usepackage[toc,page]{appendix}
\usepackage[sc]{mathpazo}
\usepackage{dsfont}
\usepackage{pifont}
\usepackage[outdir=./]{epstopdf}%
\providecommand{\U}[1]{\protect\rule{.1in}{.1in}}
\definecolor{darkblue}{rgb}{0,0,.6}
\hypersetup{
colorlinks=true,
linkcolor=darkblue,
filecolor=darkblue,
urlcolor=darkblue,
citecolor=darkblue,
}

\newtheorem{theorem}{Theorem}
\newtheorem{assumption}{Assumption}

\newtheorem{corollary}{Corollary}

\newtheorem{example}{Example}

\newtheorem{lemma}{Lemma}

\newtheorem{proposition}{Proposition}
\newtheorem{remark}{Remark}
\begin{document}

\title{Unconditional Effects of General Policy Interventions\thanks{For helpful
conversations, we thank Javier Alejo, George Bulman, and Augusto
Nieto-Barthaburu. All errors remain our own. This paper supersedes
\textit{Location-Scale and Compensated Effects in Unconditional Quantile
Regressions}. }}
\author{Juli\'an Mart\'inez-Iriarte\thanks{Department of Economics, UC Santa Cruz.
E-mail: jmart425@ucsc.edu}
\and Gabriel Montes-Rojas\thanks{CONICET and Universidad de Buenos Aires. E-mail:
gabriel.montes@fce.uba.ar}
\and Yixiao Sun\thanks{Department of Economics, UC San Diego. E-mail:
yisun@ucsd.edu}}
\maketitle

\begin{abstract}
This paper studies the unconditional effects of a general policy intervention,
which includes location-scale shifts and simultaneous shifts as special cases.
The location-scale shift is intended to study a counterfactual policy aimed at
changing not only the mean or location of a covariate but also its dispersion
or scale. The simultaneous shift refers to the situation where shifts in two
or more covariates take place simultaneously. For example, a shift in one
covariate is compensated at a certain rate by a shift in another covariate.
Not accounting for these possible scale or simultaneous shifts will result in
an incorrect assessment of the potential policy effects on an outcome variable
of interest. The unconditional policy parameters are estimated with simple
semiparametric estimators, for which asymptotic properties are studied. Monte
Carlo simulations are implemented to study their finite sample performances.
The proposed approach is applied to a Mincer equation to study the effects of
changing years of education on wages and to study the effect of smoking during
pregnancy on birth weight. \vspace{7mm}

 \vspace{3mm}


\end{abstract}

\normalem


\onehalfspacing
\newpage

\section{Introduction}

In many research areas, it is important to assess the distributional effects
of covariates on an outcome variable. Several methods have been implemented in
the literature to study this. A prolific line of research is a combination of
conditional mean and quantile regression models together with micro simulation
exercises, as in \cite{AutorKatzKearney05}, \cite{MachadoMata05}, and
\cite{Melly05} (see \citet{FortinLemieuxFirpo11} for a review). A more recent
and popular method is the recentered influence function (RIF) regression of
\cite{FirpoFortinLemieux09}, which directly estimates the effect of a change
in the covariate distribution on a functional of the unconditional
distribution of the outcome variable. The functional of interest can be the
mean, quantile, or any other aspect of the unconditional distribution.

Consider, as an example, the unconditional quantile of the outcome variable
$Y$. Let $F_{Y}$ be the unconditional distribution function of $Y,$ then the
$\tau$-quantile of $F_{Y}$ is defined by
\[
Q_{\tau}[Y]:=\arg\min\{q:\tau\leq F_{Y}(q)\}\ \ \mbox{for}\ \ \tau\in(0,1).
\]
In this paper, we seek to study how $Q_{\tau}[Y]$ changes when we induce an
infinitesimal change in a covariate $X\in\mathbb{R}$, \ allowing the presence
of other observable covariates $W$ and unobservable covariates collected in
$U.$ These covariates and the outcome variable are related via a structural or
causal function $h$ so that $Y=h(X,W,U)$. We consider a sequence of policy
experiments that change $X$ into $X_{\delta}=\mathcal{G}(X;\delta)$ for a
smooth function $\mathcal{G}(\cdot;\cdot)$. The policy experiments are indexed
by $\delta$ satisfying $\mathcal{G}(X;0)=X.$ That is, $\delta=0$ corresponds
to the \emph{status quo} policy. With this induced change in $X$, the outcome
variable becomes $Y_{\delta}=h(X_{\delta},W,U)=h(\mathcal{G}(X;\delta),W,U)$
where the distribution of $\left(  X,W,U\right)  $ is held constant. Our
policy experiment has a \emph{ceteris paribus} interpretation at the
population level: we change $X$ into $X_{\delta}$ while holding the stochastic
dependence among $X,W,$ and $U$ constant. Such a policy experiment is
implementable if the covariate $X$ is not a causal factor for either $W$ or
$U.$ In this case, when we intervene $X$ and change it into $X_{\delta}$, $W$
and $U$ will not change. This does not rule out the stochastic dependence
among $X,$ $W,$ and $U.$ In the meanwhile, the structural function
$h(\cdot,\cdot,\cdot)$ is also held constant. The main parameter of interest
is the marginal effect of the change on the unconditional quantile of the
outcome variable:
\[
\Pi_{\tau}:=\lim_{\delta\rightarrow0}\frac{Q_{\tau}[Y_{\delta}]-Q_{\tau}%
[Y]}{\delta}.
\]

\cite{FirpoFortinLemieux09} develop methods to study what corresponds to a
location shift $X_{\delta}=X+\delta$. This shift affects the entire
unconditional distribution of $Y=h(X,W,U)$, moving it towards a counterfactual
distribution of $Y_{\delta}=h(X+\delta,W,U)$. One of the main results in
\citet[][p.958,
eq. (6)]{FirpoFortinLemieux09} is that $\Pi_{\tau}$ can be represented as an
average derivative:
\[
\Pi_{\tau}=E\left[  \dot{\psi}_{x}\left(  X,W\right)  \right]  ,
\]
where
\[
\dot{\psi}_{x}\left(  x,w\right)  =\frac{\partial E\left[  \psi\left(
Y,\tau,F_{Y}\right)  |X=x,W=w\right]  }{\partial x},
\]
$\psi\left(  y,\tau,F_{Y}\right)  =\left[  \tau-1\left\{  y\leq Q_{\tau
}[Y]\right\}  \right]  /f_{Y}(Q_{\tau}[Y])$ is the influence function of the
quantile functional, and $f_{Y}(Q_{\tau}[Y])$ is the unconditional density of
$Y$ evaluated at the $\tau$-quantile $Q_{\tau}[Y].$ The unconditional quantile
effect $\Pi_{\tau}$ can then be estimated by first running an unconditional
quantile regression (henceforth, UQR), which involves regressing the influence
function $\psi\left(  Y_{i},\tau,F_{Y}\right)  $ on the covariates
$(X_{i},W_{i})$ and then taking an average of the partial derivatives of the
regression function with respect to $X.$

The same method is applicable to other functionals of interest --- we only
need to replace $\psi\left(  y,\tau,F_{Y}\right)  $ by the influence function
underlying the functional we care about. This leads to the general RIF
regression of \cite{FirpoFortinLemieux09}. The potential simplicity and
flexibility that the methodology offers motivates subsequent research to
expand the use of RIF regressions. On the empirical side, after its
introduction, RIF regressions became a popular method for analyzing and
identifying the distributional effects on outcomes in terms of changes in
observed characteristics in areas such as labor economics, income and
inequality, health economics, and public policy. On the theoretical side,
\cite{Rothe2012} provides a generalization of \cite{FirpoFortinLemieux09} for
the case of location shifts, and, more recently, \cite{SasakiUraZhang20} study
the high-dimensional setting while \cite{InoueLiXu21} focus on the two-sample
problem. An alternative estimation procedure is proposed in \cite{cquq2023}.

This paper extends the UQR and RIF regression in several ways. First, we study
general counterfactual policy changes, of which the location shift is a
special case. Our framework allows for any smooth and invertible intervention
of the target covariates. As a complement to the existing literature that
focuses on changing the \emph{marginal distribution} of the target covariates,
we consider changing the \emph{values} of the target covariates directly. An
advantage of our approach is that the changes under consideration are directly
implementable. We note that it may not be easy to induce a desired shift in
the marginal distribution, and when possible, such a shift is often achieved
via transforming the target covariates, which is what we consider here.

Second, we provide extensive discussions of a counterfactual policy that, in
addition to the location shift, affects the scale of a covariate. For example,
we may consider $X_{\delta}=X/(1+\delta)+\delta.$ We find that in this case,
the marginal effect can be decomposed as the sum of two effects: one related
to the location shift and the other related to the scale shift. In order to
interpret the scale effect, we introduce the \textit{quantile-standard
deviation elasticity}: the percentage change in the unconditional quantiles of
the outcome variable induced by a 1\% change in the standard deviation of the
target covariate.

Third, we allow the target covariates to be endogenous, and we characterize
the asymptotic bias of the unconditional effect estimator when the endogeneity
is not appropriately accounted for. We eliminate the endogeneity bias using a
control variable/function approach. Such an approach is analogous to the
method of causal inference under the unconfoundedness assumption.

Fourth, by letting the policy function depend on covariates so that
$X_{\delta}=\mathcal{G}(X,W;\delta)$, we allow the interventions to vary
across covariate-specific strata. In the Supplemental Appendix, we also
consider the case of simultaneous shifts in different covariates. We focus on
the case of simultaneous location shifts in two covariates. This happens when
a location shift in one covariate induces a location shift in another
covariate at the same time. For example, $Y=h\left(  X_{1},X_{2},W,U\right)  $
for two scalar target covariates $X_{1}$ and $X_{2}$, and the policy induces
$X_{1\delta}=X_{1}+\delta$ and $X_{2\delta}=X_{2}-\delta$. Our approach can
easily accommodate this case, and we show that the simultaneous effect can be
obtained as a linear combination of individual effects obtained by considering
one change at a time.

Finally, we propose consistent and asymptotically normal semiparametric
estimators of the location-scale effect and the simultaneous effect. The
estimators can be easily implemented in empirical work using either a probit
or logit specification of the conditional distribution function. We conduct an
extensive Monte Carlo study evaluating the finite sample performances of the
location-scale effect estimator and the accuracy of the normal approximation.
Simulation results show that the estimator works reasonably well under
different specifications and that the standard normal distribution provides a
good approximation to the finite sample distribution of a studentized test
statistic introduced in this paper.

As potential applications of our proposed approach, consider the following
empirical examples to motivate its use.


\begin{example}
\label{Example 1} \textbf{Effect of increasing education on wage inequality.}
In a Mincer equation, log wages are modeled as a function of certain
observable covariates such as years of education. A study of the effect of a
shift in education on wage inequality could be implemented using our proposed
framework. We can accommodate a counterfactual policy experiment where there
may be not only a general increase in the education level but also a change in
its dispersion.
\end{example}

\begin{example}
\label{Example_Smoking} \textbf{Smoking and birth weight.} Consider a tax
levied on the consumption of cigarettes. It is reasonable to think that the
consumption $X$ will be reduced to $X/(1+\delta)$, where $\delta$ is the tax
burden on the consumer. Thus, the tax induces a reduction in the level and
dispersion of cigarette consumption. We will use the proposed method to assess
its effect on the distribution of birth weights.
\end{example}

\begin{example}
\textbf{Wage controls and earnings distribution} During War World II, the
National War Labor Board imposed wage controls in the form of brackets: wages
below the bracket were allowed to rise, while wages above the bracket were not
allowed to rise. Importantly, these brackets differed across industries,
occupations, and regions. \cite{ziebarth2022} use the tools developed in this
paper to analyze the effect of a more uniform (less dispersion) distribution
of brackets on the distributions of earnings.

\end{example}

\begin{example}
\textbf{Trade integration and skill distribution} \cite{Gu2020} document the
impact of trade integration on both the mean and the standard deviation of the
skill distribution across municipalities in Denmark. Moreover, as argued by
\cite{Hanushek2008}, skills are related to income distribution. Thus, a
quantification of the impact of a scale effect in the skills distribution on
the quantiles of the income distribution appears to be relevant.
\end{example}

\begin{example}
\textbf{Days in a job training program.} \cite{SasakiUraZhang20} develop
high-dimensional UQR to analyze the effect on wages of counterfactual increase
in: $(i)$ the days of participation in a job training program; and $(ii)$ the
days actually taking classes in the same job training program. Our
simultaneous effect analysis can consider, for example, a reduction in $(i)$
with a simultaneous increase in $(ii)$. Thus, our paper can be used to study
the effect of a more concentrated job training program.
\end{example}

We illustrate the proposed method with two empirical applications. The first
one is related to Example \ref{Example 1}: the effect of changing education on
wage inequality, decomposing it into location and scale effects. Empirical
results reveal the contrasting nature of the two effects. The location effects
are seen to be positive and relatively similar across quantiles. On the other
hand, the scale effects are highly heterogeneous and monotonically decreasing
across quantiles. Hence, the scale effects can more than offset the location
effects. This shows that not accounting for both shifts may result in a biased
assessment of the policy effects on the quantiles of the outcome variable. The
second application is related to Example \ref{Example_Smoking} where we
estimate the unconditional effects of smoking during pregnancy on the birth
weight. The effects from reducing the mean and variance of the\ number of
cigarettes smoked are positive and are different for different quantiles of
the birth weight distribution.

The paper is organized as follows. Section \ref{locsca} studies the
unconditional effects of general policy interventions with the location-scale
shift as the main example. Section \ref{sec:dist_vs_cov} provides some further
discussion on the methodological contribution of this paper relative to
\cite{FirpoFortinLemieux09}. Section \ref{estimation} describes the estimator
of the location-scale effect and studies its asymptotic properties. Section
\ref{MC} reports the finite sample performance of the location-scale effect
estimator and the associated tests. Section \ref{app} presents the empirical
applications. Section \ref{conclusion} concludes. The proofs are in the
Appendix. The case of simultaneous changes and the details for a theoretical
example are given in the Supplementary Appendix.

A word on notation: we use $F_{Y|X}(y|x)$ and $f_{Y|X}(y|x)$ to denote the
cumulative distribution function and the probability density function of $Y,$
respectively, conditional on $X=x$. For a random variable $Z$, the
unconditional $\tau$-quantile is denoted by $Q_{\tau}[Z]$, \textit{i.e.},
$\Pr(Z\leq Q_{\tau}[Z])=\tau$, and its variance is denoted by $var(Z).$ For a
pair of random variables $Z_{1}$ and $Z_{2}$, the conditional quantile is
denoted by $Q_{\tau}[Z_{1}|z_{2}]$, \textit{i.e.}, $\Pr(Z_{1}\leq Q_{\tau
}[Z_{1}|z_{2}]|Z_{2}=z_{2})=\tau$. We adopt the following notational
conventions:
\[
\frac{\partial E(Z|X)}{\partial X}=\left.  \frac{\partial E\left(
Z|X=x\right)  }{\partial x}\right\vert _{x=X},\text{ }\frac{\partial
F_{Z|X}(z|X)}{\partial X}=\left.  \frac{\partial F_{Z|X}(z|X=x)}{\partial
x}\right\vert _{x=X}.
\]
For a column vector $v,$ $d_{v}$ stands for the number of elements in $v.$

\section{Unconditional effects of general policy interventions}

\label{locsca}

\subsection{Introducing location-scale shifts}

We start with a general structural model $Y=h(X,W,U)$, where the function $h$
is unknown, and we only observe $(X,W)$ and $Y$. Here $X$ is univariate but
the dimension of $W$ is left unrestricted. All the unobserved causal factors
of $Y$ are collected in $U$. We are concerned with the effect on the
distribution of $Y$ of general (infinitesimal) changes in $X$, the
\textit{target} variable.

Perhaps the simplest example of a counterfactual change in $X$ is a location
shift: $X_{\delta}=X+\delta$. The popular method of UQR of
\cite{FirpoFortinLemieux09} can be used to assess the effect of such changes
in the unconditional quantiles of $Y$.\footnote{See Section
\ref{sec:dist_vs_cov} for a discussion about how this paper relates to
\cite{FirpoFortinLemieux09}.} In this paper we provide results for the general
case where $X_{\delta}=\mathcal{G}(X;\delta)$ for some (suitable) policy
function $\mathcal{G}$ chosen by the researcher or policy maker. A
counterfactual change in $X$ to $X_{\delta}=\mathcal{G}(X;\delta)$ induces a
counterfactual outcome $Y_{\delta}=h(\mathcal{G}(X;\delta),W,U)=h(X_{\delta
},W,U)$. Our parameter of interest, the \textit{marginal effect for the }%
$\tau$\textit{-quantile}, is an infinitesimal contrast of unconditional
quantiles and is defined as
\begin{equation}
\Pi_{\tau}:=\lim_{\delta\rightarrow0}\frac{Q_{\tau}[Y_{\delta}]-Q_{\tau}%
[Y]}{\delta}, \label{eq:pi_tau}%
\end{equation}
whenever this limit exists.

A particular policy function that we analyze in detail is the following
\emph{location-scale shift} in $X$
\begin{equation}
X_{\delta}=\mathcal{G}(X;\delta)=\left(  X-\mu\right)  s(\delta)+\mu
+\ell(\delta). \label{eq_location_scale}%
\end{equation}
Here, $\mu$ is a \textit{known} policy parameter, and we refer to $\ell
(\delta)$ as the location shift and to $s(\delta)>0$ as the scale shift. In
order to take limits to find $\Pi_{\tau}$, we assume that $\ell(\delta)$ and
$s(\delta)$ are continuously differentiable functions of the scalar $\delta$.
Both $\ell(\delta)$ and $s(\delta)$ are chosen by the researcher or policy
maker subject to the restriction that $s(0)=1$ and $\ell(0)=0$. Note that this
choice of $\mathcal{G}$ nests the case $X_{\delta}=X+\delta$ by choosing
$\ell(\delta)=\delta$ and $s(\delta)\equiv1$.

A distinctive feature of $X_{\delta}$ in \eqref{eq_location_scale} is that
\[
var[X_{\delta}]=s(\delta)^{2}var[X],
\]
and so it allows for the study of counterfactual changes in the dispersion of
the target variable. To see this, suppose that $s(\delta)<1$, then,
realizations of $X$ that are above/below $\mu$ are \textquotedblleft
moved\textquotedblright\ towards $\mu$, followed by a location shift of
$\ell(\delta)$. Therefore, we have a constant location shift, given by
$\ell(\delta)$, and a relative location shift induced by the scale shift,
which tends to bunch observations near $\mu$. The result is a reduction of the
variance of $X$. If, on the other hand, $s(\delta)>1$, then the counterfactual
policy moves $X$ away from $\mu$ and consequently increases its variance.

Under some regularity assumptions spelled below, the marginal effect
$\Pi_{\tau}$ corresponding to the policy function $\mathcal{G}$ given in
\eqref{eq_location_scale} can be decomposed into the sum of two effects: one
associated with the location shift governed by $\ell(\delta)$, and the other
associated with the scale shift $s(\delta)$. The former corresponds to a
version of the estimand studied by \cite{FirpoFortinLemieux09}. The latter
effect is, to the best of our knowledge, new.

Subsection \ref{subsection_general_policy} contains a rigorous development of
our main results for a general policy function. Readers interested in the
location-scale shift only can skip subsection \ref{subsection_general_policy}
and focus on subsections \ref{subsection_location_scale_policy} and
\ref{Sec: elasticity} where we provide the specific results for the
location-scale shift, discuss their interpretations, and offer examples.

\subsection{Results for a general policy function}

\label{subsection_general_policy}

Central to our results is the counterfactual policy function $\mathcal{G}$,
which maps $X$ to $X_{\delta}$ and generates a counterfactual outcome
$Y_{\delta}$. As mentioned before, our parameter of interest $\Pi_{\tau}$
given in \eqref{eq:pi_tau}, compares the quantiles of
\begin{equation}
Y=h(X,W,U) \label{eq_model_0}%
\end{equation}
to the quantiles of
\begin{equation}
Y_{\delta}=h(X_{\delta},W,U)=h(\mathcal{G}(X;\delta),W,U). \label{eq_model_1}%
\end{equation}

An important assumption is that the distribution of $\left(  X,W,U\right)  $
in (\ref{eq_model_1}) is held the same as that in (\ref{eq_model_0}). To
understand the latter condition, we can consider two parallel worlds: the
worlds before and after the intervention. For each given $\delta,$ let
$\mathcal{G}^{-1}(x;\delta)$ be the inverse function of $\mathcal{G}%
(x;\delta)$ such that $\mathcal{G}(\mathcal{G}^{-1}(x;\delta);\delta)=x.$
After applying the inverse transform to the target covariate in the
post-intervention world, the distribution of $\left(  \mathcal{G}%
^{-1}\mathcal{(}X^{\delta};\delta),W^{\delta},U^{\delta}\right)  $ in the
post-intervention world is assumed to be the same as that of $\left(
X,W,U\right)  $ in the pre-intervention world. Here, no change is induced on
$W$ and $U$ and so $\left(  W^{\delta},U^{\delta}\right)  $ is actually the
same as $(W,U)$ for every individual in the population.
In essence, we keep the structural function $h\left(  \cdot,\cdot
,\cdot\right)  $ and the distribution of $\left(  X,W,U\right)  $ intact
during the policy intervention. The effect under consideration is then the
policy effect due to the policy intervention only and thus has a \emph{ceteris
paribus} causal interpretation.

For notational economy, we write $x^{\delta}=\mathcal{G}^{-1}(x;\delta)$. Then
$X_{\delta}=x$ if and only if $X=x^{\delta}.$ Define the Jacobian of the
inverse transform $x\mapsto x^{\delta}:=\mathcal{G}^{-1}(x;\delta)$ as%
\[
J(x^{\delta};\delta):=\frac{\partial x^{\delta}}{\partial x}=\left[
\frac{\partial\mathcal{G}\left(  x;\delta\right)  }{\partial x}\right]
^{-1}\bigg|_{x=x^{\delta}}.
\]
Then, the joint probability density functions\ of the covariate vector before
and after the intervention satisfy
\[
f_{X_{\delta},W}(x,w)=J(x^{\delta};\delta)\cdot f_{X,W}(x^{\delta},w).
\]

For $\varepsilon>0$, define $\mathcal{N}_{\varepsilon}:=\left\{
\delta:\left\vert \delta\right\vert \leq\varepsilon\right\}  $. We maintain
the following assumption.

\begin{assumption}
\label{Assumption:main} (i.a) For some $\varepsilon>0,$ $\mathcal{G}\left(
x;\delta\right)  $ is continuously differentiable on $\mathcal{X\otimes
N}_{\varepsilon}$, where $\mathcal{X}$ is the support of $X.$

(i.b) $\mathcal{G}\left(  x;\delta\right)  $ is strictly increasing in $x$ for
each $\delta\in\mathcal{N}_{\varepsilon}.$

(i.c) $\mathcal{G}\left(  x;0\right)  =x$ for all $x\in\mathcal{X}$.

(ii) for $\delta\in\mathcal{N}_{\varepsilon}$, the conditional density of $U$
satisfies $f_{U|X_{\delta},W}(u|x,w)=f_{U|X,W}(u|x^{\delta},w)$, and the
support $\mathcal{U}$ of $U$ given $X$ and $W$ does not depend on $\left(
X,W\right)  .$

(iii.a) $x\mapsto f_{X,W}(x,w)$ is continuously differentiable for all
$w\in\mathcal{W}$ and
\[
\int_{\mathcal{W}}\int_{\mathcal{X}}\sup_{\delta\in\mathcal{N}_{\varepsilon}%
}\left\vert \frac{\partial\left[  J\left(  x^{\delta};\delta\right)
f_{X,W}(x^{\delta},w)\right]  }{\partial\delta}\right\vert dxdw<\infty
\]
where $\mathcal{W}$ is the support of $W.$

(iii.b) $x\mapsto f_{U|X,W}(u|x,w)$ is continuously differentiable for all
$\left(  u,w\right)  $ and
\begin{align*}
\int_{\mathcal{W}}\int_{\mathcal{X}}\int_{\mathcal{U}}\sup_{\delta
\in\mathcal{N}_{\varepsilon}}\left\vert \frac{\partial}{\partial\delta}\left[
f_{U|X,W}(u|x^{\delta},w)f_{X,W}(x^{\delta},w)\right]  \right\vert dudxdw  &
<\infty,\\
\int_{\mathcal{W}}\int_{\mathcal{X}}\int_{\mathcal{U}}\sup_{\delta
\in\mathcal{N}_{\varepsilon}}\left\vert \frac{\partial f_{X,W}(x^{\delta}%
,w)}{\partial\delta}\right\vert f_{U|X,W}(u|x,w)dudxdw  &  <\infty.
\end{align*}

(iv) $f_{X,W}(x,w)$ is equal to $0$ on the boundary of the support of $X$
given $W=w$ for all $w\in\mathcal{W}.$

(v) $f_{Y}(Q_{\tau}[Y])>0.$
\end{assumption}

\begin{remark}
Assumption \ref{Assumption:main}(i) imposes some restrictions on the policy
function $\mathcal{G}\left(  x;\delta\right)  .$ It is reasonable that
$\mathcal{G}\left(  x;\delta\right)  $ is strictly increasing in $x,$ as a
non-monotonic and non-invertible function does not seem to be practically
relevant. The strictly increasing property implies that $J\left(
x;\delta\right)  >0$ for all $x\in\mathcal{X}$ and $\delta\in\mathcal{N}%
_{\varepsilon}.$ The condition that $\mathcal{G}\left(  x;0\right)  =x$ says
that there is no intervention when $\delta=0,$ and it implies that $J\left(
x;0\right)  =1$ for all $x\in\mathcal{X}.$ Assumption \ref{Assumption:main}%
(ii) assumes that how $U$ depends on the covariate vector is maintained when
we induce a change in the covariate vector. Note that Assumption
\ref{Assumption:main}(ii) is different from $f_{U|X_{\delta},W}%
(u|x,w)=f_{U|X,W}(u|x,w)$, which in general cannot hold when $U$ depends on
$X$ and $W$. The counterfactual model in (\ref{eq_model_1}) says that we
maintain the structure of the causal system. Assumption \ref{Assumption:main}%
(ii) says that we also maintain how the unobservable depends on the
observables. As discussed above, we also implicitly assume that $\left(
\mathcal{G}^{-1}\mathcal{(}X^{\delta};\delta),W^{\delta}\right)  $ has the
same distribution as $\left(  X,W\right)  .$ The rest of Assumption
\ref{Assumption:main} consists of regularity conditions.
\end{remark}

\begin{remark}
Assumption \ref{Assumption:main} does not assume that $U$ is independent of
$\left(  X,W\right)  .$ It does not assume that $U$ is conditionally
independent of $X$ given $W$ either. Assumption \ref{Assumption ID} below will
impose identification assumptions.
\end{remark}

The following theorem characterizes the effects of the policy change on the
distribution of $Y_{\delta}$ and its quantiles.

\begin{theorem}
\label{th:uqpe_scale} Let Assumption \ref{Assumption:main} hold.

(i) For each $\left(  x,w\right)  \in\mathcal{X}\otimes\mathcal{W},$%
\[
\lim_{\delta\rightarrow0}\frac{f_{X_{\delta},W}(x,w)-f_{X,W}(x,w)}{\delta
}=-\frac{\partial}{\partial x}\left[  \kappa\left(  x\right)  f_{X,W}%
(x,w)\right]  ,
\]
where
\[
\kappa\left(  x\right)  :=\frac{\partial\mathcal{G}(x;\delta)}{\partial\delta
}\bigg|_{\delta=0}.
\]

(ii) As $\delta\rightarrow0$, we have
\begin{align*}
&  \frac{F_{Y_{\delta}}(y)-F_{Y}\left(  y\right)  }{\delta}\\
&  \rightarrow E\left[  \left(  \frac{\partial F_{Y|X,W}(y|X,W)}{\partial
X}-\mathds1\left\{  h(X,W,U)\leq y\right\}  \frac{\partial\ln f_{U|X,W}%
(U|X,W)}{\partial X}\right)  \kappa\left(  X\right)  \right]
\end{align*}
uniformly in $y\in\mathcal{Y}$, the support of $Y$.

(iii) The marginal effect of the intervention $X_{\delta}=\mathcal{G}%
(X;\delta)$ on the $\tau$-quantile of the outcome variable $Y$ can be
represented by
\begin{equation}
\Pi_{\tau}=A_{\tau}-B_{\tau} \label{Pi_theorem}%
\end{equation}
where%
\begin{align*}
A_{\tau}  &  =E\left[  \frac{\partial E\left[  \psi\left(  Y,\tau
,F_{Y}\right)  |X,W\right]  }{\partial X}\kappa\left(  X\right)  \right]  ,\\
B_{\tau}  &  =E\left[  \psi\left(  Y,\tau,F_{Y}\right)  \frac{\partial\ln
f_{U|X,W}(U|X,W)}{\partial X}\kappa\left(  X\right)  \right]  ,
\end{align*}
and
\[
\psi\left(  y,\tau,F_{Y}\right)  =\frac{\tau-1\left(  y<Q_{\tau}[Y]\right)
}{f_{Y}(Q_{\tau}[Y])}.
\]

\end{theorem}

\begin{remark}
To understand Theorem \ref{th:uqpe_scale}(i), we can write%
\[
f_{X_{\delta},W}(x,w)-f_{X,W}(x,w)=f_{X_{\delta},W}(x,w)-f_{X,W}(x^{\delta
},w)+f_{X,W}(x^{\delta},w)-f_{X,W}(x,w).
\]
It is quite intuitive that the second term is approximately $\delta\cdot
\frac{\partial x^{\delta}}{\partial\delta}|_{\delta=0}\cdot\frac{\partial
f_{X,W}(x,w)}{\partial x}=-\delta\cdot\kappa\left(  x\right)  \cdot
\frac{\partial f_{X,W}(x,w)}{\partial x}$ when $\delta$ is small. Here we have
used the result that $\kappa\left(  x\right)  $ also equals $-\frac{\partial
x^{\delta}}{\partial\delta}|_{\delta=0}$ (see the proof of Theorem
\ref{th:uqpe_scale} in the appendix). The first term reflects the effect from
the Jacobian of the transformation. Indeed, $f_{X_{\delta},W}(x,w)-f_{X,W}%
(x^{\delta},w)=\left[  J(x^{\delta};\delta)-J(x^{\delta};0)\right]
f_{X,W}(x^{\delta},w)$ as $J(x^{\delta};0)=1.$ The first term is then
approximately equal to $\delta\cdot f_{X,W}(x,w)\cdot\frac{\partial J\left(
x,\delta\right)  }{\partial\delta}\big|_{\delta=0}.$ But
\[
\frac{\partial J\left(  x,\delta\right)  }{\partial\delta}\bigg|_{\delta
=0}=\frac{\partial}{\partial\delta}J\left(  x,\delta\right)  \bigg|_{\delta
=0}=\frac{\partial}{\partial\delta}\frac{\partial x^{\delta}}{\partial
x}\bigg|_{\delta=0}=\frac{\partial}{\partial x}\frac{\partial x^{\delta}%
}{\partial\delta}\bigg|_{\delta=0}=-\frac{\partial\kappa\left(  x\right)
}{\partial x},
\]
and hence the first term is approximately $-\delta\cdot f_{X,W}(x,w)\cdot
\frac{\partial\kappa\left(  x\right)  }{\partial x}.$ Combining these two
approximations yields Theorem \ref{th:uqpe_scale}(i).
\end{remark}

\begin{remark}
By definition, $\kappa\left(  x\right)  $ measures the marginal change of
$\mathcal{G}(x;\delta)$ as we increase $\delta$ from zero infinitesimally.
Theorems \ref{th:uqpe_scale} (ii) and (iii) show that only $\kappa\left(
x\right)  $ appears in the marginal effect and the Jacobian does not. This is
not surprising, as what matters for the marginal effect is the marginal change
in the policy function.
\end{remark}

\begin{remark}
Theorem \ref{th:uqpe_scale}(iii) represents the structural parameter
$\Pi_{\tau}$ in terms of statistical objects. While the first term $A_{\tau}$
is identifiable, the second term $B_{\tau}$, which involves the conditional
density of $U$ given $X$ and $W,$ is not. If we use $\hat{A}_{\tau},$ a
consistent estimator of $A_{\tau},$ as an estimator of $\Pi_{\tau},$ then the
second term $B_{\tau}$ is the asymptotic bias of $\hat{A}_{\tau}.$ This bias
is an endogeneity bias, as it is in general not equal to zero when $X$ is not
independent of $U$ (conditioning on $W$). Similar results have been
established in \cite{yixiao2021} but only for location shifts. If we do not
have the identification condition such as what is given in Assumption
\ref{Assumption ID} below, Theorem \ref{th:uqpe_scale}(iii) allows us to use a
bound approach to bound $B_{\tau}$ and infer the range of the policy effect or
conduct a sensitivity analysis similar to that in \cite{martinez2020}.
\end{remark}

\begin{remark}
\label{Remark: more general functional}While the paper focuses on the quantile
functional, Theorem \ref{th:uqpe_scale}(iii) is formulated in a general way.
The result holds for any Hadamard differentiable functional and for the mean
functional. We only need to replace $\psi\left(  y,\tau,F_{Y}\right)  $ by the
influence function of the functional that we are interested in. For example,
for the mean functional, we can replace $\psi\left(  y,\tau,F_{Y}\right)  $ by
$y-E(Y)$, and Theorem \ref{th:uqpe_scale}(iii) remains valid.
\end{remark}

To identify $\Pi_{\tau},$ we make the following independence or conditional
independence assumption.


\begin{assumption}
\label{Assumption ID}For $\delta\in\mathcal{N}_{\varepsilon}$, the
unobservable $U$ satisfies either $f_{U|X,W}(u|x,w)=f_{U|X,W}(u|x^{\delta
},w)=f_{U}(u)$ or $f_{U|X,W}(u|x,w)=f_{U|X,W}(u|x^{\delta},w)=f_{U|W}(u|w).$
\end{assumption}

Under the above assumption, $\partial\ln f_{U|X,W}(u|x,w)/\partial x=0$ and
the second term $B_{\tau}$ in (\ref{Pi_theorem}) vanishes. In this case,
$\Pi_{\tau}=A_{\tau}$ and hence is identified. The corollary below then
follows directly from Theorem \ref{th:uqpe_scale}(iii).

\begin{corollary}
\label{Corrollary:uqpe_scale}Let Assumption \ref{Assumption:main} hold with
Assumption \ref{Assumption:main} (ii) strengthened to Assumption
\ref{Assumption ID}. Then \
\begin{equation}
\Pi_{\tau}=E\left[  \frac{\partial E\left[  \psi\left(  Y,\tau,F_{Y}\right)
|X,W\right]  }{\partial X}\kappa\left(  X\right)  \right]  =\frac{1}%
{f_{Y}(Q_{\tau}[Y])}E\left[  \frac{\partial\mathcal{S}_{Y|X,W}\left(  Q_{\tau
}[Y]|X,W\right)  }{\partial X}\kappa\left(  X\right)  \right]
\label{eq_pi_tau_general}%
\end{equation}
where $\mathcal{S}_{Y|X,W}\left(  \cdot|x,w\right)  :=1-F_{Y|X,W}\left(
\cdot|x,w\right)  $ is the\ conditional survival function.
\end{corollary}

\begin{remark}
Both conditions in Assumption \ref{Assumption ID} require that $f_{U|X,W}%
(u|x,w)=f_{U|X,W}(u|x^{\delta},w)$. This is related to the assumption in
\citet[][pp.955-957]{FirpoFortinLemieux09}, framed as \textquotedblleft
maintaining the conditional distribution of Y given X
unaffected.\textquotedblright\ In essence, \citet{FirpoFortinLemieux09}
requires\ $f_{U|X}(u|x)=f_{U|X}(u|x^{\delta}).$ When this condition fails, we
may still have $f_{U|X,W}(u|x,w)=f_{U|X,W}(u|x^{\delta},w).$ Such a condition
has also been used in \cite{Lieli2020} and \cite{Spini2021} in a context of
extrapolation to populations with different distributions of the covariates.
\end{remark}

\begin{remark}
\label{control_variable} The first condition in Assumption \ref{Assumption ID}
is satisfied if $U$ is independent of $(X,W)$.
In our view, this condition is hard to achieve in empirical applications. The
second condition in Assumption \ref{Assumption ID}, which is commonly used to
achieve identification in applied work, is a conditional independence
assumption. Such a condition is often referred to as the unconfoundedness
condition in the causal inference literature. {The assumption is more general
than $Y(x)\perp X|W$ for any }$x\in\mathcal{X}.${ It is a \textquotedblleft
local\textquotedblright\ unconfoundedness condition in the sense that
$Y(x)|\left(  X,W\right)  =_{d}Y(x^{\delta})|$}$\left(  {X,W}\right)  ${ for
$\delta\in\mathcal{N}_{\varepsilon}$, a small $\varepsilon$-radius
neighborhood around 0. A more stringent condition would require $Y(x)|\left(
X,W\right)  =_{d}Y(\tilde{x})|$}$\left(  {X,W}\right)  ${ for any $x,\tilde
{x}\in$}$\mathcal{X}${. }


\end{remark}


We note in passing that Corollary \ref{Corrollary:uqpe_scale} has the
following alternative representation:
\[
\Pi_{\tau}=\left\langle E\left[  \frac{\partial E\left[  \psi\left(
Y,\tau,F_{Y}\right)  |X,W\right]  }{\partial X}\bigg| X\right]  ,\frac
{\partial\mathcal{G}(X;\delta)}{\partial\delta}\bigg|_{\delta=0}\right\rangle
,
\]
where $\left\langle \cdot,\cdot\right\rangle $ is the inner product defined by
$\left\langle h\left(  X\right)  ,g\left(  X\right)  \right\rangle :=E\left[
h\left(  X\right)  g\left(  X\right)  \right]  $ in the space $\mathcal{L}%
_{2}\left(  X\right)  .$ By the Cauchy-Schwarz inequality, $|\langle h\left(
X\right)  ,g\left(  X\right)  \rangle|\leq\left\Vert h\left(  X\right)
\right\Vert \left\Vert g\left(  X\right)  \right\Vert $, where $\left\Vert
\cdot\right\Vert $ is the norm defined by $\left\Vert h\left(  X\right)
\right\Vert :=\sqrt{\left\langle h\left(  X\right)  , h\left(  X\right)
\right\rangle }$. Consider the class of policy functions with a unit norm,
namely $\left\Vert \frac{\partial\mathcal{G}(X;\delta)}{\partial\delta
}\bigg|_{\delta=0}\right\Vert =1.$ Then
\[
|\Pi_{\tau}|\leq\left\Vert E\left[  \frac{\partial E\left[  \psi\left(
Y,\tau,F_{Y}\right)  |X,W\right]  }{\partial X}\bigg|X\right]  \right\Vert .
\]
Thus, if a policy function satisfies
\[
\frac{\partial\mathcal{G}(X;\delta)}{\partial\delta}\bigg|_{\delta=0}=E\left[
\frac{\partial E\left[  \psi\left(  Y,\tau,F_{Y}\right)  |X,W\right]
}{\partial X}\bigg|X\right]  \cdot\left\Vert E\left[  \frac{\partial E\left[
\psi\left(  Y,\tau,F_{Y}\right)  |X,W\right]  }{\partial X}\bigg|X\right]
\right\Vert ^{-1},
\]
then it achieves the highest $\Pi_{\tau}$ (in magnitude) in this class. We
leave optimal policy designs based on a cost-benefit analysis for future research.

\subsection{Results for the location-scale shift}

\label{subsection_location_scale_policy}

In this subsection we obtain a representation for $\Pi_{\tau}$ for the
particular case of the location-scale shift given in (\ref{eq_location_scale}%
):
\[
X_{\delta}=\mathcal{G}(X;\delta) = \left(  X-\mu\right)  s(\delta)+\mu
+\ell(\delta).
\]
The corollary below also follows directly from Theorem \ref{th:uqpe_scale}(iii).

\begin{corollary}
\label{Corrollary:uqpe_scale_2}Let Assumption \ref{Assumption:main} hold with
Assumption \ref{Assumption:main} (ii) strengthened to Assumption
\ref{Assumption ID}. Then, for the location-scale shift in
(\ref{eq_location_scale}) with $\ell(0)=0,$ $s(0)=1,$ and $s(\delta)>0,$ the
marginal effect can be decomposed as
\begin{equation}
\Pi_{\tau}=\Pi_{\tau,L}+\Pi_{\tau,S}, \label{eq_pi_tau}%
\end{equation}
where%
\begin{align*}
\Pi_{\tau,L}  &  =\frac{\dot{\ell}\left(  0\right)  }{f_{Y}(Q_{\tau}[Y])}%
\int_{\mathcal{W}}\int_{\mathcal{X}}\frac{\partial\mathcal{S}_{Y|X,W}(Q_{\tau
}[Y]|x,w)}{\partial x}f_{X,W}(x,w)dxdw,\\
\Pi_{\tau,S}  &  =\frac{\dot{s}\left(  0\right)  }{f_{Y}(Q_{\tau}[Y])}%
\int_{\mathcal{W}}\int_{\mathcal{X}}\frac{\partial\mathcal{S}_{Y|X,W}(Q_{\tau
}[Y]|x,w)}{\partial x}\left(  x-\mu\right)  f_{X,W}(x,w)dxdw,
\end{align*}
and $\mathcal{S}_{Y|X,W}\left(  \cdot|x,w\right)  :=1-F_{Y|X,W}\left(
\cdot|x,w\right)  $ is the\ conditional survival function.
\end{corollary}

Corollary \ref{Corrollary:uqpe_scale_2} shows that the overall effect
$\Pi_{\tau}$ can be decomposed into the sum of $\Pi_{\tau,L}$ and $\Pi
_{\tau,S}.$ Here $\Pi_{\tau,L}$ is the location effect and is the estimand in
\cite{FirpoFortinLemieux09} when we set $\dot{\ell}(0)=1$ and $s\left(
\delta\right)  \equiv1$. $\Pi_{\tau,S}$ is the scale effect and is present
whenever $s(\delta)$ is not identically 1 and $\dot{s}\left(  0\right)  \neq
0$.\footnote{It can be seen that $\Pi_{\tau,S}$ depends on $\mu$. However, we
suppress this dependence from the notation for simplicity.}


To better understand the location and scale effects in Corollary
\ref{Corrollary:uqpe_scale_2}, consider the case that $X$ and $U$ are
independent and there is no $W.$ Then
\begin{align}
\Pi_{\tau,L}  &  =\frac{\dot{\ell}\left(  0\right)  }{f_{Y}(Q_{\tau}[Y])}%
\int_{\mathcal{X}}\frac{\partial\mathcal{S}_{Y|X}(Q_{\tau}[Y]|x)}{\partial
x}f_{X}(x)dx,\label{Pi_L_independence_case}\\
\Pi_{\tau,S}  &  =\frac{\dot{s}\left(  0\right)  }{f_{Y}(Q_{\tau}[Y])}%
\int_{\mathcal{X}}\frac{\partial\mathcal{S}_{Y|X}(Q_{\tau}[Y]|x)}{\partial
x}\left(  x-\mu\right)  f_{X}(x)dx.\nonumber
\end{align}
To sign the location effect $\Pi_{\tau,L}$, we can assess whether
$\mathcal{S}_{Y|X}(Q_{\tau}[Y]|x)$ is increasing in $x$ or not. If $\dot{\ell
}\left(  0\right)  \geq0$ and $\mathcal{S}_{Y|X}(Q_{\tau}[Y]|x)$ is increasing
in $x$ on average, more precisely, $\int_{\mathcal{X}}\frac{\partial
\mathcal{S}_{Y|X}(Q_{\tau}[Y]|x)}{\partial x}f_{X}(x)dx\geq0,$ then $\Pi
_{\tau,L}\geq0.$ As an example, consider the case that $h\left(  x,u\right)  $
is increasing in $x$ for each $u.$ Then, $\mathcal{S}_{Y|X}(Q_{\tau}[Y]|x)$ is
increasing in $x$ for all $x\in\mathcal{X}$, and so $\Pi_{\tau,L}\geq0$ if
$\dot{\ell}\left(  0\right)  \geq0.$

It is a bit more challenging to determine the sign of the scale effect
$\Pi_{\tau,S}$, which depends on, not only the function form of $\frac
{\partial\mathcal{S}_{Y|X}(Q_{\tau}[Y]|x)}{\partial x}$, but also the
distribution of $X.$ The next example provides some insight into the scale effect.

\begin{example}
\textbf{Normal Covariate.}\label{example_stein} Consider the linear model
$Y=\lambda+X\gamma+U$ where $X$ and $U$ are independent and $X\sim N(\mu
_{X},\sigma_{X}^{2})$. We can use Stein's lemma (see, for example,
\citet[][pp.124-125]{casella} and references therein) to gain some insight
into the scale effect. The lemma states that for a differentiable function
$m\left(  \cdot\right)  $ such that $E[|m^{\prime}(X)|]<\infty$,
$E[m(X)(X-\mu_{X})]=\sigma^{2}E[m^{\prime}(X)]$ whenever $X\sim N(\mu
_{X},\sigma_{X}^{2})$. Taking $m(x)=\partial\mathcal{S}_{Y|X}(Q_{\tau
}[Y]|x)/\partial x$ and using Stein's lemma, we can express the scale effect
for $\mu=\mu_{X}$ as
\[
\Pi_{\tau,S}=\frac{\dot{s}\left(  0\right)  }{f_{Y}(Q_{\tau}[Y])}E\left[
\frac{\partial\mathcal{S}_{Y|X}(Q_{\tau}[Y]|X)}{\partial X}\left(  X-\mu
_{X}\right)  \right]  =\frac{\dot{s}\left(  0\right)  \sigma_{X}^{2}}%
{f_{Y}(Q_{\tau}[Y])}E\left[  \frac{\partial^{2}\mathcal{S}_{Y|X}(Q_{\tau
}[Y]|X)}{\partial X^{2}}\right]  .
\]
Therefore, when $X$ is normal and $\dot{s}(0)>0,$ the scale effect is
non-negative (non-positive) if $\mathcal{S}_{Y|X}(Q_{\tau}[Y]|x)$ is a convex
(concave) function of $x$. It is interesting to see that the location effect
depends on the first order derivative of $\mathcal{S}_{Y|X}(Q_{\tau}[Y]|x)$
(see equation (\ref{Pi_L_independence_case})) while the scale effect depends
on its second-order derivative.
\end{example}

In the next example, we simplify $\Pi_{\tau,S}$ under the additional
assumption that $U$ is also normal.

\begin{example}
\textbf{Normal Covariate and Normal Noise. }%
\label{example_normal_location_model} Consider a linear model $Y=\lambda
+X\gamma+U$ where $X$ and $U$ are independent. We have: $\Pi_{\tau,L}%
=\dot{\ell}\left(  0\right)  \gamma$. In addition to the normal covariate
assumption $X\sim N(\mu_{X},\sigma^{2}_{X})$, suppose $U$ is also normal
$U\sim N(0,\sigma_{U}^{2})$. Then, for $\mu=\mu_{X},$
\[
\Pi_{\tau,S}=\dot{s}\left(  0\right)  \sqrt{R_{YX}^{2}}Q_{\tau}[X_{\gamma
}^{\circ}]
\]
where $R_{YX}^{2}$ is the population R-squared defined by $R_{YX}%
^{2}:=var(\lambda+X\gamma)/var(Y)$ and $X_{\gamma}^{\circ}=\left(  X-\mu
_{X}\right)  \gamma.$ While the location effect is constant across quantiles,
the scale effect varies across quantiles.

\end{example}

In Example \ref{example_normal_location_model}, the scale effect, when
$\mu=\mu_{X}$, does not depend on $\mu_{X}$ or \textrm{sign}$\left(
\gamma\right)  $. To understand this and obtain a more general result, we note
that $\Pi_{\tau,S}$ is proportional to the following covariance:
\begin{align}
cov\left(  \frac{\partial\mathcal{S}_{Y|X}(Q_{\tau}[Y]|X)}{\partial
X},X\right)   &  =cov\left(  f_{U}(Q_{\tau}[Y]-\delta-X\gamma),X\right)
\gamma\nonumber\\
&  =cov\left(  f_{U}(Q_{\tau}[U^{\circ}]-X_{\gamma}^{\circ}),X_{\gamma}%
^{\circ}\right)  , \label{signing_covariance}%
\end{align}
where $U^{\circ}:=U+X_{\gamma}^{\circ}$ and we have used%
\begin{align*}
Q_{\tau}[Y]  &  =Q_{\tau}[U+\delta+X\gamma]=Q_{\tau}\left[  U+\left(
X-\mu_{X}\right)  \gamma+\delta+\mu_{X}\gamma\right] \\
&  =Q_{\tau}[U^{\circ}]+\delta+\mu_{X}\gamma.
\end{align*}
Now, if $X-\mu_{X}$ is symmetrically distributed around zero, then $X_{\gamma
}^{\circ}$ also shares this property. In this case, the covariance in
(\ref{signing_covariance}) does not depend on \textrm{sign}$\left(
\gamma\right)  $ as the distributions of $X_{\gamma}^{\circ}$ and $U^{\circ}$
remain the same if we flip the sign of $\gamma.$ Also, since the distributions
of $X_{\gamma}^{\circ}$ and $U^{\circ}$ do not depend on $\mu_{X},$ the
covariance in (\ref{signing_covariance}) does not depend on $\mu_{X}.$ On the
other hand, for the denominator of the scale effect, we have%
\[
f_{Y}(Q_{\tau}[Y])=f_{Y}(Q_{\tau}[U^{\circ}]+\delta+\mu_{X}\gamma
)=f_{U^{\circ}}\left(  Q_{\tau}[U^{\circ}]\right)  .
\]
If $X-\mu_{X}$ is symmetrically distributed around zero, then the distribution
of $U^{\circ}$ does not depend on $\mu_{X}$ or \textrm{sign}$\left(
\gamma\right)  .$ Hence, $f_{Y}(Q_{\tau}[Y])$ does not depend on $\mu_{X}$ or
\textrm{sign}$\left(  \gamma\right)  .$

Since both the numerator and the denominator of $\Pi_{\tau,S}$ are invariant
to $\mu_{X}$ and \textrm{sign}$\left(  \gamma\right)  $, we obtain the
following proposition immediately.

\begin{proposition}
Consider the linear model $Y=\lambda+X\gamma+U$ where $X$ and $U$ are
independent. If $X-E\left[  X\right]  $ is symmetrically distributed around
zero, then the scale effect computed for $\mu=E\left[  X\right]  $ does not
depend on either $E\left[  X\right]  $ or \textrm{sign}$\left(  \gamma\right)
.$
\end{proposition}

\subsection{Interpretation of the scale effects\label{Sec: elasticity}}

Consider a situation where we only care about the scale effect, that is, we
set $\ell(\delta)\equiv0$. Then, we have $X_{\delta}=\mu+\left(  X-\mu\right)
s(\delta)$. If we denote by $\sigma_{X}$ and $\sigma_{X_{\delta}}$ the
standard deviations of $X$ and $X_{\delta}$, respectively, then $\sigma
_{X_{\delta}}=\sigma_{X}s\left(  \delta\right)  $. To interpret $\Pi_{\tau,S}%
$, we assume $Q_{\tau}[Y_{\delta}]\neq0$ and consider the following
\textit{quantile-standard deviation elasticity}
\[
\mathcal{E}_{\tau,\delta}:=\frac{dQ_{\tau}[Y_{\delta}]}{Q_{\tau}[Y_{\delta}%
]}\left(  \frac{d\sigma_{X_{\delta}}}{\sigma_{X_{\delta}}}\right)  ^{-1}.
\]
By straightforward calculations, we have%
\[
\mathcal{E}_{\tau,\delta}=\frac{1}{Q_{\tau}[Y]}\frac{dQ_{\tau}[Y_{\delta}%
]}{d\delta}\left(  \frac{1}{s\left(  \delta\right)  }\frac{ds\left(
\delta\right)  }{d\delta}\right)  ^{-1}.
\]
When $s(0)=1$ and $\dot{s}\left(  0\right)  \neq0$, the elasticity at
$\delta=0$ is
\begin{equation}
\mathcal{E}_{\tau,\delta=0}=\frac{\Pi_{\tau,S}}{\dot{s}\left(  0\right)
Q_{\tau}[Y]}. \label{eq:elasticity_0}%
\end{equation}
Therefore, a $1\%$ increase in the standard deviation of $X$ results in a
$\Pi_{\tau,S}/\{\dot{s}\left(  0\right)  Q_{\tau}[Y]\}\%$ change in the $\tau
$-quantile of $Y$.

\addtocounter{example}{-1}

\begin{example}
[Continued]Plugging $\Pi_{\tau,S}=\dot{s}\left(  0\right)  \sqrt{R_{YX}^{2}%
}Q_{\tau}[X_{\gamma}^{\circ}]$, we obtain the \textit{quantile-standard
deviation elasticity} at $\delta=0$ as
\[
\mathcal{E}_{\tau,\delta=0}=\sqrt{R_{YX}^{2}}\frac{Q_{\tau}[X_{\gamma}^{\circ
}]}{Q_{\tau}[Y]}.
\]
So, $\mathcal{E}_{\tau,\delta=0}$ is positive if $Q_{\tau}[X_{\gamma}^{\circ
}]$ and $Q_{\tau}[Y]$ have the same sign. When $\alpha=0$, $\mu_{X}=0$, and
$X$ and $U$ are independent normals, we have $Q_{\tau}[X_{\gamma}^{\circ
}]/Q_{\tau}[Y]=\sqrt{R_{YX}^{2}}$ and so $\mathcal{E}_{\tau,\delta=0}%
=R_{YX}^{2}$. Interestingly, the \textit{quantile-standard deviation
elasticity} is equal to the population R-squared for all quantile levels.
\end{example}

Often times, when the outcome of interest (e.g., price and wage) is strictly
positive, we are interested in $\log Y$. In such a case, we denote the scale
effect by $\tilde{\Pi}_{\tau,S}$. Since we set $\ell(\delta)\equiv0$ and there
is no location effect, the new scale effect is given by
\[
\tilde{\Pi}_{\tau,S}:=\lim_{\delta\rightarrow0}\frac{Q_{\tau}[\log Y_{\delta
}]-Q_{\tau}[\log Y]}{\delta}.
\]
Since $\log\left(  \cdot\right)  $ is a strictly increasing transformation, we
have
\[
\tilde{\Pi}_{\tau,S}=\lim_{\delta\rightarrow0}\frac{\log Q_{\tau}[Y_{\delta
}]-\log Q_{\tau}[Y]}{\delta},
\]
and we can relate $\tilde{\Pi}_{\tau,S}$ to ${\Pi}_{\tau,S}$ by
\[
\tilde{\Pi}_{\tau,S}=\frac{1}{Q_{\tau}[Y]}{\Pi}_{\tau,S}.
\]
Comparing this to \eqref{eq:elasticity_0}, we obtain that the elasticity at
$\delta=0$ is
\[
\mathcal{E}_{\tau,\delta=0}=\frac{\tilde{\Pi}_{\tau,S}}{\dot{s}(0)}.
\]
This says that a $1\%$ increase in the standard deviation of $X$ results in a
$\tilde{\Pi}_{\tau,S}/\dot{s}\left(  0\right)  \%$ change in the $\tau
$-quantile of $Y$. When $\dot{s}\left(  0\right)  =1$ (\emph{e.g.}, $s\left(
\delta\right)  =1+\delta),$ the scale effect $\tilde{\Pi}_{\tau,S}$ (based on
$\log\left(  Y)\right)  $ can be interpreted directly as the quantile-standard
deviation elasticity. When $\dot{s}\left(  0\right)  =-1$ (\emph{e.g.},
$s\left(  \delta\right)  =1/(1+\delta)),$ the scale effect $\tilde{\Pi}%
_{\tau,S}$ has the same magnitude as the quantile-standard deviation
elasticity but with an opposite sign.

{\color{red} }

\subsection{Other potential applications}

The framework developed here can be extended in several directions. In the
Supplementary Appendix \ref{comp_appendix}, we consider a case where a
location shift in one covariate is compensated or amplified by a location
shift in another covariate, allowing for simultaneous changes in different covariates.

Our framework is also useful for evaluating heterogeneous interventions.
Specifically, we can accommodate cases where interventions vary across
covariate-specific strata.\footnote{We thank an anonymous referee for
suggesting this possibility.} For instance, a plausible intervention could
involve increasing $X$ among units with $W\in\mathcal{W}_{1}$ and decreasing
$X$ among units with $W\in\mathcal{W}_{2}$ where $\mathcal{W}_{1}$ and
$\mathcal{W}_{2}$ are non-overlapping subsets of $\mathcal{W}.$ One possible
implementation of this is through the following $\mathcal{G}$ function, which
now depends on $W$:%
\[
X_{\delta}=\mathcal{G}(X,W,\delta)=(X+\delta)\mathds1\{W\in\mathcal{W}%
_{1}\}+(X-\delta)\mathds1\{W\in\mathcal{W}_{2}\}.
\]
In this case, individuals with characteristics $W\in\mathcal{W}_{1}$
experience an upperward shift in $X$, while those with $W\in\mathcal{W}_{2}$
experience a downward shift. 

For a general $\mathcal{G}$ function that depends on $W,$ we need to replace
Assumption \ref{Assumption:main}(i) by the following:

\begin{assumption}
\label{Assumption:main_updated}(i.a) For some $\varepsilon>0$ and for each
$w\in\mathcal{W},$ $\mathcal{G}\left(  x,w;\delta\right)  $ is continuously
differentiable in $\left(  x,\delta\right)  $ on $\mathcal{X\otimes
N}_{\varepsilon}$, where $\mathcal{X}$ represents the support of $X$.

(i.b) $\mathcal{G}\left(  x,w;\delta\right)  $ is strictly increasing in $x$
for each $\delta\in\mathcal{N}_{\varepsilon}$ and $w\in\mathcal{W}$. 

(i.c) $\mathcal{G}\left(  x,w;0\right)  =x$ for all $x\in\mathcal{X}$ and
$w\in\mathcal{W}$.
\end{assumption}

With the above assumption in place, we redefine $\kappa\left(  \cdot\right)  $
as
\[
\kappa\left(  x,w\right)  =\frac{\partial\mathcal{G}(x,w;\delta)}%
{\partial\delta}\bigg|_{\delta=0}.
\]
Then Theorem \ref{th:uqpe_scale} remains valid with $\kappa\left(  x\right)  $
replaced by $\kappa\left(  x,w\right)  $ and Assumption \ref{Assumption:main}%
(i) by Assumption\ \ref{Assumption:main_updated}. It is noteworthy that the
differentiability of $\mathcal{G}(x,w;\delta)$ with respect to $w$ is not
required for the theorem to hold. Rather, the previously stated assumptions
need to hold for each $w\in\mathcal{W}$. 

Since $G$ can take a general form, our framework is not only applicable to the
scenarios mentioned above but can also be further extended in other
directions. 

\section{Distribution intervention vs. value intervention}

\label{sec:dist_vs_cov}

The seminal paper by \cite{FirpoFortinLemieux09} (FFL hereafter in this
section) considers the effect of a change in the marginal distribution of $X$
from $F_{X}$ to either $(i)$ a \emph{fixed} $G_{X}$ or $(ii)$ a
\textquotedblleft variable\textquotedblright\ $G_{X,\delta}$ which depends on
$\delta$. \cite{Rothe2012} also focuses on these two cases.

In the first case, FFL considers a change from $F_{X}$ to a \emph{fixed}
$G_{X}$. Keeping $F_{Y|X}$ the same, a counterfactual distribution can be
obtained by $F_{Y}^{\ast}(y)=\int_{\mathcal{X}}F_{Y|X}(y|x)dG_{X}(x)$. For
$\delta\in\lbrack0,1]$, the convex combination $F_{Y,\delta}:=(1-\delta
)F_{Y}+\delta F_{Y}^{\ast}$ is a cdf and can be interpreted as a perturbation
of $F_{Y}$ in the direction of $F_{Y}^{\ast}-F_{Y}$. For a certain statistic
$\rho(F)$ of interest, such as a particular quantile of $Y$, we have
\begin{equation}
\frac{\partial\rho(F_{Y,\delta})}{\partial\delta}\bigg\vert_{\delta=0}%
=\int_{\mathcal{Y}}\psi_{\rho}(y,F_{Y})d(F_{Y}^{\ast}-F_{Y})(y)
\end{equation}
where $\psi_{\rho}(y,F_{Y})$ is the influence function of $\rho$ at $F_{Y}$.
See Chapter 20 in \cite{vanderVaart98} or Section 2.1 in
\cite{NeweyIchimura2022}. Since $F_{Y}^{\ast}(y)-F_{Y}(y)=\int_{\mathcal{X}%
}F_{Y|X}(y|x)d(G_{X}-F_{X})(x)$, we have
\begin{align}
\frac{\partial\rho(F_{Y,\delta})}{\partial\delta}\bigg\vert_{\delta=0}  &
=\int_{\mathcal{Y}}\int_{\mathcal{X}}\psi_{\rho}(y,F_{Y})f_{Y|X}%
(y|x)dyd(G_{X}-F_{X})(x)\nonumber\\
&  =\int_{\mathcal{X}}E[\psi_{\rho}(Y,F_{Y})|X=x]d(G_{X}-F_{X})(x).
\end{align}
This is essentially Theorem 1 in FFL, which provides a characterization of a
directional derivative of the functional $\rho\left(  \cdot\right)  $ in the
direction induced by a change in the marginal distribution of $X.$ The theorem
is silent on how the change in the marginal distribution is implemented.

The second case, covered in Corollary 1 in FFL, is closer to what we consider
here. In this case, $G_{X,\delta}$ is the distribution induced by the location
shift $X+\delta$. The counterfactual distribution is $F_{Y,\delta}^{\ast
}(y)=\int_{\mathcal{X}}F_{Y|X}(y|x)dG_{X,\delta}(x)$. The parameter of
interest is $\lim_{\delta\rightarrow0}\left[  \rho(F_{Y,\delta}^{\ast}%
)-\rho(F_{Y})\right]  /\delta.$ Corollary 1 in FFL shows that
\begin{equation}
\lim_{\delta\rightarrow0}\frac{\rho(F_{Y,\delta}^{\ast})-\rho(F_{Y})}{\delta
}=\int_{\mathcal{X}}\frac{\partial E[\psi_{\rho}\left(  y,F_{Y}\right)
|X=x]}{\partial x}dF_{X}(x). \label{eq:equivalence}%
\end{equation}

Our general intervention $X_{\delta}=\mathcal{G}(X;\delta)$ includes the above
location shift as a special case. To see this, we assume that $W$ is not
present and set $\mathcal{G}(X;\delta)=X+\delta$, in which case $\kappa\left(
x\right)  =1,$ and it follows from Remark
\ref{Remark: more general functional} and Corollary
\ref{Corrollary:uqpe_scale} that $\Pi_{\rho}:=E\left[  \frac{\partial E\left[
\psi_{\rho}\left(  Y,F_{Y}\right)  |X\right]  }{\partial X}\right]
=\int_{\mathcal{X}}\frac{\partial E[\psi_{\rho}(Y,F_{Y})|X=x]}{\partial
x}dF_{X}(x)$, which is identical to the right-hand side of
(\ref{eq:equivalence}). This shows that our approach is strictly more general
than the second case considered by FFL.

There is another main difference between FFL and our paper. From a broad point
of view, FFL considers the scenario where the conditional distribution of $Y$
given $X$ is fixed, and ask how the unconditional distribution of $Y$ would
change if the marginal distribution of $X$ had changed. This is largely a
predictive exercise unless the conditional distribution of $Y$ given $X$ has a
structural or causal interpretation, that is, $X$ is exogenous. In our paper,
we allow for an endogenous $X$ in the sense that $X$ and $U$ may be
correlated. This could arise, for example, when a common factor causes both
$X$ and $U$. As discussed in Remark \ref{control_variable}, $X$ and $U$ may be
dependent even after conditioning on the causal variable $W$. In such a case,
we need to find additional control variables that do not necessarily enter the
structural function $h$ such that $X$ and $U$ become conditionally independent
conditional on $W$ and these additional control variables. The endogeneity
problem is then addressed by using the control variable approach.

At the conceptual level, we consider the policy experiment where both the
structural function $h$ and the distribution of ($X,W,U$) are kept intact.
Given that $h$ is the same, we can say that the effect is causal and have a
\emph{ceteris paribus} interpretation. Given that the distribution of
($X,W,U$) is the same, the policy experiment applies to the current population
under consideration and is fully implementable. Hence the effect is what a
policy maker can achieve under the current environment and is therefore fully policy-relevant.




Furthermore, our counterfactual exercise focuses on manipulating the value of
the target covariate, while the bulk of the literature focuses more on
manipulating its marginal distribution and often uses a value intervention as
an example of how the marginal distribution may be shifted. The advantage of
using a value intervention is that the policy function $\mathcal{G}%
(\cdot;\delta)$ defines clearly how the policy can be implemented. This is in
contrast to the intervention of the marginal distribution where the policy
maker is not given a clear recipe to achieve such an intervention. In
addition, it seems to be easier to attach a cost implication to the value
intervention. A policy maker may want to trade off the cost with the policy
goal they hope to achieve. A marginal distribution intervention seems to be
more of theoretical interest unless it can be implemented empirically via a
value intervention as considered in this paper.\footnote{An important example
of value interventions is the literature on policy relevant treatment effects
where an instrumental variable is manipulated in order to shift the program
participation rate. See, for example, \cite{Heckman2005}.}

\section{Estimation and asymptotic results}

\label{estimation}


In this section, we focus on the estimation of $\Pi_{\tau}$ given in
\eqref{eq_pi_tau}. The estimator involves several preliminary steps. Firstly,
for a given quantile, we need to estimate $Q_{\tau}[Y]$. This is given by
\begin{equation}
\hat{q}_{\tau}=\arg\min_{q}\sum_{i=1}^{n}\left(  \tau-\mathds{1}\left\{
Y_{i}\leq q\right\}  \right)  (Y_{i}-q).\label{eq_hat_q}%
\end{equation}
Next, we need to estimate the density of $Y$ evaluated at $Q_{\tau}[Y]$. This
can be estimated by
\begin{equation}
\hat{f}_{Y}\left(  \hat{q}_{\tau}\right)  =\frac{1}{n}\sum_{i=1}%
^{n}\mathcal{K}_{h}\left(  Y_{i}-\hat{q}_{\tau}\right)  \label{eq_hat_f}%
\end{equation}
where $\mathcal{K}_{h}(u)=h^{-1}\mathcal{K}(h^{-1}u)$ for a given kernel
$\mathcal{K}$ and a bandwidth $h$. For the average derivative of the
conditional cdf, we propose either a logit model as in
\cite{FirpoFortinLemieux09} or a probit model. Note that $\mathcal{S}%
_{Y|X,W}(Q_{\tau}[Y]|x,w)=1-F_{Y|X,W}(Q_{\tau}[Y]|x,w).$ We model
$\mathcal{S}_{Y|X,W}(Q_{\tau}[Y]|x,w)$ via $F_{Y|X,W}(Q_{\tau}[Y]|x,w)$ by
assuming that
\begin{equation}
F_{Y|X,W}(Q_{\tau}[Y]|x,w)=G(\phi_{\mathrm{x}}\left(  x\right)  ^{\prime
}\alpha_{\tau}+\phi_{\mathrm{w}}(w)^{\prime}\beta_{\tau}%
)\label{eq:model_probit}%
\end{equation}
where $\phi_{\mathrm{x}}\left(  \cdot\right)  $ and $\phi_{\mathrm{w}}\left(
\cdot\right)  $ are column vectors of smooth basis functions and $G(\cdot)$ is
either the cdf of a logistic random variable (logit) or a standard normal
random variable (probit). Note that the subscripts \textquotedblleft%
$\mathrm{x}$\textquotedblright\ and \textquotedblleft$\mathrm{w}%
$\textquotedblright\ serve only to distinguish $\phi_{\mathrm{x}}\left(
\cdot\right)  $ from $\phi_{\mathrm{w}}\left(  \cdot\right)  .$ They are not
related to the arguments of these functions. For the choices of $\phi
_{\mathrm{x}}\left(  \cdot\right)  $ and $\phi_{\mathrm{w}}\left(
\cdot\right)  ,$ we may take $\phi_{\mathrm{x}}\left(  x\right)  =x$ or
($x,x^{2})^{\prime}$ and $\phi_{\mathrm{w}}\left(  w\right)  =\left(
1,w\right)  ^{\prime}.$ By default, we include the constant in the vector $w.$
Other more flexible choices are possible, but it is beyond the scope of this
paper to consider a fully nonparametric specification.

Let $Z_{i}=[\phi_{\mathrm{x}}\left(  X_{i}\right)  ^{\prime},\phi_{\mathrm{w}%
}(W_{i})^{\prime}]^{\prime}$ and $\theta_{\tau}=(\alpha_{\tau}^{\prime}%
,\beta_{\tau}^{\prime})^{\prime}.$ We estimate $\theta_{\tau}$ by the maximum
likelihood estimator:%
\begin{align}
\hat{\theta}_{\tau}  &  :=(\hat{\alpha}_{\tau},\hat{\beta}_{\tau}^{\prime
})^{\prime}=\arg\max_{\theta\in\Theta}\sum_{i=1}^{n}l_{i}(\theta;\hat{q}%
_{\tau})\nonumber\\
&  =\arg\max_{\theta\in\Theta}\sum_{i=1}^{n}\bigg\{\mathds1\left\{  Y_{i}%
\leq\hat{q}_{\tau}\right\}  \log\left[  G(Z_{i}^{\prime}\theta)\right]
+\mathds1\left\{  Y_{i}>\hat{q}_{\tau}\right\}  \log\left[  1-G(Z_{i}^{\prime
}\theta)\right]  \bigg\}, \label{eq_likelihood}%
\end{align}
where $\Theta$ is a compact parameter space that contains $\theta_{\tau}$ as
an interior point. The estimator of $\Pi_{\tau}$ is then
\[
\hat{\Pi}_{\tau}=\hat{\Pi}_{\tau,L}+\hat{\Pi}_{\tau,S}%
\]
where
\begin{align}
\hat{\Pi}_{\tau,L}  &  =-\frac{\dot{\ell}(0)}{\hat{f}_{Y}\left(  \hat{q}%
_{\tau}\right)  }\frac{1}{n}\sum_{i=1}^{n}g(Z_{i}^{\prime}\hat{\theta}_{\tau
})\dot{\phi}_{\mathrm{x}}\left(  X_{i}\right)  ^{\prime}\hat{\alpha}_{\tau
},\label{eq:est_pi_l}\\
\hat{\Pi}_{\tau,S}  &  =-\frac{\dot{s}(0)}{\hat{f}_{Y}\left(  \hat{q}_{\tau
}\right)  }\frac{1}{n}\sum_{i=1}^{n}g(Z_{i}^{\prime}\hat{\theta}_{\tau}%
)\dot{\phi}_{\mathrm{x}}\left(  X_{i}\right)  ^{\prime}\hat{\alpha}_{\tau
}\left(  X_{i}-\mu\right)  . \label{eq:est_pi_s}%
\end{align}
In the above, $g$ is the derivative of $G$, that is, the logistic density or
the standard normal density and $\dot{\phi}_{\mathrm{x}}\left(  x\right)
=\partial\phi_{\mathrm{x}}\left(  x\right)  /\partial x$, which has the same
dimension as $\phi_{\mathrm{x}}\left(  x\right)  $. In order to establish the
asymptotic distribution of $\hat{\Pi}_{\tau}$, we need the following three
sets of assumptions, one for each preliminary estimation step.

\begin{assumption}
\label{assumption_quantile}\textbf{Quantile.} The density of $Y$ is positive,
continuous, and differentiable at $Q_{\tau}[Y]$.
\end{assumption}

\begin{assumption}
\label{assumption_logit_probit}\textbf{Logit/Probit.} For $G$ either the cdf
of a logistic or a standard normal random variable, we have

\begin{enumerate}
\item[(i)] $F_{Y|Z}(Q_{\tau}[Y]|z)=G(z^{\prime}\theta_{\tau})$ for an interior
point $\theta_{\tau}\in\Theta$ and $\hat{\theta}_{\tau}=\theta_{\tau}%
+o_{p}\left(  1\right)  .$

\item[(ii)] For
\[
H_{i}\left(  \theta;q\right)  =\frac{\partial^{2}l_{i}\left(  \theta;q\right)
}{\partial\theta\partial\theta^{\prime}},
\]
which is the Hessian of observation $i$, the following holds
\[
\sup_{(\theta,q)\in\mathcal{N}}\left\Vert \frac{1}{n}\sum_{i=1}^{n}%
H_{i}(\theta;q)-E[H_{i}(\theta;q)]\right\Vert \overset{p}{\rightarrow}0,
\]
where $\mathcal{N}$ is a neighborhood of $(\theta_{\tau}^{\prime},Q_{\tau
}[Y]^{\prime})^{\prime}$, and $H:=E[H_{i}(\theta_{\tau};Q_{\tau}[Y])]$ is
negative definite.

\item[(iii)] For the score $s_{i}$ defined by%
\[
s_{i}\left(  \theta,q\right)  =\frac{\partial l_{i}\left(  \theta;q\right)
}{\partial\theta},
\]
the following stochastic equicontinuity assumption holds:
\[
\frac{1}{n}\sum_{i=1}^{n}\left\{  s_{i}(\theta_{\tau};\hat{q}_{\tau})-E\left[
s_{i}(\theta_{\tau};q)\right]  |_{q=\hat{q}_{\tau}}\right\}  =\frac{1}{n}%
\sum_{i=1}^{n}s_{i}(\theta_{\tau};Q_{\tau}[Y])+o_{p}(n^{-1/2}),
\]

and the map $q\mapsto E\left[  s_{i}(\theta_{\tau};q)\right]  $ is
continuously differentiable at $Q_{\tau}[Y]$ with
\[
\frac{\partial E\left[  s_{i}(\theta_{\tau};q)\right]  }{\partial
q}\bigg\vert_{q=Q_{\tau}[Y]}=:H_{Q}.
\]

\item[(iv)] For $\tilde{X}_{i}=(1,X_{i})^{\prime}$,
\begin{align*}
M_{1}\left(  \theta\right)   &  :=E\left\{  [\dot{g}(Z_{i}^{\prime}\theta
)\dot{\phi}_{\mathrm{x}}\left(  X_{i}\right)  ^{\prime}\alpha]\tilde{X}%
_{i}Z_{i}^{\prime}\right\}  \in\mathbb{R}^{2\times d_{Z}}\\
M_{2}\left(  \theta\right)   &  :=E\left[  g(Z_{i}^{\prime}\theta)\tilde
{X}_{i}\dot{\phi}_{\mathrm{x}}(X_{i})^{\prime}\right]  \in\mathbb{R}^{2\times
d_{\phi_{\mathrm{x}}}}%
\end{align*}
are well defined for any $\theta\in\mathcal{N}_{\theta_{\tau}}$, a
neighborhood of $\theta_{\tau}$; and the following uniform law of large
numbers holds:
\begin{align*}
&  \sup_{\theta\in\mathcal{N}_{\theta_{\tau}}}\bigg\|\frac{1}{n}\sum_{i=1}%
^{n}[\dot{g}(Z_{i}^{\prime}\theta)\dot{\phi}_{\mathrm{x}}\left(  X_{i}\right)
^{\prime}\alpha]\tilde{X}_{i}Z_{i}^{\prime}-M_{1}\left(  \theta\right)
\bigg\|\overset{p}{\rightarrow}0,\\
&  \sup_{\theta\in\mathcal{N}_{\theta_{\tau}}}\bigg\|\frac{1}{n}\sum_{i=1}%
^{n}g(Z_{i}^{\prime}\theta)\tilde{X}_{i}\dot{\phi}_{\mathrm{x}}(X_{i}%
)^{\prime}-M_{2}\left(  \theta\right)  \bigg\|\overset{p}{\rightarrow}0,
\end{align*}
where $\dot{g}$ is the derivative of $g$.
\end{enumerate}
\end{assumption}

In the above assumption, we assume that $F_{Y|Z}(Q_{\tau}[Y]|z)=G(z^{\prime
}\theta_{\tau})$ with $G$ being either the cdf of a logistic or a standard
normal random variable. It is important to note that other cdfs can also be
utilized. For instance, when the interest lies in the lowest quantiles with
$\tau$ very close to 0, the cdf of a Gumbel distribution (also known as a Type
I extreme value distribution) can be employed. This choice leads to a
complementary log-log model, wherein $F_{Y|Z}(Q_{\tau}[Y]|z)$ is modeled by
$1-\exp(-\exp(z^{\prime}\theta_{\tau}))$ and the index $z^{\prime}\theta
_{\tau}$ can be written in the complementary log-log form $\log(-\log
(1-F_{Y|Z}(Q_{\tau}[Y]|z)).$\footnote{We thank an anonymous referee for
suggesting the complementary log-log or log-log link when our focus is on
extreme quantiles.}

\begin{assumption}
\label{assumption_density}\textbf{Density.}

\begin{enumerate}
\item[(i)] The kernel function $K(\cdot)$ satisfies (i) $\int_{-\infty
}^{\infty}K(u)du=1$, (ii) $\int_{-\infty}^{\infty}u^{2}K(u)du<\infty$, and
(iii) $K(u)=K(-u)$, and it is twice differentiable with Lipschitz continuous
second-order derivative $K^{\prime\prime}\left(  u\right)  $ satisfying (i)
$\int_{-\infty}^{\infty}K^{\prime\prime}(u)udu<\infty$ and $\left(  ii\right)
$ there exist positive constants $C_{1}$ and $C_{2}$ such that $\left\vert
K^{\prime\prime}\left(  u_{1}\right)  -K^{\prime\prime}\left(  u_{2}\right)
\right\vert \leq C_{2}\left\vert u_{1}-u_{2}\right\vert ^{2}$ for $\left\vert
u_{1}-u_{2}\right\vert \geq C_{1}.$

\item[(ii)] As $n\uparrow\infty$, the bandwidth satisfies: $h\downarrow0$,
$nh^{3}\uparrow\infty$, and $nh^{5}=O(1)$.
\end{enumerate}
\end{assumption}

Under Assumption \ref{assumption_quantile}, $\hat{q}_{\tau}$ given in
\eqref{eq_hat_q} is asymptotically linear with
\[
\hat{q}_{\tau}-Q_{\tau}[Y]=\frac{1}{n}\sum_{i=1}^{n}\frac{\tau
-\mathds{1}\left\{  Y_{i}\leq Q_{\tau}[Y]\right\}  }{f_{Y}(Q_{\tau}[Y])}%
+o_{p}(n^{-1/2})=\frac{1}{n}\sum_{i=1}^{n}\psi(Y_{i},\tau,F_{Y})+o_{p}%
(n^{-1/2}).
\]
See, for example, \cite{serfling1980}. Assumption
\ref{assumption_logit_probit} is mostly necessary to deal with the preliminary
estimator $\hat{q}_{\tau}$ that enters the likelihood in
\eqref{eq_likelihood}. Assumption \ref{assumption_density} is taken from
\cite{yixiao2020}.

The following lemma contains the influence function for the maximum likelihood
estimator $\hat{\theta}_{\tau}$.

\begin{lemma}
\label{lemma_mle_alpha_beta} Under Assumptions \ref{assumption_quantile} and
\ref{assumption_logit_probit}, we have
\[
\hat{\theta}_{\tau}-\theta_{\tau}=-H^{-1}\frac{1}{n}\sum_{i=1}^{n}s_{i}%
(\theta_{\tau};Q_{\tau}[Y])-H^{-1}H_{Q}\frac{1}{n}\sum_{i=1}^{n}\psi
(Y_{i},\tau,F_{Y})+o_{p}(n^{-1/2}).
\]

\end{lemma}

\begin{theorem}
\label{theorem_est_pi} Under Assumptions \ref{assumption_quantile},
\ref{assumption_logit_probit}, and \ref{assumption_density}, the estimators
given in \eqref{eq:est_pi_l} and \eqref{eq:est_pi_s} satisfy
\[%
\begin{pmatrix}
\hat{\Pi}_{\tau,L}\\
\hat{\Pi}_{\tau,S}%
\end{pmatrix}
-%
\begin{pmatrix}
\Pi_{\tau,L}\\
\Pi_{\tau,S}%
\end{pmatrix}
=\frac{1}{n}\sum_{i=1}^{n}\Phi_{i,\tau}+O\left(  h^{2}\right)  +o_{p}%
(n^{-1/2})+o_{p}(n^{-1/2}h^{-1/2}),
\]
where
\begin{align*}
\Phi_{i,\tau}  &  =\frac{1}{f_{Y}\left(  Q_{\tau}[Y]\right)  }D_{\mu}\left\{
g(Z_{i};\theta_{\tau})\dot{\phi}_{\mathrm{x}}\left(  X_{i}\right)  ^{\prime
}\alpha_{\tau}\tilde{X}_{i}-E\left[  g(Z_{i};\theta_{\tau})\dot{\phi
}_{\mathrm{x}}\left(  X_{i}\right)  ^{\prime}\alpha_{\tau}\tilde{X}%
_{i}\right]  \right\} \\
&  -\frac{1}{f_{Y}\left(  Q_{\tau}[Y]\right)  }D_{\mu}MH^{-1}s_{i}%
(\theta_{\tau};Q_{\tau}[Y])\\
&  -\left[
\begin{pmatrix}
\Pi_{\tau,L}\\
\Pi_{\tau,S}%
\end{pmatrix}
\frac{\dot{f}_{Y}(Q_{\tau}[Y])}{f_{Y}\left(  Q_{\tau}[Y]\right)  }+\frac
{1}{f_{Y}\left(  Q_{\tau}[Y]\right)  }D_{\mu}MH^{-1}H_{Q}\right]  \psi
(Y_{i},\tau,F_{Y})\\
&  -%
\begin{pmatrix}
\Pi_{\tau,L}\\
\Pi_{\tau,S}%
\end{pmatrix}
\frac{1}{f_{Y}\left(  Q_{\tau}[Y]\right)  }\left\{  \mathcal{K}_{h}\left(
Y_{i}-Q_{\tau}[Y]\right)  -E\mathcal{K}_{h}\left(  Y_{i}-Q_{\tau}[Y]\right)
\right\}  ,
\end{align*}
$\dot{f}_{Y}\left(  \cdot\right)  $ is the derivative of $f_{Y}\left(
\cdot\right)  ,$
\[
D_{\mu}=%
\begin{pmatrix}
D_{L}^{\prime}\\
D_{\mu,S}^{\prime}%
\end{pmatrix}
=%
\begin{pmatrix}
-\dot{\ell}(0) & 0\\
\mu\dot{s}(0) & -\dot{s}(0)
\end{pmatrix}
,
\]%
\[
M=M_{1}\left(  \theta_{\tau}\right)  +%
\begin{pmatrix}
M_{2}\left(  \theta_{\tau}\right)  , & O
\end{pmatrix}
\in\mathbb{R}^{2\times d_{Z}},
\]
and $O\in\mathbb{R}^{2\times d_{\phi_{\mathrm{w}}}}$ is a matrix of zeros.
\end{theorem}

Theorem \ref{theorem_est_pi} establishes the contribution from each estimation
step. In particular, the last term in $n^{-1}\sum_{i=1}^{n}\Phi_{i,\tau}$ is
the contribution from estimating the density of $Y$ non-parametrically. This
term converges at a non-parametric rate, which is slower than other terms. As
a result, the asymptotic distribution of the location-scale effect estimator
is determined by the last term in $n^{-1}\sum_{i=1}^{n}\Phi_{i,\tau}$.
However, we do not recommend dropping all other terms. Instead, we write the
asymptotic normality result in the form
\begin{equation}
\left[  \frac{1}{n^{2}}\sum_{i=1}^{n}\hat{\Phi}_{i,\tau}\hat{\Phi}_{i,\tau
}^{\prime}\right]  ^{-1/2}\left[
\begin{pmatrix}
\hat{\Pi}_{\tau,L}\\
\hat{\Pi}_{\tau,S}%
\end{pmatrix}
-%
\begin{pmatrix}
\Pi_{\tau,L}\\
\Pi_{\tau,S}%
\end{pmatrix}
\right]  \overset{d}{\rightarrow}N(0,I_{2})\label{asym_norm_full_form2}%
\end{equation}
as $n\uparrow\infty$, $nh^{3}\uparrow\infty$, and $nh^{5}\downarrow0$ where
$\hat{\Phi}_{i,\tau}$ is a plug-in estimator of $\Phi_{i,\tau}.$ In
particular,
\begin{align}
&  \left[  n^{-2}\sum_{i=1}^{n}(l_{1}^{\prime}\hat{\Phi}_{i,\tau})^{2}\right]
^{-1/2}\left(  \hat{\Pi}_{\tau,L}-\Pi_{\tau,L}\right)  \overset{d}{\rightarrow
}N(0,1),\nonumber\\
&  \left[  n^{-2}\sum_{i=1}^{n}(l_{2}^{\prime}\hat{\Phi}_{i,\tau})^{2}\right]
^{-1/2}\left(  \hat{\Pi}_{\tau,S}-\Pi_{\tau,S}\right)  \overset{d}{\rightarrow
}N(0,1),\label{asym_norm_individual_form2}%
\end{align}
as $n\uparrow\infty$, $nh^{3}\uparrow\infty$, and $nh^{5}\downarrow0$ where
$l_{1}=\left(  1,0\right)  ^{\prime}$ and $l_{2}=\left(  0,1\right)  ^{\prime
}$. Note that Theorem \ref{theorem_est_pi} has shown that the estimation error
in $\hat{\Pi}_{\tau,L}$ or $\hat{\Pi}_{\tau,S}$ is an average of independent
observations. The above asymptotic normality results can be proved using a
Lyapunov CLT under the following conditions (see the proof of Theorem 2.9 in
\cite{pagan_ullah_1999}): 

(i) $n^{-1}h\sum_{i=1}^{n}E\left[  \Phi_{i,\tau}\Phi_{i,\tau}^{\prime}\right]
$ is nonsingular for all large enough $n.$

(ii) $\left(  n^{-1}h\sum_{i=1}^{n}E\Phi_{i,\tau}\Phi_{i,\tau}^{\prime
}\right)  ^{-1/2}\left(  n^{-1}h\sum_{i=1}^{n}\hat{\Phi}_{i,\tau}\hat{\Phi
}_{i,\tau}^{\prime}\right)  \left(  n^{-1}h\sum_{i=1}^{n}E\Phi_{i,\tau}%
\Phi_{i,\tau}^{\prime}\right)  ^{-1/2}=I_{2}+o_{p}\left(  1\right)  .$

(ii) Assumption \ref{assumption_density} holds, $\int_{-\infty}^{\infty
}\left\vert K(u)\right\vert ^{2+\Delta}du<\infty$ for some $\Delta>0,$ and
$\left\vert f_{Y}^{\prime\prime}(Q_{\tau}[Y]\right\vert <C$ for some constant
$C.$

Inferences based on our asymptotic results account for the estimation errors
from all estimation steps and are more reliable in finite samples. This is
supported by simulation evidence not reported here, but available upon
request. On the other hand, if we parametrize the density of $Y$ and estimate
it at the parametric $\sqrt{n}$-rate, then the last term in $n^{-1}\sum
_{i=1}^{n}\Phi_{i,\tau}$ will take a different form and will be of the same
order as the other terms. In this case, the location-scale effect estimator is
$\sqrt{n}$-asymptotically normal, and all the terms in Theorem
\ref{theorem_est_pi} will contribute to the asymptotic variance. With an
obvious modification of the last term in $\Phi_{i,\tau},$ the asymptotic
normality can be presented in the same way as in (\ref{asym_norm_full_form2}).

Let
\[
\Gamma_{\tau,S}=D_{\mu,S}^{\prime}E\left[  \frac{\partial F_{Y|X,W}(Q_{\tau
}[Y]|X,W)}{\partial X}\tilde{X}\right]
\]
be the numerator of $\Pi_{\tau,S}.$ Then the scale effect $\Pi_{\tau,S}$ is
zero if and only if $\Gamma_{\tau,S}=0.$ To test the null hypothesis
$H_{0}:\Pi_{\tau,S}=0,$ we can equivalently test the null hypothesis
$H_{0}:\Gamma_{\tau,S}=0.$ Unlike $\Pi_{\tau,S},$ $\Gamma_{\tau,S}$ can be
estimated at the parametric rate even if $f_{Y}\left(  \cdot\right)  $ is not
parametrically specified. More specifically, under Assumption
\ref{assumption_logit_probit}, we can estimate $\Gamma_{\tau,S}$ by
\[
\hat{\Gamma}_{\tau,S}:=D_{\mu,S}^{\prime}\frac{1}{n}\sum_{i=1}^{n}%
[g(Z_{i}^{\prime}\hat{\theta}_{\tau})\dot{\phi}_{\mathrm{x}}\left(
X_{i}\right)  ^{\prime}\hat{\alpha}_{\tau}]\tilde{X}_{i},
\]
where $D_{\mu,S}^{\prime}=\left(  \mu,-1\right)  $ upon setting $\dot{s}(0)=1$
without loss of generality.

Under the assumptions of Theorem \ref{theorem_est_pi}, we can show that
\[
\hat{\Gamma}_{\tau,S}-\Gamma_{\tau,S}=D_{\mu,S}^{\prime}\frac{1}{n}\sum
_{i=1}^{n}\Phi_{i,\tau}^{\Gamma}+o_{p}\left(  \frac{1}{\sqrt{n}}\right)  ,
\]
where
\begin{align*}
\Phi_{i,\tau}^{\Gamma}  &  =g(Z_{i}^{\prime}\theta_{\tau})\dot{\phi
}_{\mathrm{x}}\left(  X_{i}\right)  ^{\prime}\alpha_{\tau}\tilde{X}%
_{i}-E\left\{  [g(Z_{i}^{\prime}\theta_{\tau})\dot{\phi}_{\mathrm{x}}\left(
X_{i}\right)  ^{\prime}\alpha_{\tau}]\tilde{X}_{i}\right\} \\
&  -{MH}^{-1}s_{i}(\theta_{\tau};Q_{\tau}[Y])-{MH}^{-1}H_{Q}\psi(Y_{i}%
,\tau,F_{Y}).
\end{align*}
Define
\[
V_{\tau}=\lim_{n\rightarrow\infty}\frac{1}{n}\sum_{i=1}^{n}E(D_{\mu,S}%
^{\prime}\Phi_{i,\tau}^{\Gamma})^{2}.
\]
If $D_{\mu,S}^{\prime}\Phi_{i,\tau}^{\Gamma}$ has a finite second moment and
$V_{\tau}>0,$ then a standard CLT yields $V_{\tau}^{-1/2}\sqrt{n}(\hat{\Gamma
}_{\tau,S}-\Gamma_{\tau,S}^{\mu})\overset{d}{\rightarrow}N\left(  0,1\right)
$. To test $H_{0}:\Gamma_{\tau,S}=0,$ we construct the test statistic
\[
t_{\tau,S}:=\frac{\sqrt{n}\hat{\Gamma}_{\tau,S}}{\sqrt{\hat{V}_{\tau}}}\text{
for }\hat{V}_{\tau}=\frac{1}{n}\sum_{i=1}^{n}(D_{\mu,S}^{\prime}\hat{\Phi
}_{i,\tau}^{\Gamma})^{2},
\]
where
\begin{align}
\hat{\Phi}_{i,\tau}^{\Gamma}  &  =g(Z_{i}\hat{\theta}_{\tau})\dot{\phi
}_{\mathrm{x}}\left(  X_{i}\right)  ^{\prime}\hat{\alpha}_{\tau}\tilde{X}%
_{i}-\frac{1}{n}\sum_{i=1}^{n}g(Z_{i}\hat{\theta}_{\tau})\dot{\phi
}_{\mathrm{x}}\left(  X_{i}\right)  ^{\prime}\hat{\alpha}_{\tau}\tilde{X}%
_{i}\nonumber\\
&  -{\hat{M}\hat{H}}^{-1}s_{i}(\hat{\theta}_{\tau};\hat{q}_{\tau})-{\hat
{M}\hat{H}}^{-1}\hat{H}_{Q}\hat{\psi}(Y_{i},\tau,F_{Y}). \label{eq:var_t_test}%
\end{align}
In the above, $\hat{\psi}(Y_{i},\tau,F_{Y})=\left[  \tau-1\left\{  Y_{i}%
\leq\hat{q}_{\tau}\right\}  \right]  /\hat{f}_{Y}\left(  \hat{q}_{\tau
}\right)  $ and the score $s_{i}(\hat{\theta}_{\tau};\hat{q}_{\tau})$ is
obtained by evaluating the expression given in \eqref{eq_score} at
$\theta=\hat{\theta}_{\tau}$ and $q=\hat{q}_{\tau}$. ${\hat{M}},$ ${\hat{H},}$
and $\hat{H}_{Q}$ are the sample versions of ${M},$ ${H,}$ and $H_{Q},$
respectively. Details are given in the proof of the corollary below.

\begin{corollary}
\label{corollary_est_pi} Let the assumptions of Theorem \ref{theorem_est_pi}
hold. Assume that $D_{\mu,S}^{\prime}\Phi_{i,\tau}^{\Gamma}$ has a finite
second moment and $\hat{V}_{\tau}/V_{\tau}\overset{p}{\rightarrow}1$ for some
$V_{\tau}>0.$ Then, under the null hypothesis $H_{0}:\Pi_{\tau,S}=0,$
\[
t_{\tau,S}\overset{d}{\rightarrow}N(0,1).
\]

\end{corollary}

\section{Monte Carlo experiments}

\label{MC}


In this section, we use Monte Carlo simulations to evaluate the finite sample
performances of the proposed estimators and tests of location and scale
effects. We employ the same data generating process as in Example
\ref{example_normal_location_model} for which we have derived the closed-form
expressions for the location and scale effects. In particular, we let
\[
Y=\lambda+X\gamma+U,
\]
where $X\sim N(\mu_{X},\sigma_{X}^{2})$ and $U\sim N(0,\sigma_{U}^{2})$. We
set $\lambda=0,$ $\sigma_{U}^{2}=1,$ $\dot{\ell}(0)=1$ and $\dot{s}(0)=-1$.
The last derivative corresponds to, for example, $s(\delta)=(1+\delta)^{-1}$.
Then, from the results in Example \ref{example_normal_location_model}, the
true location effect is $\Pi_{\tau,L}=\gamma,$ and the true scale effect is
\[
\Pi_{\tau,S}^{\mu_{X}}=-\sqrt{R_{YX}^{2}}Q_{\tau}[X_{\gamma}^{\circ}%
]=-\sqrt{R_{YX}^{2}}\sqrt{var(X_{\gamma}^{\circ})}Q_{\tau}[\varepsilon
]=-\sqrt{R_{YX}^{2}}\cdot\sigma_{X}\cdot\left\vert \gamma\right\vert \cdot
Q_{\tau}[\varepsilon]
\]
where $\varepsilon$ is standard normal.

We consider quantiles $\tau\in\{0.10,0.25,0.50,0.75,0.90\}$ and sample sizes
$n=500$ and $n=1000$. The number of simulations is set to $10,000$ for each experiment.

We implement our estimators in \texttt{Matlab}. The unconditional quantile
estimator in equation \eqref{eq_hat_q} is easily computed as an order
statistic. The density function is estimated as a kernel density estimator as
in equation \eqref{eq_hat_f} using a standard normal kernel. For the bandwidth
choice in the kernel density estimation, we use a modified version of
Silverman's rule of thumb. More specifically, since we require $nh^{3}%
\uparrow\infty$ and $nh^{5}\downarrow0$ as $n\uparrow\infty$, we take
$h=1.06\hat{\sigma}_{Y}n^{-1/4}$, where $\hat{\sigma}_{Y}$ is the sample
standard deviation of $Y$.

\subsection{Bias, variance, and mean squared error}

In this subsection, we consider the bias, variance, and mean-squared error
(MSE) of the proposed location and scale effects estimators. For each effect
estimator, we consider either a probit or a logit specification for the
conditional cdf $F_{Y|X}(Q_{\tau}[Y]|X).$ Under our data generating process,
the probit with $F_{Y|X}(Q_{\tau}[Y]|X)=\boldsymbol{\Phi}(X\alpha_{\tau}%
+\beta_{\tau})$ for the standard normal CDF $\boldsymbol{\Phi}$ is correctly
specified while the logit with $F_{Y|X}(Q_{\tau}[Y]|X)=\left[  1+\exp\left(
X\alpha_{\tau}+\beta_{\tau}\right)  \right]  ^{-1}$ is misspecified.

The bias, variance, and MSE are reported in Table \ref{table:MC} when $\mu
_{X}=0$, $\gamma=1$ and $\sigma_{X}^{2}=1$ so that the true location effect is
$1$ for any $\tau$ and the true scale effect is $-\sqrt{0.5}Q_{\tau
}[\varepsilon]\approx-0.707Q_{\tau}[\varepsilon].$ To save space, simulation
results for other values of $\gamma$ and $\sigma_{X}^{2}$ are omitted.

Table \ref{table:MC} shows that the estimator based on the probit
specification outperforms that based on the logit one. This is consistent with
the correct specification of probit. For each estimator, the bias decreases as
the sample size $n$ increases. The variance also decreases as the sample size
$n$ increase, and as a result, the MSE also becomes smaller when the sample
size grows. For our purposes, the scale effect estimator performs well. For
non-central quantiles, the difference in the scale effect estimates under the
probit and logit specifications is in general larger than the difference in
the location effect estimates. For central quantiles, the probit and logit
specifications lead to more or less the same estimates for both the scale
effect and the location effect.

\begin{center}
{\scriptsize \begin{table}[ptbh]
\caption{The biases, variances, and mean-squared errors of the location and
scale effects estimators with $\gamma=1$ and $\sigma_{X}^{2}=1$.}%
\label{table:MC}%
{\scriptsize  \centering\hspace{2cm}
\begin{tabular}
[c]{l|cccccc}\hline\hline
&  & {$\tau=0.1$} & {$\tau=0.25$} & {$\tau=0.50$} & {$\tau=0.75$} &
{$\tau=0.90$}\\\hline
&  &  &  & $n=500$ &  & \\\hline
Bias & {$\Pi_{L}$ (probit)} & -0.015 & 0.013 & 0.023 & 0.012 & -0.016\\
& {$\Pi_{L}$ (logit)} & -0.016 & 0.012 & 0.023 & 0.012 & -0.016\\
& {$\Pi_{S}$ (probit)} & -0.008 & 0.008 & 0.000 & -0.007 & 0.008\\
& {$\Pi_{S}$ (logit)} & 0.039 & 0.034 & 0.000 & -0.034 & -0.039\\\hline
Variance & {$\Pi_{L}$ (probit)} & 0.019 & 0.010 & 0.008 & 0.010 & 0.019\\
& {$\Pi_{L}$ (logit)} & 0.019 & 0.010 & 0.008 & 0.010 & 0.020\\
& {$\Pi_{S}$ (probit)} & 0.032 & 0.007 & 0.003 & 0.008 & 0.033\\
& {$\Pi_{S}$ (logit)} & 0.033 & 0.007 & 0.003 & 0.008 & 0.034\\\hline
MSE & {$\Pi_{L}$ (probit)} & 0.019 & 0.010 & 0.009 & 0.010 & 0.019\\
& {$\Pi_{L}$ (logit)} & 0.020 & 0.011 & 0.009 & 0.010 & 0.020\\
& {$\Pi_{S}$ (probit)} & 0.033 & 0.007 & 0.003 & 0.008 & 0.033\\
& {$\Pi_{S}$ (logit)} & 0.035 & 0.009 & 0.003 & 0.009 & 0.035\\\hline\hline
&  &  &  & $n=1000$ &  & \\\hline
Bias & {$\Pi_{L}$ (probit)} & -0.011 & 0.009 & 0.017 & 0.008 & -0.013\\
& {$\Pi_{L}$ (logit)} & -0.011 & 0.009 & 0.017 & 0.008 & -0.013\\
& {$\Pi_{S}$ (probit)} & -0.007 & 0.005 & -0.000 & -0.004 & 0.010\\
& {$\Pi_{S}$ (logit)} & 0.041 & 0.032 & -0.000 & -0.031 & -0.038\\\hline
Variance & {$\Pi_{L}$ (probit)} & 0.011 & 0.006 & 0.005 & 0.006 & 0.011\\
& {$\Pi_{L}$ (logit)} & 0.011 & 0.006 & 0.005 & 0.006 & 0.011\\
& {$\Pi_{S}$ (probit)} & 0.018 & 0.004 & 0.001 & 0.004 & 0.017\\
& {$\Pi_{S}$ (logit)} & 0.018 & 0.004 & 0.001 & 0.004 & 0.018\\\hline
MSE & {$\Pi_{L}$ (probit)} & 0.011 & 0.006 & 0.005 & 0.006 & 0.011\\
& {$\Pi_{L}$ (logit)} & 0.011 & 0.006 & 0.005 & 0.006 & 0.011\\
& {$\Pi_{S}$ (probit)} & 0.018 & 0.004 & 0.001 & 0.004 & 0.017\\
& {$\Pi_{S}$ (logit)} & 0.020 & 0.005 & 0.001 & 0.005 & 0.019\\\hline\hline
\end{tabular}
}\end{table}}
\end{center}

\subsection{Accuracy of the normal approximation}

\begin{figure}[ptb]
\centering
\includegraphics[
	height=3.4411in,
	width=4.4529in
	]	{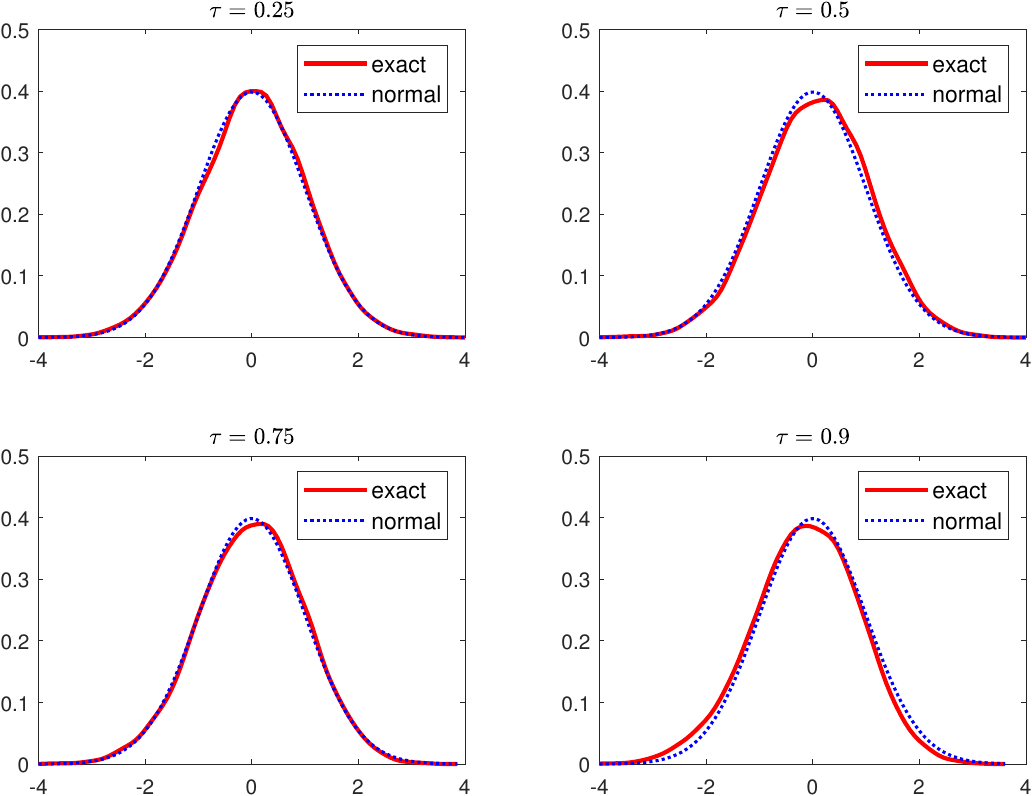} \caption{Finite sample exact
distribution of the studentized location effect statistic when $\gamma=0.25$,
$\sigma_{X}^{2}=1$, and $n=1000.$}%
\label{fig:beta_025_location}%
\end{figure}

\begin{figure}[ptb]
\centering
\includegraphics[
	height=3.4601in,
	width=4.4815in
	]	{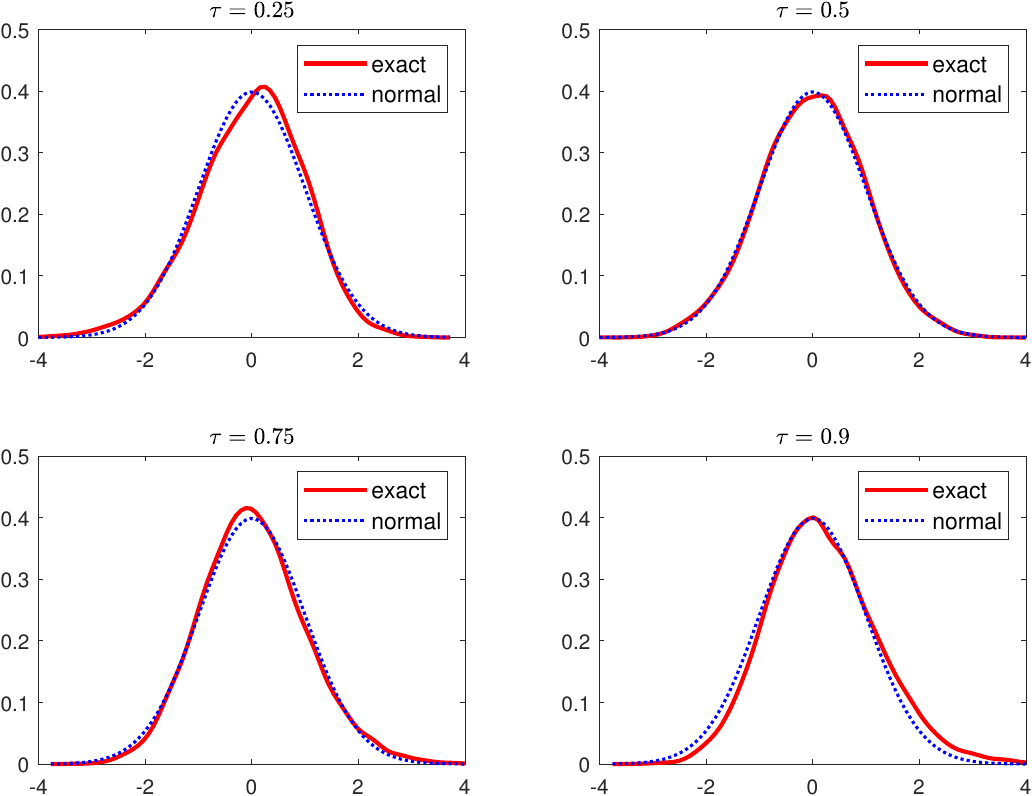} \caption{Finite sample exact
distribution of the studentized scale effect statistic when $\gamma=0.75$,
$\sigma_{X}^{2}=1$, and $n=1000.$}%
\label{fig:beta_075_scale}%
\end{figure}

In this subsection, we investigate the finite sample accuracy of the normal
approximation given in (\ref{asym_norm_individual_form2}). Using the same data
generating process as in the previous subsection and employing the probit
specification, we simulate the distributions of the studentized statistics
\[
\left[  n^{-2}\sum_{i=1}^{n}(l_{1}^{\prime}\hat{\Phi}_{i,\tau})^{2}\right]
^{-1/2}(\hat{\Pi}_{\tau,L}-\Pi_{\tau,L})
\]
and
\[
\left[  n^{-2}\sum_{i=1}^{n}(l_{2}^{\prime}\hat{\Phi}_{i,\tau})^{2}\right]
^{-1/2}(\hat{\Pi}_{\tau,S}-\Pi_{\tau,S}).,
\]
for the location and scale effects, respectively. We plot each distribution
and compare it with the standard normal distribution. We consider $\gamma
\in\left\{  0.25,0.50,0.75,1\right\}  $ and use the same $\tau$ values as in
the previous subsection. Simulation results for the two sample sizes $n=500$
and $n=1000$ are qualitatively similar, and we report only the case when
$n=1000$ here. Figures \ref{fig:beta_025_location}--\ref{fig:beta_075_scale}
report the (simulated) finite sample distributions when $\sigma_{X}^{2}=1$ and
$n=1000$ for some selected values of $\gamma$ and $\tau$ together with a
standard normal density that is superimposed on each figure. It is clear from
these figures that the standard normal distribution provides an accurate
approximation to the distribution of the studentized test statistic for both
the location and scale effects.

\begin{figure}[ptb]
\centering
\includegraphics[
	height=3.4247in,
	width=4.4322in
	]	{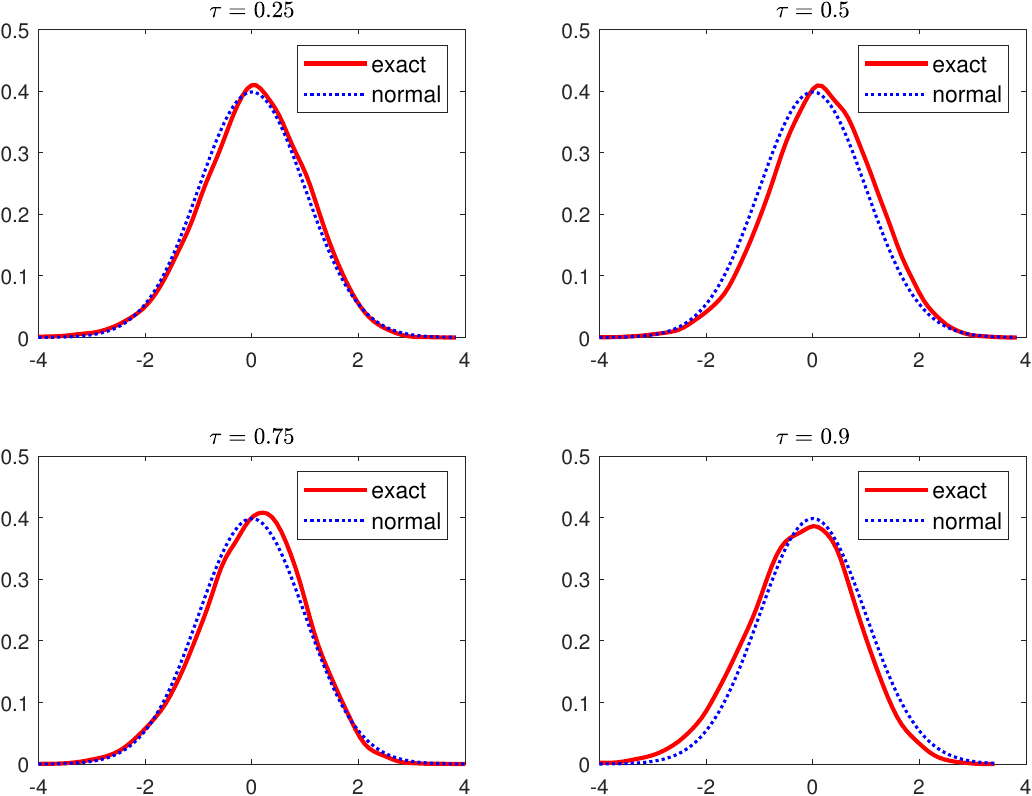}\caption{Finite sample exact
distribution of the studentized location effect statistic when $\gamma=0.75$,
$\sigma_{X}^{2}=1$, and $n=1000.$}%
\end{figure}

\bigskip

\begin{figure}[ptb]
\centering
\includegraphics[
	height=3.4541in,
	width=4.4745in
	]	{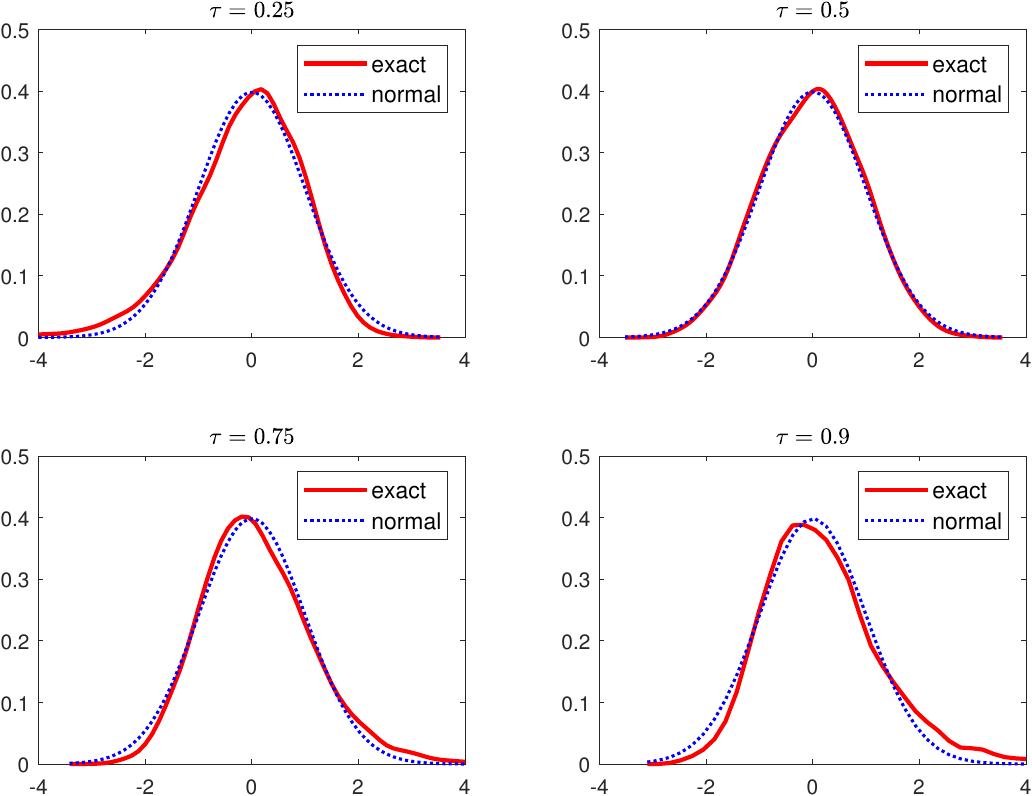}\caption{Finite sample exact
distribution of the studentized scale effect statistic when $\gamma=0.25$,
$\sigma_{X}^{2}=1$, and $n=1000.$}%
\end{figure}

Table \ref{table:emp_cov_location_scale} reports the empirical coverage of
95\% confidence intervals for the location and scale effects. The empirical
coverage is close to the nominal coverage in all cases. This is consistent
with Figures \ref{fig:beta_025_location}--\ref{fig:beta_075_scale}. We may
then conclude that the normal approximation can be reliably used for inference
on the location and scale effects.%

\begin{table}[ptb]
\caption
{Empirical coverage of 95\% confidence intervals for the location and scale effects when  $\sigma
_{X}^2=1.$}
\centering
{\scriptsize
\begin{tabular}
[c]{l|lccccc}\hline\hline
& $\gamma$ & $\tau=0.1$ & $\tau=0.25$ & $\tau=0.50$ & $\tau=0.75$ &
$\tau=0.90$\\\hline
& \multicolumn{6}{|c}{$n=500$}\\\hline
Location & ${0.25}$ & $0.946$ & $0.950$ & $0.951$ & $0.950$ & $0.947$\\
& ${0.5}$ & $0.942$ & $0.952$ & $0.950$ & $0.953$ & $0.938$\\
& ${0.75}$ & $0.940$ & $0.954$ & $0.952$ & $0.956$ & $0.937$\\
& ${1}$ & $0.937$ & $0.957$ & $0.950$ & $0.957$ & $0.935$\\\hline
Scale & ${0.25}$ & $0.900$ & $0.921$ & $0.973$ & $0.916$ & $0.902$\\
& ${0.5}$ & $0.930$ & $0.943$ & $0.957$ & $0.939$ & $0.928$\\
& ${0.75}$ & $0.937$ & $0.950$ & $0.954$ & $0.946$ & $0.933$\\
& ${1}$ & $0.939$ & $0.952$ & $0.951$ & $0.945$ & $0.933$\\\hline\hline
& \multicolumn{6}{|c}{$n=1000$}\\\hline
Location & $0.25$ & $0.948$ & $0.951$ & $0.951$ & $0.954$ & $0.945$\\
& $0.5$ & $0.946$ & $0.950$ & $0.952$ & $0.957$ & $0.943$\\
& $0.75$ & $0.945$ & $0.952$ & $0.953$ & $0.957$ & $0.940$\\
& $1$ & $0.941$ & $0.952$ & $0.952$ & $0.958$ & $0.942$\\\hline
Scale & $0.25$ & $0.922$ & $0.939$ & $0.965$ & $0.940$ & $0.921$\\
& $0.5$ & $0.938$ & $0.949$ & $0.955$ & $0.950$ & $0.933$\\
& $0.75$ & $0.942$ & $0.951$ & $0.952$ & $0.952$ & $0.938$\\
& $1$ & $0.939$ & $0.952$ & $0.950$ & $0.953$ & $0.940$\\\hline\hline
\end{tabular}
}%

\label{table:emp_cov_location_scale}
\end{table}%

\subsection{Power of the t-test of a zero scale effect}

To investigate the power of the t-test proposed in Corollary
\ref{corollary_est_pi}, we simulate the following model:
\[
Y=\lambda+X\gamma+U,
\]
where
\[%
\begin{pmatrix}
X\\
U
\end{pmatrix}
\sim N\left(
\begin{pmatrix}
1\\
0
\end{pmatrix}
,%
\begin{pmatrix}
1 & 0\\
0 & 1
\end{pmatrix}
\right)  .
\]
Here we set $\lambda=0,$ $\mu_{X}=1$ and $\dot{s}(0)=-1$. When $\gamma=0$, $X$
is excluded from the outcome equation and thus the scale effect is 0. The null
hypothesis of a zero scale effect corresponds to the case that $\gamma=0$. The
power of the test is obtained by varying $\gamma$ around 0 in a grid from
$-0.4$ to $0.4$ with an increment of $0.01$.

\begin{figure}[ptb]
\hspace*{-1.9cm} \centering
\includegraphics[scale=0.65]{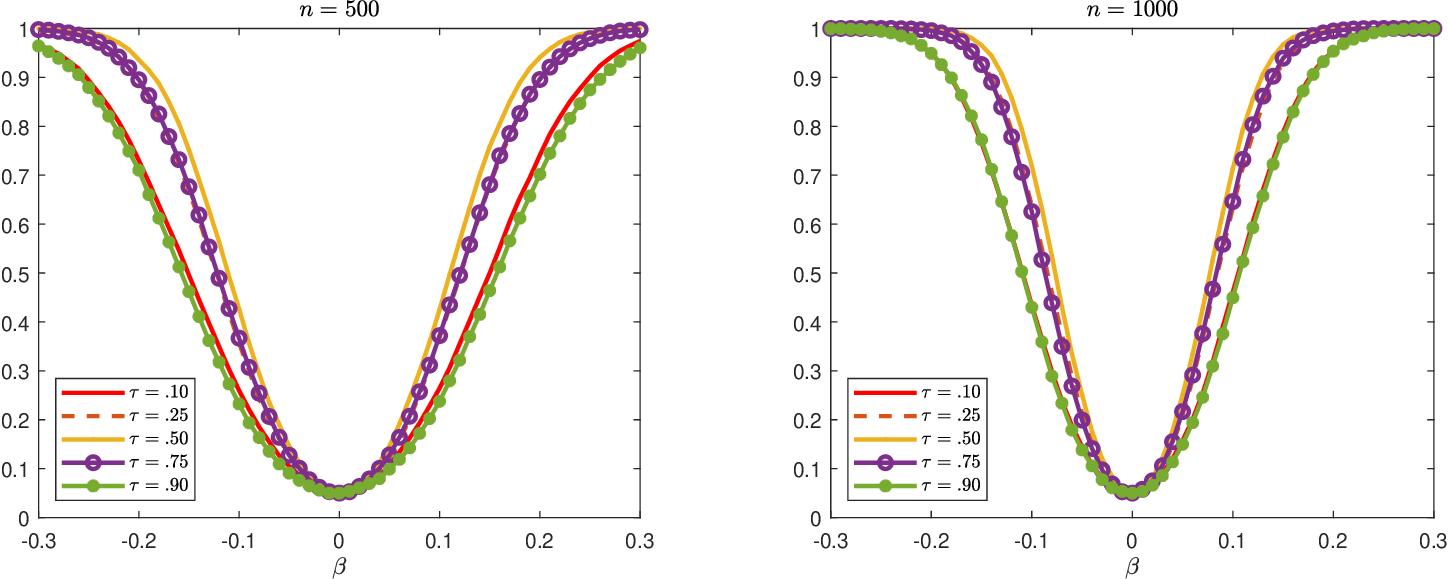}\caption{Size-adjusted power of
the t-test for a zero scale effect.}%
\label{fig:power}%
\end{figure}

Figure \ref{fig:power} graphs the size-adjusted power of the t-test for
different quantile levels when $n=500$ and when $n=1000$. The power is
calculated using the probit specification, namely $F_{Y|X}(Q_{\tau
}[Y]|X)=\boldsymbol{\Phi}(X\alpha_{\tau}+\beta_{\tau})$. The size adjustment
is based on the empirical critical value such that the test rejects the null
5\% of the time. Figure \ref{fig:power} shows that the power increases as
$\gamma$ deviates more from its null value of zero, and that for a given
nonzero value of $\gamma,$ the power increases with the sample size. Results
not reported here show that the test has a quite accurate size in that the
empirical rejection probability under the null is close to 5\%, the nominal
level of the test.


\section{Empirical application}

\label{app}

In this section, we consider two applications: education and wages, and
smoking and birth weights.

\subsection{Education and wages}

Our first application is based on a household labor survey from
\cite{wooldridge} that can be accessed online for replication.\footnote{See
\url{http://fmwww.bc.edu/ec-p/data/wooldridge/wage1.des} and
\url{http://fmwww.bc.edu/ec-p/data/wooldridge/wage1.dta} for the data in the
Stata data file format.} The idea is to evaluate the effects of education on
the quantile of the unconditional distribution of log wages. In this
application, $Y=lwage,$ which is log hourly wage, and $X=educ$, which is years
of education is our target variable. The controls are:
$W=[exper\ tenure\ nonwhite\ female]$, where $exper$ is years of working
experience, $tenure$ is years with current employer, $nonwhite$ is a dummy
that equals 1 if the individual is non-white, and $female$ is a dummy that
equals 1 if the individual is female. We assume that Assumption
\ref{Assumption ID} holds for this choice of $W.$

While the main goal is to study the scale effect, we also present results for
the location effect. We set $\dot{\ell}(0)=1$ and $\dot{s}(0)=-1$. Note that
when $\dot{s}(0)=-1,$ the estimated effects we present below are the
unconditional scale effects when the variance of the covariate is
\emph{reduced} by a small amount. For the mean of years of education $\mu_{X}%
$, we let $\mu_{X}=12.29$ based on the Barro-Lee Data on Educational
Attainment.\footnote{The dataset is available from
\url{https://databank.worldbank.org/reports.aspx?source=Education Statistics}
We use the series \textquotedblleft Barro-Lee: Average years of total
schooling, age 25+, total\textquotedblright\ for the US between 1970-2010 and
find that the average years of schooling is 12.29.} We set $\mu=\mu_{X}=12.29$
to study the location and scale effects. In a similar fashion to the Monte
Carlo analysis, we consider $\tau\in\{0.10,0.25,0.50,0.75,0.90\}$. The sample
size for the household labor survey is $n=526$, which is comparable to $n=500$
in the simulation exercises. We compute the standard errors using the
approximation in \eqref{asym_norm_individual_form2}.

\begin{table}[ptb]
\caption{Effects of location-scale shifts in education on the unconditional
quantiles of log-wage.}%
\label{tab:wage1}%
\centering
{\scriptsize {\
\begin{tabular}
[c]{l|cccccc}\hline\hline
&  & \textbf{$\tau=0.10$} & \textbf{$\tau=0.25$} & \textbf{$\tau=0.50$} &
\textbf{$\tau=0.75$} & \textbf{$\tau=0.90$}\\\hline
Location (probit) & {Estimate} & 0.039 & 0.062 & 0.101 & 0.101 & 0.118\\
& {} & (0.008) & (0.011) & (0.015) & (0.016) & (0.021)\\
& {$95\% \ CI_{L}$} & 0.025 & 0.041 & 0.072 & 0.069 & 0.076\\
& {$95\% \ CI_{U}$} & 0.054 & 0.083 & 0.129 & 0.132 & 0.160\\\hline
Location (logit) & {Estimate} & 0.038 & 0.065 & 0.103 & 0.100 & 0.120\\
& {} & (0.007) & (0.010) & (0.015) & (0.016) & (0.021)\\
& {$95\% \ CI_{L}$} & 0.024 & 0.044 & 0.074 & 0.069 & 0.080\\
& {$95\% \ CI_{U}$} & 0.053 & 0.085 & 0.131 & 0.132 & 0.160\\\hline\hline
Scale (probit) & {Estimate} & 0.045 & 0.029 & -0.025 & -0.103 & -0.203\\
& {} & (0.014) & (0.011) & (0.013) & (0.028) & (0.065)\\
& {$95\% \ CI_{L}$} & 0.018 & 0.007 & -0.051 & -0.158 & -0.330\\
& {$95\% \ CI_{U}$} & 0.071 & 0.052 & 0.001 & -0.049 & -0.077\\\hline
Scale (logit) & {Estimate} & 0.045 & 0.034 & -0.024 & -0.110 & -0.227\\
& {} & (0.014) & (0.012) & (0.014) & (0.029) & (0.066)\\
& {$95\% \ CI_{L}$} & 0.017 & 0.011 & -0.051 & -0.167 & -0.356\\
& {$95\% \ CI_{U}$} & 0.072 & 0.058 & 0.002 & -0.053 & -0.099\\\hline\hline
\end{tabular}
} }
\par
{\scriptsize Notes: standard errors are in parentheses. }\end{table}

\begin{figure}[ptb]
\centering
\includegraphics[scale=0.65]{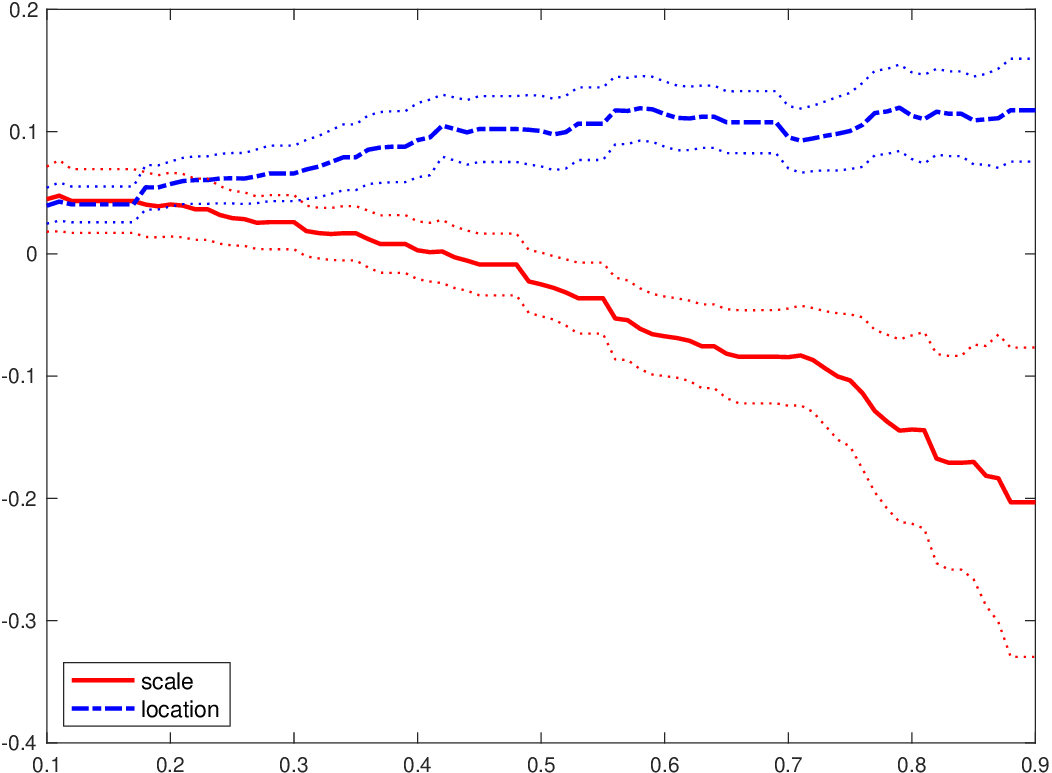}\caption{Point and interval
estimates of the location and scale effects of education on the unconditional
quantiles of log-wage based on the probit specification: $\Pi_{\tau,L}$
(dashed blue) and $\Pi_{\tau,S}$ (solid red). }%
\label{fig:emp_results}%
\end{figure}


The most interesting results in Table \ref{tab:wage1} appear in the
unconditional scale effects. As discussed in Section \ref{Sec: elasticity},
the scale effects can be interpreted as percentage changes of the
unconditional quantiles. Consider the scale effect for $\tau=0.10$. Both the
probit and logit specifications suggest an effect of about .045. Then, using
the quantile-standard deviation elasticity, a $1\%$ decrease in the standard
deviation of education would produce a positive effect of $.045\%$ on the
unconditional quantile at the quantile level $\tau=0.10$. Given that the
sample standard deviation of $educ$ is $2.77$, the $1\%$ decrease is
approximately a change in the standard deviation from $2.77$ to $2.74$.
Consider now the scale effect for $\tau=0.50$. In this case, both probit and
logit specifications provide a statistically insignificant effect (at the
$5\%$ level). Confront this with the results of Example
\ref{example_normal_location_model} where in the linear model $Y=\lambda
+X\gamma+U$, the scale effect $\Pi_{0.50,S}=0$ if both $X$ and $U$ are
symmetrically distributed around 0. Thus, $\hat{\Pi}_{0.50,S}\approx0$ is
consistent with a linear model with symmetrically distributed $X$ and $U.$
Finally, consider the scale effect for $\tau=0.90$, again using both probit
and logit specifications. In this case, the effects are negative, suggesting a
$1\%$ decrease in the standard deviation would reduce the upper $\tau=0.90$
quantile by $.20\%$ (probit) and $.23\%$ (logit). Overall this analysis shows
that the scale effects are monotonically decreasing in $\tau$. This can be
seen in Figure \ref{fig:emp_results} that plots, for a finer grid of $\tau
$,\footnote{For Figure \ref{fig:emp_results} we use $\tau
=0.10,0.11,...,0.89,0.90$.} the probit estimates for both the location (dashed
blue) and scale (solid red) effects.

How can this be interpreted? The location effects suggest that the marginal
contribution of one more year of education benefits more the upper parts of
the unconditional distribution of wages. The scale effects suggest the
contrary. Reducing the overall dispersion of education would increase the
lower quantile wages, but reduce the upper ones.

\begin{table}[ptb]
\caption{Effects of location-scale shifts in education on the unconditional
quantiles of log-wage.}%
\label{tab:wage2}%
\centering
{\scriptsize {\
\begin{tabular}
[c]{l|cccccc}\hline\hline
&  & \textbf{$\tau=0.10$} & \textbf{$\tau=0.25$} & \textbf{$\tau=0.50$} &
\textbf{$\tau=0.75$} & \textbf{$\tau=0.90$}\\\hline
Location (probit) & {Estimate} & 0.039 & 0.062 & 0.101 & 0.101 & 0.118\\
& {} & (0.008) & (0.011) & (0.015) & (0.016) & (0.021)\\
& {$95\% \ CI_{L}$} & 0.025 & 0.041 & 0.072 & 0.069 & 0.076\\
& {$95\% \ CI_{U}$} & 0.054 & 0.083 & 0.129 & 0.132 & 0.160\\\hline
Location (logit) & {Estimate} & 0.038 & 0.065 & 0.103 & 0.100 & 0.120\\
& {} & (0.007) & (0.010) & (0.015) & (0.016) & (0.021)\\
& {$95\% \ CI_{L}$} & 0.024 & 0.044 & 0.074 & 0.069 & 0.080\\
& {$95\% \ CI_{U}$} & 0.053 & 0.085 & 0.131 & 0.132 & 0.160\\\hline\hline
Scale (probit) & {Estimate} & 0.045 & 0.029 & -0.025 & -0.103 & -0.203\\
& {} & (0.014) & (0.011) & (0.013) & (0.028) & (0.065)\\
& {$95\% \ CI_{L}$} & 0.018 & 0.007 & -0.051 & -0.158 & -0.330\\
& {$95\% \ CI_{U}$} & 0.071 & 0.052 & 0.001 & -0.049 & -0.077\\\hline
Scale (logit) & {Estimate} & 0.045 & 0.034 & -0.024 & -0.110 & -0.227\\
& {} & (0.014) & (0.012) & (0.014) & (0.029) & (0.066)\\
& {$95\% \ CI_{L}$} & 0.017 & 0.011 & -0.051 & -0.167 & -0.356\\
& {$95\% \ CI_{U}$} & 0.072 & 0.058 & 0.002 & -0.053 & -0.099\\\hline\hline
\end{tabular}
} }
\par
{\scriptsize Notes: standard errors are in parentheses. }\end{table}

\subsection{Smoking and birth weight}

This second application considers the relationship between smoking during
pregnancy and the child's birth weight. This was previously studied by
\cite{Abrevaya2001}, \cite{KoenkerHallock01}, \cite{Rothe2010}, and
\cite{ChernozhukovFernandezVal11}. We use the natality data from the National
Vital Statistics System for the year 2018.\footnote{Available here:
\url{https://www.nber.org/research/data/vital-statistics-natality-birth-data}.}
The outcome variable is \emph{birth weight} in grams, while the target
variable is the average number of cigarettes smoked daily during pregnancy. We
focus on the sample of mothers who are smokers. The sample consists of 219,667 observations.

For this model $Y$ is \emph{birth weight} in grams and $X$ is the mother's
reported average number of cigarettes smoked per day during pregnancy. We use
the same covariates as \cite{Abrevaya2001}:\footnote{We omit the dummy of
whether the mother smoked during pregnancy because we focus on the sample of
smoking mothers.} $(i)$ a dummy for whether the mother is black; $(ii)$ a
dummy for marital status; $(iii)$ age and age squared; $(iv)$ a set of dummies
for education attainment: high school graduate, some college, and college
graduate; $(v)$ weight gain during pregnancy, $(vi)$ a set of dummies for
prenatal visit: visit during the second trimester, visit during the third
trimester, and no visit at all; and $(vii)$ a dummy for the sex of the child.

For this application, we set $\mu=0$, $l(\delta)\equiv0$, and $s(\delta
)=1/\left(  1+\delta\right)  $, so that according to
\eqref{eq_location_scale}, counterfactual cigarette consumption is now
$X_{\delta}=X/(1+\delta)$, which has a smaller mean and variance than $X.$
Note, again, that $\dot s(0)=-1$. To motivate such a counterfactual policy, we
can think of a tax on the price of cigarettes, which induces the consumer to
reduce cigarette consumption from $X$ to $X/(1+\delta)$. .\footnote{Suppose
that $\alpha_{x}$ is the exponent of $X$ in the Cobb-Douglas utility function.
Suppose further that the exponents are normalized to sum to 1. Then, if $M$ is
the income, and $p_{x}$ is the price of $X$, we have that $\alpha_{x}M=p_{x}%
X$. Similarly, under the proposed counterfactual tax $\alpha_{x}%
M=p_{x}(1+\delta)X_{\delta}$. It follows that $X_{\delta}=X/(1+\delta)$.}

\begin{figure}[ptb]
\caption{Point and interval estimates of the location-scale effects of smoking
on the unconditional quantiles of \emph{birth weight} based on the probit
specification.}%
\label{fig:emp_results_smoking}
\centering
\includegraphics[scale=0.65]{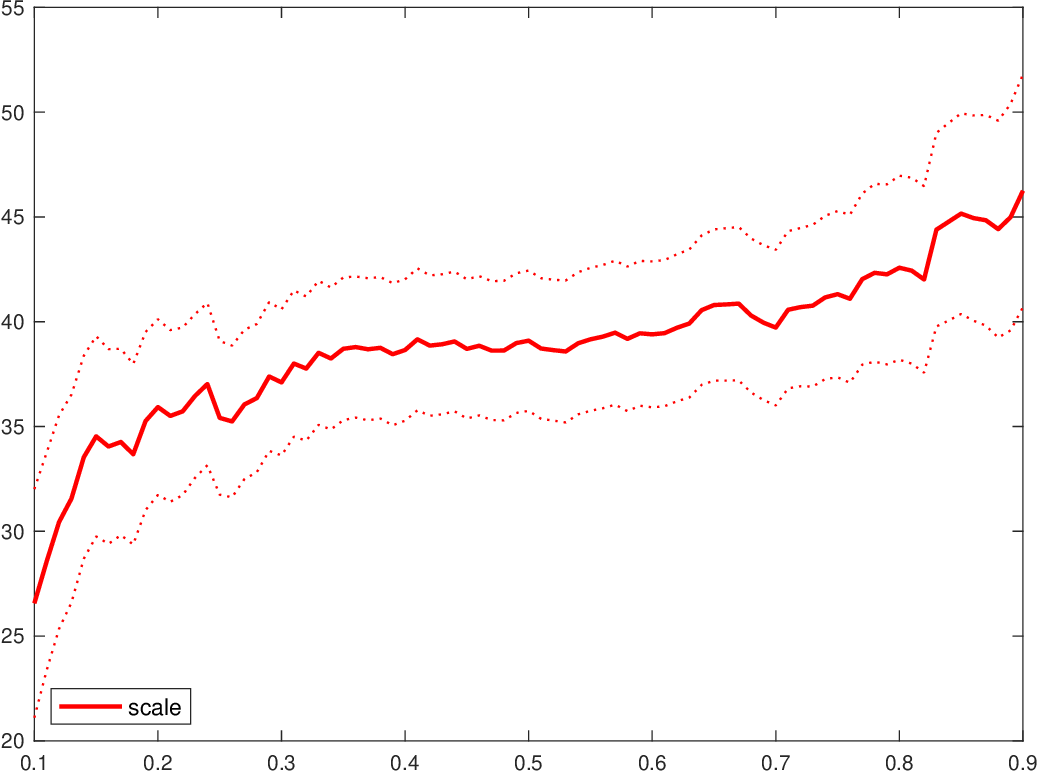}
\end{figure}

Table \ref{tab:smoking1} and Figure \ref{fig:emp_results_smoking} show the
results. The effects are positive and monotonically increasing across
quantiles. This means that the marginal impact on the \emph{birth weight} of a
tax on cigarettes is positive. The effects are stronger for upper quantiles of
the distribution of \emph{birth weight}. In order to interpret the magnitudes,
we use the quantile-standard deviation elasticity. According to
\eqref{eq:elasticity_0}, the elasticity can be calculated as $\mathcal{E}%
_{\tau,\delta=0}=-\Pi_{\tau,S}/Q_{\tau}[Y]$ as $\dot{s}\left(  0\right)  =-1.$
For example, for $\tau=0.50$, $\mathcal{E}_{0.50,\delta=0}=-0.0128$. This
means that a $1\%$ decrease in the standard deviation of the consumption of
cigarettes increases the median birth weight by $0.0128\%$.

\begin{table}[ptb]
\caption{Effects of a negative location-scale shift in smoking ($X_{\delta
}=X/\left(  1+\delta)\right)  $ on the unconditional quantiles of \emph{birth
weight}.}%
\label{tab:smoking1}
\centering
{\scriptsize {\
\begin{tabular}
[c]{l|cccccc}\hline\hline
&  & \textbf{$\tau=0.10$} & \textbf{$\tau=0.25$} & \textbf{$\tau=0.50$} &
\textbf{$\tau=0.75$} & \textbf{$\tau=0.90$}\\\hline
Scale (probit) & {Estimate} & 26.562 & 35.412 & 39.096 & 41.316 & 46.249\\
& {} & (2.784) & (1.870) & (1.708) & (2.024) & (2.834\\
& {$95\% \ CI_{L}$} & 21.106 & 31.746 & 35.749 & 37.349 & 40.694\\
& {$95\% \ CI_{U}$} & 32.018 & 39.078 & 42.443 & 45.282 & 51.803\\\hline
Scale (logit) & {Estimate} & 25.242 & 34.412 & 40.038 & 44.232 & 50.077\\
& {} & (2.722) & (1.848) & (1.711) & (1.984) & (2.695)\\
& {$95\% \ CI_{L}$} & 19.908 & 30.790 & 36.684 & 40.343 & 44.795\\
& {$95\% \ CI_{U}$} & 30.577 & 38.034 & 43.392 & 48.122 & 55.360\\\hline\hline
\end{tabular}
} }
\par
{\scriptsize Notes: standard errors are in parentheses.}\end{table}

\section{Conclusion}

\label{conclusion}

This paper has provided a general procedure to analyze the distributional
impact of changes in covariates on an outcome variable. The standard
unconditional quantile regression analysis focuses on a particular impact
coming from a location shift. We have provided a framework to study the
unconditional policy effects generated by a smooth and invertible intervention
of one or more target variables, allowing them to be possibly endogeneous. We
focus particularly on a location-scale shift and show how to additively
decompose the total effect into a location effect and a scale effect. They can
be analyzed and estimated separately. Additionally, we consider the case of
simultaneous changes in different covariates. We show how this can be obtained
from the usual vector-valued unconditional quantile regressions.

\bibliographystyle{econometrica}
\bibliography{CQRUQR.bib}

@string{jae = {Journal of Applied Econometrics}}

@article{Rothe2010,
	author = {Rothe, Christoph},
	journal = {Journal of Econometrics},
	number = {1},
	pages = {56--70},
	publisher = {Elsevier B.V.},
	title = {{Nonparametric Estimation of Distributional Policy Effects}},
	volume = {155},
	year = {2010}}

@article{Heckman2005,
	author = {Heckman, James J. and Vytlacil, Edward},
	journal = {Econometrica},
	number = {3},
	pages = {669--738},
	title = {{Structural Equations, Treatment Effects, and Econometric Policy Evaluation}},
	volume = {73},
	year = {2005}}

@article{Abrevaya2001,
	author = {Jason Abrevaya},
	journal = {Empirical Economics},
	pages = {247--257},
	title = {{The effects of Demographics and Maternal Behavior on the Distribution of birth outcomes}},
	volume = {26(1)},
	year = {2001}}

@article{Lieli2020,
	author = {Hsu, Yu-Chin and Lai, Tsung-Chih and Lieli, Robert P.},
	journal = {Journal of Business and Economic Statistics},
	title = {{Counterfactual Treatment Effects: Estimation and Inference}},
	volume = {Forthcoming},
	year = {2020}}

@unpublished{Spini2021,
	author = {Spini, Pietro},
	note = {Working Paper},
	title = {{Robustness, Heterogeneous Treatment Effects and Covariate Shifts}},
	year = {2021}}

@article{ChernozhukovFernandezVal11,
	author = {Victor Chernozhukov and Iv\'{a}n Fern\'{a}ndez-Val},
	journal = {Review of Economic Studies},
	pages = {559--589},
	title = {Inference for Extremal Conditional Quantile Models, with an Application to Market and Birthweight Risks},
	volume = {79},
	year = {2011}}

@unpublished{AutorKatzKearney05,
	author = {David H. Autor and Lawrence S. Katz and Melissa S. Kearney},
	note = {NBER Working Paper 11628},
	title = {Rising wage inequality: The role of composition and prices},
	year = {2005}}

@article{FirpoFortinLemieux09,
	author = {Sergio Firpo and Nicole Fortin and Thomas Lemieux},
	journal = {Econometrica},
	pages = {953-973},
	title = {Unconditional quantile regression},
	volume = {77(3)},
	year = {2009}}

@incollection{FortinLemieuxFirpo11,
	author = {Nicole Fortin and Thomas Lemieux and Sergio Firpo},
	booktitle = {Handbook of Labor Economics},
	editor = {O. Ashenfelter and D. Card},
	pages = {1--12},
	publisher = {Amsterdam: Elsevier},
	title = {Decomposition methods in economics},
	volume = {4},
	year = {2011}}

@unpublished{InoueLiXu21,
	author = {Atsushi Inoue and Tong Li and Qi Xu},
	note = {ARXIV: https://arxiv.org/pdf/2105.09445.pdf},
	title = {Two Sample Unconditional Quantile Effect},
	year = {2021}}

@article{KoenkerHallock01,
	author = {Roger Koenker and Kevin Hallock},
	journal = {Journal of Economic Perspectives},
	pages = {143--156},
	title = {Quantile Regression},
	volume = {15(4)},
	year = {2001}}

@article{MachadoMata05,
	author = {Jos\'{e} A. F. Machado and Jos\'{e} Mata},
	journal = JAE,
	pages = {445--465},
	title = {Counterfactual decomposition of changes in wage distributions using quantile},
	volume = {20},
	year = {1995}}

@article{Melly05,
	author = {Blaise Melly},
	journal = {Labour Economics},
	pages = {577--590},
	title = {Decomposition of differences in distribution using quantile regressions},
	volume = {12},
	year = {2005}}

@book{vanderVaart98,
	address = {Cambridge},
	author = {Aad van der Vaart},
	publisher = {Cambridge University Press},
	title = {Asymptotic Statistics},
	year = {1998}}

@book{serfling1980,
	author = {Robert J. Serfling},
	publisher = {New York: Wiley},
	title = {Approximation Theorems of Mathematical Statistics},
	year = {1980}}

@book{wooldridge,
	address = {Cambridge, MA},
	author = {Jeffrey M. Wooldridge},
	publisher = {MIT Press},
	title = {Econometric Analysis of Cross Section and Panel Data},
	year = {2002}}

@unpublished{yixiao2020,
	author = {Martinez-Iriarte, Julian and Sun, Yixiao},
	note = {Working Paper},
	title = {{Identification and Estimation of Unconditional Policy Effects of an Endogenous Binary Treatment: an Unconditional MTE Approach}},
	year = {2023}}

@unpublished{yixiao2021,
	author = {Martinez-Iriarte, Julian and Sun, Yixiao},
	note = {Working Paper},
	title = {{Characterizing Asymptotic Biases of Unconditional Regression Estimators of Policy Effects Under Endogeneity}},
	year = {2021}}

@book{casella,
	address = {Pacific Grove, CA},
	author = {Casella, George and Berger, Roger L.},
	publisher = {Duxbury},
	title = {Statistical Inference, 2nd. edition},
	year = {2001}}

@unpublished{martinez2020,
	author = {Martinez-Iriarte, Julian},
	note = {Working Paper},
	title = {Sensitivity Analysis in Unconditional Quantile Effects},
	year = {2023}}

@article{Rothe2012,
	author = {Rothe, Christoph},
	journal = {Econometrica},
	number = {5},
	pages = {2269--2301},
	title = {{Partial Distributional Policy Effects}},
	volume = {80},
	year = {2012}}

@article{Gu2020,
	author = {Gu, Grace W. and Malik, Samreem and Pozzoli, Dario and Rocha, Vera},
	journal = {Economic Inquiry},
	number = {1},
	pages = {241--259},
	title = {{Trade-induced Skill Polarization}},
	volume = {58},
	year = {2019}}

@article{Hanushek2008,
	author = {Hanushek, Eric A. and Woessmann, Ludger},
	journal = {Journal of Economic Literature},
	number = {46},
	pages = {607--668},
	title = {{The Role of Cognitive Skills in Economic Development}},
	volume = {3},
	year = {2008}}

@unpublished{SasakiUraZhang20,
	author = {Y. Sasaki and T. Ura and Y. Zhang},
	note = {Working Paper},
	title = {Unconditional Quantile Regression with High Dimensional Data},
	year = {2020}}

@article{NeweyIchimura2022,
	author = {Whitney K. Newey and Hidehiko Ichimura},
	issue = {1},
	journal = {Quantitative Economics},
	pages = {29--61},
	publisher = {Elsevier B.V.},
	title = {{The Influence Function of Semiparametric Estimators}},
	volume = {13},
	year = {2022}}

@unpublished{ziebarth2022,
	author = {Chris Vickers and Nicolas L. Ziebarth},
	note = {Working Paper},
	title = {The Effects of the National War Labor Board on Labor Income Inequality},
	year = {2022}}

@unpublished{cquq2023,
	author = {Javier Alejo and Antonio Galvao and Julian Martinez-Iriarte and Gabriel Montes-Rojas},
	note = {Working Paper},
	title = {Unconditional Quantile Partial Effects via Conditional Quantile Regression},
	year = {2023}}

@book{pagan_ullah_1999,
	author = {Pagan, Adrian and Ullah, Aman},
	collection = {Themes in Modern Econometrics},
	doi = {10.1017/CBO9780511612503},
	place = {Cambridge},
	publisher = {Cambridge University Press},
	series = {Themes in Modern Econometrics},
	title = {Nonparametric Econometrics},
	year = {1999}}

\bigskip

\bigskip

\section*{Appendix}

\renewcommand{\thesubsection}{A.\arabic{subsection}}

\setcounter{equation}{0} \renewcommand\theequation{A.\arabic{equation}}

\subsection{Proof of Theorem \ref{th:uqpe_scale}}

\textbf{Part (i).} To obtain the joint density of $\left(  X_{\delta
},W\right)  $, we note that
\[
F_{X_{\delta},W}(x,w)=\Pr(X_{\delta}\leq x,W\leq w)=\Pr(X\leq x^{\delta},W\leq
w)=F_{X,W}(x^{\delta},w),
\]
and so
\[
f_{X_{\delta},W}(x,w)=\frac{\partial x^{\delta}}{\partial x}\cdot
f_{X,W}(x^{\delta},w)=J(x^{\delta};\delta)f_{X,W}(x^{\delta},w).
\]
Evaluated at $\delta=0,$ $J\left(  x^{\delta};\delta\right)  $ is $1$ and
$f_{X_{\delta},W}(x,w)$ is $f_{X,W}(x,w)$. Given this, we expand
$f_{X_{\delta},W}(x,w)-f_{X,W}(x,w)$ around $\delta=0$, which is possible
under Assumptions \ref{Assumption:main}(i) and (iii.a). First, we observe that%
\begin{align*}
\frac{\partial J(x^{\delta};\delta)}{\partial x^{\delta}}  &  =\frac{\partial
}{\partial x^{\delta}}\frac{\partial x^{\delta}}{\partial x}=\frac{\partial
}{\partial x}\frac{\partial x^{\delta}}{\partial x^{\delta}}=0,\\
\frac{\partial J(x^{\delta};\delta)}{\partial\delta}\bigg|_{\delta=0}  &
=\frac{\partial}{\partial\delta}\frac{\partial x^{\delta}}{\partial
x}\bigg|_{\delta=0}=\frac{\partial}{\partial x}\frac{\partial x^{\delta}%
}{\partial\delta}\bigg|_{\delta=0}=-\frac{\partial\kappa\left(  x\right)
}{\partial x},
\end{align*}
where the last line follows from the fact that
\[
\kappa\left(  x\right)  :=\frac{\partial\mathcal{G}(x;\delta)}{\partial\delta
}\bigg|_{\delta=0}=-\frac{\partial x^{\delta}}{\partial\delta}\bigg|_{\delta
=0}.
\]
Differentiating both sides of $\mathcal{G}\left(  x^{\delta};\delta\right)
=x$ with respect to $\delta,$ we obtain that
\[
\frac{\partial x^{\delta}}{\partial\delta}=-\left(  \frac{\partial
\mathcal{G}\left(  x^{\delta};\delta\right)  }{\partial x^{\delta}}\right)
^{-1}\frac{\partial\mathcal{G}\left(  x^{\delta};\delta\right)  }%
{\partial\delta}%
\]
and so%
\[
-\frac{\partial x^{\delta}}{\partial\delta}\bigg|_{\delta=0}=\left(
\frac{\partial\mathcal{G}\left(  x;0\right)  }{\partial x}\right)  ^{-1}%
\frac{\partial\mathcal{G}\left(  x;\delta\right)  }{\partial\delta
}\bigg|_{\delta=0}=\frac{\partial\mathcal{G}\left(  x;\delta\right)
}{\partial\delta}\bigg|_{\delta=0}:=\kappa(x),
\]
where we have used $\mathcal{G}\left(  x;0\right)  =x.$ Now we have
\begin{align*}
&  f_{X_{\delta},W}(x,w)-f_{X,W}(x,w)\\
&  =J(x^{\delta};\delta)f_{X,W}(x^{\delta},w)-f_{X,W}(x,w)\\
&  =\delta\left[  \frac{\partial J\left(  x^{\delta};\delta\right)  }{\partial
x^{\delta}}\frac{\partial x^{\delta}}{\partial\delta}+\frac{\partial J\left(
x^{\delta};\delta\right)  }{\partial\delta}\right]  \bigg|_{\delta=0}%
f_{X,W}(x,w)\\
&  +\delta J\left(  x;0\right)  \frac{\partial f_{X,W}(x,w)}{\partial x}%
\frac{\partial x^{\delta}}{\partial\delta}\bigg|_{\delta=0}+\delta
R_{1}(x,w,\delta)\\
&  =\delta\frac{\partial J\left(  x^{\delta};\delta\right)  }{\partial\delta
}\bigg|_{\delta=0}f_{X,W}(x,w)+\delta J\left(  x;0\right)  \frac{\partial
f_{X,W}(x,w)}{\partial x}\frac{\partial x^{\delta}}{\partial\delta
}\bigg|_{\delta=0}+\delta R_{1}(x,w,\delta)\\
&  =\delta\left[  -\frac{\partial\kappa\left(  x\right)  }{\partial x}%
f_{X,W}(x,w)-\frac{\partial f_{X,W}(x,w)}{\partial x}\kappa\left(  x\right)
\right]  +\delta R_{1}(x,w,\delta)\\
&  =-\delta\frac{\partial}{\partial x}\left[  \kappa\left(  x\right)
f_{X,W}(x,w)\right]  +\delta R_{1}(x,w,\delta),
\end{align*}
where, for $\tilde{\delta}(x,w)$ between $0$ and $\delta$,
\begin{equation}
R_{1}(x,w,\delta)=\left\{  \frac{\partial\left[  J\left(  x^{\delta}%
;\delta\right)  f_{X,W}(x^{\delta},w)\right]  }{\partial\delta}\bigg|_{\delta
=\tilde{\delta}(x,w)}-\frac{\partial\left[  J\left(  x^{\delta};\delta\right)
f_{X,W}(x^{\delta},w)\right]  }{\partial\delta}\bigg|_{\delta=0}\right\}  .
\end{equation}
By the continuity of the derivative of $J\left(  x^{\delta};\delta\right)
f_{X,W}(x^{\delta},w)$ with respect to $\delta$, we have $R_{1}(x,w,\delta
)=o(1)$ for each $\left(  x,w\right)  $ as $\delta\rightarrow0.$

\textbf{Part (ii).} Consider first the counterfactual distribution
$F_{Y_{\delta}}$:
\[
F_{Y_{\delta}}(y)=\int_{\mathcal{W}}\int_{\mathcal{X}}\int_{\mathcal{U}%
}\mathds1\left\{  h(x,w,u)\leq y\right\}  f_{U|X_{\delta},W}%
(u|x,w)f_{X_{\delta},W}(x,w)dudxdw,
\]
where for simplicity we have assumed that the support of $X$ conditional on
any $W=w$ does not depend on $w$ and we have denoted the support by
$\mathcal{X}$. By Assumption \ref{Assumption:main}(ii), $f_{U|X_{\delta}%
,W}(u|x,w)=f_{U|X,W}(u|x^{\delta},w)$. So we can write
\begin{align*}
&  F_{Y_{\delta}}(y)\\
&  =\int_{\mathcal{W}}\int_{\mathcal{X}}\int_{\mathcal{U}}\mathds1\left\{
h(x,w,u)\leq y\right\}  f_{U|X_{\delta},W}(u|x,w)f_{X_{\delta},W}(x,w)dudxdw\\
&  =\int_{\mathcal{W}}\int_{\mathcal{X}}\int_{\mathcal{U}}\mathds1\left\{
h(x,w,u)\leq y\right\}  f_{U|X,W}(u|x^{\delta},w)f_{X_{\delta},W}(x,w)dudxdw\\
&  =\underbrace{\int_{\mathcal{W}}\int_{\mathcal{X}}\int_{\mathcal{U}%
}\mathds1\left\{  h(x,w,u)\leq y\right\}  f_{U|X,W}(u|x,w)f_{X,W}%
(x,w)dudxdw}_{=F_{Y}(y)}\\
&  +\int_{\mathcal{W}}\int_{\mathcal{X}}\int_{\mathcal{U}}\mathds1\left\{
h(x,w,u)\leq y\right\}  f_{U|X,W}(u|x,w)\left[  f_{X_{\delta},W}%
(x,w)-f_{X,W}(x,w)\right]  dudxdw\\
&  +\int_{\mathcal{W}}\int_{\mathcal{X}}\int_{\mathcal{U}}\mathds1\left\{
h(x,w,u)\leq y\right\}  \left[  f_{U|X,W}(u|x^{\delta},w)-f_{U|X,W}%
(u|x,w)\right]  f_{X_{\delta},W}(x,w)dudxdw.
\end{align*}
Hence, we have
\[
\frac{F_{Y_{\delta}}(y)-F_{Y}(y)}{\delta}:=G_{1,\delta}\left(  y\right)
+G_{2,\delta}\left(  y\right)  ,
\]
where%
\begin{align}
G_{1,\delta}\left(  y\right)   &  =\int_{\mathcal{W}}\int_{\mathcal{X}}%
\int_{\mathcal{U}}\mathds1\left\{  h(x,w,u)\leq y\right\}  f_{U|X,W}%
(u|x,w)\frac{1}{\delta}\left[  f_{X_{\delta},W}(x,w)-f_{X,W}(x,w)\right]
dudxdw\nonumber\\
&  =\int_{\mathcal{W}}\int_{\mathcal{X}}F_{Y|X,W}(y|x,w)\frac{1}{\delta
}\left[  f_{X_{\delta},W}(x,w)-f_{X,W}(x,w)\right]  dxdw,\label{eq:quotien_G1}%
\end{align}
and
\begin{align}
G_{2,\delta}\left(  y\right)   &  =\int_{\mathcal{W}}\int_{\mathcal{X}}%
\int_{\mathcal{U}}\mathds1\left\{  h(x,w,u)\leq y\right\}  \nonumber\\
&  \times\frac{1}{\delta}\left[  f_{U|X,W}(u|x^{\delta},w)-f_{U|X,W}%
(u|x,w)\right]  f_{X_{\delta},W}(x,w)dudxdw.\label{eq:quotien_G2}%
\end{align}

We first consider the term $G_{1,\delta}\left(  y\right)  .$ Using Part (i)
and Assumption \ref{Assumption:main}(iv), we have
\begin{align*}
G_{1,\delta}\left(  y\right)   &  =-\int_{\mathcal{W}}\int_{\mathcal{X}%
}F_{Y|X,W}\left(  y|x,w\right)  \frac{\partial\left[  \kappa\left(  x\right)
f_{X,W}(x,w)\right]  }{\partial x}\\
&  +\int_{\mathcal{W}}\int_{\mathcal{X}}F_{Y|X,W}(y|x,w)R_{1}(x,w,\delta
)dxdw\\
&  =\int_{\mathcal{W}}\int_{\mathcal{X}}\frac{\partial F_{Y|X,W}%
(y|x,w)}{\partial x}\kappa\left(  x\right)  f_{X,W}(x,w)dxdw\\
&  +\int_{\mathcal{W}}\int_{\mathcal{X}}F_{Y|X,W}(y|x,w)R_{1}(x,w,\delta)dxdw,
\end{align*}
where the second equality follows from integration by parts. Under Assumption
\ref{Assumption:main}(iii.a), we can use the dominated convergence theorem to
obtain%
\[
\lim_{\delta\rightarrow0}\sup_{y\in\mathcal{Y}}\bigg |\int_{\mathcal{W}}%
\int_{\mathcal{X}}F_{Y|X,W}(y|x,w)R_{1}(x,w,\delta)dxdw\bigg |=0.
\]
Thus, we have that $G_{1,\delta}\left(  y\right)  $ converges to
$G_{1,0}\left(  y\right)  $, given by
\[
G_{1,0}\left(  y\right)  :=\int_{\mathcal{W}}\int_{\mathcal{X}}\frac{\partial
F_{Y|X,W}(y|x,w)}{\partial x^{\prime}}\kappa\left(  x\right)  f_{X,W}%
(x,w)dxdw
\]
uniformly in $y\in\mathcal{Y}$, as $\delta\rightarrow0$.

Next, we consider $G_{2,\delta}\left(  y\right)  .$ Using Assumption
\ref{Assumption:main}(iii.b), we have
\begin{align*}
&  \left[  f_{U|X,W}(u|x^{\delta},w)-f_{U|X,W}(u|x,w)\right]  f_{X,W}%
(x^{\delta},w)\\
&  =\frac{\partial f_{U|X,W}(u|x^{\delta},w)}{\partial x^{\delta\prime}%
}f_{X,W}(x^{\delta},w)\frac{\partial x^{\delta}}{\partial\delta}%
\bigg|_{\delta=0}\cdot\delta+\delta R_{2}(u,x,w,\delta)\\
&  =-\frac{\partial f_{U|X,W}(u|x,w)}{\partial x^{\prime}}f_{X,W}%
(x,w)\kappa\left(  x\right)  \delta+\delta R_{2}(u,x,w,\delta),
\end{align*}
where%
\begin{align*}
&  R_{2}(u,x,w,\delta)\\
&  =\frac{\partial\left[  f_{U|X,W}(u|x^{\delta},w)f_{X,W}(x^{\delta
},w)\right]  }{\partial\delta}\bigg|_{\delta=\tilde{\delta}(u,x,w)}%
-\frac{\partial\left[  f_{U|X,W}(u|x^{\delta},w)f_{X,W}(x^{\delta},w)\right]
}{\partial\delta}\bigg|_{\delta=0}\\
&  -\left[  f_{U|X,W}(u|x,w)\frac{\partial f_{X,W}(x^{\delta},w)}%
{\partial\delta}\bigg|_{\delta=\tilde{\delta}(u,x,w)}-f_{U|X,W}(u|x,w)\frac
{\partial f_{X,W}(x^{\delta},w)}{\partial\delta}\bigg|_{\delta=0}\right]  .
\end{align*}
Note that in the above, the transpose on $x$ is not relevant but we keep it so
that the same lines of arguments can be used for proving Theorem
\ref{th:comp_change}. Hence%
\begin{align*}
G_{2,\delta}\left(  y\right)   &  =-\int_{\mathcal{W}}\int_{\mathcal{X}}%
\int_{\mathcal{U}}\mathds1\left\{  h(x,w,u)\leq y\right\}  \frac{\partial
f_{U|X,W}(u|x,w)}{\partial x^{\prime}}f_{X,W}(x,w)\kappa\left(  x\right)
dudxdw\\
&  +\int_{\mathcal{W}}\int_{\mathcal{X}}\int_{\mathcal{U}}\mathds1\left\{
h(x,w,u)\leq y\right\}  R_{2}(u,x,w,\delta)dudxdw.
\end{align*}
Under Assumption \ref{Assumption:main}(iii.b), we can invoke the dominated
convergence theorem to get
\[
\lim_{\delta\rightarrow0}\sup_{y\in\mathcal{Y}}\left\vert \int_{\mathcal{W}%
}\int_{\mathcal{X}}\int_{\mathcal{U}}\mathds1\left\{  h(x,w,u)\leq y\right\}
R_{2}(u,x,w,\delta)dudxdw\right\vert =0.
\]
Hence, uniformly in $y\in\mathcal{Y}$, as $\delta\rightarrow0$, $G_{2,\delta
}\left(  y\right)  $ converges to
\begin{align*}
G_{2,0}\left(  y\right)   &  =-\int_{\mathcal{W}}\int_{\mathcal{X}}%
\int_{\mathcal{U}}\mathds1\left\{  h(x,w,u)\leq y\right\}  \frac{\partial
f_{U|X,W}(u|x,w)}{\partial x^{\prime}}\kappa\left(  x\right)  f_{X,W}%
(x,w)dudxdw\\
&  =-\int_{\mathcal{W}}\int_{\mathcal{X}}\int_{\mathcal{U}}\mathds1\left\{
h(x,w,u)\leq y\right\}  \frac{\partial\ln f_{U|X,W}(u|x,w)}{\partial
x^{\prime}}\kappa\left(  x\right)  f_{U|X,W}(u|x,w)f_{X,W}(x,w)dudxdw\\
&  =-E\left[  \mathds1\left\{  h(X,W,U)\leq y\right\}  \frac{\partial\ln
f_{U|X,W}(U|X,W)}{\partial X^{\prime}}\kappa\left(  X\right)  \right]  .
\end{align*}

Combining the above results yields
\begin{align*}
&  \frac{F_{Y_{\delta}}(y)-F_{Y}\left(  y\right)  }{\delta}\\
&  \rightarrow G_{10}\left(  y\right)  +G_{20}(y)\\
&  =E\left[  \left(  \frac{\partial F_{Y|X,W}(y|X,W)}{\partial X^{\prime}%
}-\mathds1\left\{  h(X,W,U)\leq y\right\}  \frac{\partial\ln f_{U|X,W}%
(U|X,W)}{\partial X^{\prime}}\right)  \kappa\left(  X\right)  \right] \\
&  :=G\left(  y\right)
\end{align*}
uniformly over $y\in\mathcal{Y}$ as $\delta\rightarrow0.$

\textbf{Part (iii). }Note that $\psi\left(  y,\tau,F_{Y}\right)  $ is the
influence function of the quantile functional. Using Part (ii) and Assumption
\ref{Assumption:main}(v), we have%
\[
\Pi_{\tau}=\int_{\mathcal{Y}}\psi\left(  y,\tau,F_{Y}\right)  dG\left(
y\right)  =\int_{\mathcal{Y}}\psi\left(  y,\tau,F_{Y}\right)  dG_{1,0}\left(
y\right)  +\int_{\mathcal{Y}}\psi\left(  y,\tau,F_{Y}\right)  dG_{2,0}\left(
y\right)
\]
by Lemma 21.3 in \cite{vanderVaart98}. Now
\begin{align*}
\int_{\mathcal{Y}}\psi\left(  y,\tau,F_{Y}\right)  dG_{1,0}\left(  y\right)
&  =\int_{\mathcal{Y}}\psi\left(  y,\tau,F_{Y}\right)  dG_{1,0}\left(
y\right) \\
&  =\int_{\mathcal{W}}\int_{\mathcal{X}}\left[  \int_{\mathcal{Y}}\psi\left(
y,\tau,F_{Y}\right)  \frac{\partial f_{Y|X,W}(y|x,w)}{\partial x^{\prime}%
}dy\right]  \kappa\left(  x\right)  f_{X,W}(x,w)dxdw\\
&  =\int_{\mathcal{W}}\int_{\mathcal{X}}\frac{\partial}{\partial x^{\prime}%
}\left[  \int_{\mathcal{Y}}\psi\left(  y,\tau,F_{Y}\right)  f_{Y|X,W}%
(y|x,w)dy\right]  \kappa\left(  x\right)  f_{X,W}(x,w)dxdw\\
&  =\int_{\mathcal{W}}\int_{\mathcal{X}}\frac{\partial E\left[  \psi\left(
Y,\tau,F_{Y}\right)  |X=x,W=w\right]  }{\partial x^{\prime}}\kappa\left(
x\right)  f_{X,W}(x,w)dxdw
\end{align*}
and
\begin{align*}
&  \int_{\mathcal{Y}}\psi\left(  y,\tau,F_{Y}\right)  dG_{2,0}\left(  y\right)
\\
&  =-\int_{\mathcal{W}}\int_{\mathcal{X}}\int_{\mathcal{U}}\left[
\int_{\mathcal{Y}}\psi\left(  y,\tau,F_{Y}\right)  d\mathds1\left\{
h(x,w,u)\leq y\right\}  \right]  \frac{\partial\ln f_{U|X,W}(u|x,w)}{\partial
x^{\prime}}\kappa\left(  x\right) \\
&  \times f_{U|X,W}(u|x,w)f_{X,W}(x,w)dudxdw\\
&  =-\int_{\mathcal{W}}\int_{\mathcal{X}}\int_{\mathcal{U}}\psi\left(
h(x,w,u),\tau,F_{Y}\right)  \frac{\partial\ln f_{U|X,W}(u|x,w)}{\partial
x^{\prime}}\kappa\left(  x\right) \\
&  \times f_{U|X,W}(u|x,w)f_{X,W}(x,w)dudxdw.
\end{align*}
Therefore,
\begin{align*}
\Pi_{\tau}  &  =\int_{\mathcal{W}}\int_{\mathcal{X}}\frac{\partial E\left[
\psi\left(  y,\tau,F_{Y}\right)  |X=x,W=w\right]  }{\partial x^{\prime}}%
\kappa\left(  x\right)  f_{X,W}(x,w)dxdwdy\\
&  -\int_{\mathcal{W}}\int_{\mathcal{X}}\int_{\mathcal{U}}\psi\left(
h(x,w,u),\tau,F_{Y}\right)  \frac{\partial\ln f_{U|X,W}(u|x,w)}{\partial
x^{\prime}}\kappa\left(  x\right) \\
&  \times f_{U|X,W}(u|x,w)f_{X,W}(x,w)dudxdw\\
&  =A_{\tau}-B_{\tau}.
\end{align*}

\subsection{Proof of Lemma \ref{lemma_mle_alpha_beta}}

The main complication in this lemma is that the dependent variable is
$\mathds 1\left\{  Y_{i}\leq\hat q_{\tau}\right\}  $. This means that the
preliminary estimator $\hat q_{\tau}$ might affect the asymptotic distribution
of $\hat\alpha_{\tau}$ and $\hat\beta_{\tau}$.

As mentioned in the main text, under Assumption \ref{assumption_quantile},
\[
\hat{q}_{\tau}-Q_{\tau}[Y]=\frac{1}{n}\sum_{i=1}^{n}\frac{\tau
-\mathds{1}\left\{  Y_{i}\leq Q_{\tau}[Y]\right\}  }{f_{Y}(Q_{\tau}[Y])}%
+o_{p}(n^{-1/2})=\frac{1}{n}\sum_{i=1}^{n}\psi(Y_{i},\tau,F_{Y})+o_{p}%
(n^{-1/2}).
\]

Recall that
\[
\hat{\theta}_{\tau}=\arg\max_{\theta\in\Theta}\sum_{i=1}^{n}%
\bigg\{\mathds1\left\{  Y_{i}\leq\hat{q}_{\tau}\right\}  \log\left[
G(Z_{i}^{\prime}\theta)\right]  +\mathds1\left\{  Y_{i}>\hat{q}_{\tau
}\right\}  \log\left[  1-G(Z_{i}^{\prime}\theta)\right]  \bigg\}.
\]
Let $s_{i}(\theta;\hat{q}_{\tau})$ denote the score for observation $i$. Then,
under Assumption \ref{assumption_logit_probit}(i), we have
\[
\frac{1}{n}\sum_{i=1}^{n}s_{i}(\hat{\theta}_{\tau};\hat{q}_{\tau})=0.
\]
Taking a mean-value expansion (element-by-element), we obtain
\[
\underbrace{\frac{1}{n}\sum_{i=1}^{n}s_{i}(\hat{\theta}_{\tau};\hat{q}_{\tau
})}_{=0}=\frac{1}{n}\sum_{i=1}^{n}s_{i}(\theta_{\tau};\hat{q}_{\tau})+\frac
{1}{n}\sum_{i=1}^{n}H_{i}(\tilde{\theta}_{\tau};\hat{q}_{\tau})\left(
\hat{\theta}_{\tau}-\theta_{\tau}\right)  ,
\]
where $\tilde{\theta}_{\tau}$ is between $\theta_{\tau}$ and $\hat{\theta
}_{\tau}$ and can be different for different rows of $H_{i}.$ Under the
assumption of the uniform law of large numbers for the Hessian (i.e.,
Assumption \ref{assumption_logit_probit}(ii)), we obtain
\[
\frac{1}{n}\sum_{i=1}^{n}H_{i}(\tilde{\theta}_{\tau},\hat{q}_{\tau
})\overset{p}{\rightarrow}E[H_{i}(\theta_{\tau};Q_{\tau}[Y])]=:H.
\]
We have then
\begin{equation}
0=\frac{1}{n}\sum_{i=1}^{n}s_{i}(\theta_{\tau};\hat{q}_{\tau})+H\left(
\hat{\theta}_{\tau}-\theta_{\tau}\right)  +o_{p}\left(  \left\Vert \hat
{\theta}_{\tau}-\theta_{\tau}\right\Vert \right)  . \label{eq_score_mv}%
\end{equation}

Now, we use the stochastic equicontinuity in Assumption
\ref{assumption_logit_probit}(iii):
\[
\frac{1}{n}\sum_{i=1}^{n}\left(  s_{i}(\theta_{\tau};\hat{q}_{\tau})-E\left[
s_{i}(\theta_{\tau};q)\right]  |_{q=\hat{q}_{\tau}}\right)  =\frac{1}{n}%
\sum_{i=1}^{n}s_{i}(\theta_{\tau};Q_{\tau}[Y])+o_{p}(n^{-1/2}).
\]
Here we have used that $E[s_{i}(\theta_{\tau};Q_{\tau}[Y])]=0$: the score
evaluated at the true quantile has expected value 0. Plugging this back into
\eqref{eq_score_mv}, we obtain
\begin{equation}
0=E\left[  s_{i}(\theta_{\tau};q)\right]  |_{q=\hat{q}_{\tau}}+\frac{1}{n}%
\sum_{i=1}^{n}s_{i}(\theta_{\tau};Q_{\tau}[Y])+H\left(  \hat{\theta}_{\tau
}-\theta_{\tau}\right)  +o_{p}\left(  \left\Vert \hat{\theta}_{\tau}%
-\theta_{\tau}\right\Vert \right)  . \label{eq_score_mv_2}%
\end{equation}
Here $E\left[  s_{i}(\theta_{\tau};q)\right]  |_{q=\hat{q}_{\tau}}$ is random
because we first compute the expectation $E\left[  s_{i}(\theta_{\tau
};q)\right]  $ for a fixed $q$ and then replace $q$ by $\hat{q}_{\tau}$, which
is random. To show that $E\left[  s_{i}(\theta_{\tau};q)\right]  |_{q=\hat
{q}_{\tau}}$ is $O_{p}(n^{-1/2})$, we observe that (see equation 15.18 in
\cite{wooldridge})
\begin{equation}
s_{i}(\theta;q)=\dfrac{g(Z_{i}^{\prime}\theta)Z_{i}\left[  \mathds1\left\{
Y_{i}\leq q\right\}  -G\left(  Z_{i}^{\prime}\theta\right)  \right]
}{G\left(  Z_{i}^{\prime}\theta\right)  \left[  1-G\left(  Z_{i}^{\prime
}\theta\right)  \right]  }. \label{eq_score}%
\end{equation}
Therefore, using the law of iterated expectations, we obtain
\[
E\left[  s_{i}(\theta;q)\right]  =E\left[  \dfrac{g(Z_{i}^{\prime}\theta
)Z_{i}\left[  F_{Y|X,W}(q|X_{i},W_{i})-G\left(  Z_{i}^{\prime}\theta\right)
\right]  }{G\left(  Z_{i}^{\prime}\theta\right)  \left[  1-G\left(
Z_{i}^{\prime}\theta\right)  \right]  }\right]  .
\]
So
\begin{equation}
H_{Q}=\frac{\partial E\left[  s_{i}(\theta_{\tau};q)\right]  }{\partial
q}\bigg\vert_{q=Q_{\tau}[Y]}=E\left[  \dfrac{g(Z_{i}^{\prime}\theta_{\tau
})Z_{i}\left[  f_{Y|X,W}(Q_{\tau}[Y]|X_{i},W_{i})\right]  }{G\left(
Z_{i}^{\prime}\theta_{\tau}\right)  \left[  1-G\left(  Z_{i}^{\prime}%
\theta_{\tau}\right)  \right]  }\right]  . \label{eq_exp_score}%
\end{equation}
We have
\begin{align*}
E\left[  s_{i}(\theta_{\tau};q)\right]  |_{q=\hat{q}_{\tau}}  &
=\underbrace{E\left[  s_{i}(\theta_{\tau};Q_{\tau}[Y])\right]  }%
_{=0}+\underbrace{\frac{\partial E\left[  s_{i}(\theta_{\tau};q)\right]
}{\partial q}\bigg\vert_{q=Q_{\tau}[Y]}}_{=H_{Q}}\left(  \hat{q}_{\tau
}-Q_{\tau}[Y]\right)  +o_{p}(n^{-1/2})\\
&  =H_{Q}\left(  \hat{q}_{\tau}-Q_{\tau}[Y]\right)  +o_{p}(n^{-1/2}),
\end{align*}
which implies that $E\left[  s_{i}(\theta_{\tau};q)\right]  |_{q=\hat{q}%
_{\tau}}=O_{p}(n^{-1/2})$. Going back to \eqref{eq_score_mv_2}, we obtain
\[
\left\Vert H\left(  \hat{\theta}_{\tau}-\theta_{\tau}\right)  +o_{p}\left(
\left\Vert \hat{\theta}_{\tau}-\theta_{\tau}\right\Vert \right)  \right\Vert
\leq\left\Vert E\left[  s_{i}(\theta_{\tau};q)\right]  |_{q=\hat{q}_{\tau}%
}\right\Vert +\left\Vert \frac{1}{n}\sum_{i=1}^{n}s_{i}(\theta_{\tau};Q_{\tau
}[Y])\right\Vert ,
\]
which implies that
\[
\hat{\theta}_{\tau}-\theta_{\tau}=O_{p}(n^{-1/2}).
\]
Furthermore, since $H$ is negative definite, then we have
\begin{align}
\hat{\theta}_{\tau}-\theta_{\tau}  &  =\underbrace{-H^{-1}\frac{1}{n}%
\sum_{i=1}^{n}s_{i}(\theta_{\tau};Q_{\tau}[Y])}_{\text{Usual influence
function}}-\underbrace{H^{-1}E\left[  s_{i}(\theta_{\tau};q)\right]
|_{q=\hat{q}_{\tau}}}_{\text{Contribution of }\hat{q}_{\tau}}+o_{p}%
(n^{-1/2})\nonumber\\
&  =-H^{-1}\frac{1}{n}\sum_{i=1}^{n}s_{i}(\theta_{\tau};Q_{\tau}%
[Y])-H^{-1}H_{Q}\left(  \hat{q}_{\tau}-Q_{\tau}[Y]\right)  +o_{p}%
(n^{-1/2})\nonumber\\
&  =-H^{-1}\frac{1}{n}\sum_{i=1}^{n}s_{i}(\theta_{\tau};Q_{\tau}%
[Y])-H^{-1}H_{Q}\frac{1}{n}\sum_{i=1}^{n}\psi(Y_{i},\tau,F_{Y})+o_{p}%
(n^{-1/2}). \label{eq:probit_coef}%
\end{align}

\subsection{Proof of Theorem \ref{theorem_est_pi}}

To establish the joint asymptotic distribution of the estimators of the
location and scale effect, we need to obtain the asymptotic distribution of
$\hat{f}_{Y}\left(  \hat{q}_{\tau}\right)  $. By Lemma 6 in \cite{yixiao2020},
we have that
\begin{equation}
\hat{f}_{Y}(y)-f_{Y}(y)=\frac{1}{n}\sum_{i=1}^{n}\mathcal{K}_{h}\left(
Y_{i}-y\right)  -E\left[  K_{h}\left(  Y-y\right)  \right]  +B_{f}%
(y)+o_{p}(h^{2}), \label{density_error}%
\end{equation}
where the bias is
\[
B_{f_{Y}}(y)=\frac{1}{2}h^{2}f_{Y}^{\prime\prime2}(y)\int_{-\infty}^{\infty
}u^{2}\mathcal{K}(u)du.
\]
Moreover, we can write
\[
\hat{f}_{Y}(\hat{q}_{\tau})-\hat{f}_{Y}(Q_{\tau}[Y])=\dot{f}_{Y}(Q_{\tau
}[Y])\left(  \hat{q}_{\tau}-Q_{\tau}[Y]\right)  +o_{p}(n^{-1/2}h^{-1/2}),
\]
where $\dot{f}_{Y}$ is the derivative of the density. Thus, we have that
\begin{align}
&  \hat{f}_{Y}\left(  \hat{q}_{\tau}\right)  -f_{Y}\left(  Q_{\tau}[Y]\right)
\nonumber\\
&  =\hat{f}_{Y}\left(  \hat{q}_{\tau}\right)  -\hat{f}_{Y}\left(  Q_{\tau
}[Y]\right)  +\hat{f}_{Y}\left(  Q_{\tau}[Y]\right)  -f_{Y}\left(  Q_{\tau
}[Y]\right) \nonumber\\
&  =\dot{f}_{Y}(Q_{\tau}[Y])\left(  \hat{q}_{\tau}-Q_{\tau}[Y]\right)
+\hat{f}_{Y}\left(  Q_{\tau}[Y]\right)  -f_{Y}\left(  Q_{\tau}[Y]\right)
+o_{p}(n^{-1/2}h^{-1/2}). \label{density_exp}%
\end{align}
The first term captures the uncertainty associated with estimating the
quantile, and the second term captures the uncertainty associated with
estimating the density.

Next, we can write the location and scale effects as
\[%
\begin{pmatrix}
\hat{\Pi}_{\tau,L}\\
\hat{\Pi}_{\tau,S}%
\end{pmatrix}
-%
\begin{pmatrix}
\Pi_{\tau,L}\\
\Pi_{\tau,S}%
\end{pmatrix}
=D_{\mu}\left[  \frac{n^{-1}\sum_{i=1}^{n}g_{\dot{\phi}}(Z_{i};\hat{\theta
}_{\tau})\hat{\alpha}_{\tau}\tilde{X}_{i}}{\hat{f}_{Y}\left(  \hat{q}_{\tau
}\right)  }-\frac{E\left[  g_{\dot{\phi}}(Z_{i};\theta_{\tau})\alpha_{\tau
}\tilde{X}_{i}\right]  }{f_{Y}\left(  Q_{\tau}[Y]\right)  }\right]  .
\]
where
\[
g_{\dot{\phi}}\left(  Z_{i};\theta_{\tau}\right)  =g(Z_{i}^{\prime}%
\theta_{\tau})\dot{\phi}_{\mathrm{x}}\left(  X_{i}\right)  ^{\prime}.
\]
Now%
\begin{align*}
&  \frac{n^{-1}\sum_{i=1}^{n}g_{\dot{\phi}}(Z_{i};\hat{\theta}_{\tau}%
)\hat{\alpha}_{\tau}\tilde{X}_{i}}{\hat{f}_{Y}\left(  \hat{q}_{\tau}\right)
}-\frac{E\left[  g_{\dot{\phi}}(Z_{i};\theta_{\tau})\alpha_{\tau}\tilde{X}%
_{i}\right]  }{f_{Y}\left(  Q_{\tau}[Y]\right)  }\\
&  =\frac{n^{-1}\sum_{i=1}^{n}\left[  g_{\dot{\phi}}(Z_{i};\hat{\theta}_{\tau
})\hat{\alpha}_{\tau}\tilde{X}_{i}\right]  -E\left[  g_{\dot{\phi}}%
(Z_{i};\theta_{\tau})\alpha_{\tau}\tilde{X}_{i}\right]  }{\hat{f}_{Y}\left(
\hat{q}_{\tau}\right)  }-E\left[  g_{\dot{\phi}}(Z_{i};\theta_{\tau}%
)\alpha_{\tau}\tilde{X}_{i}\right]  \frac{\hat{f}_{Y}\left(  \hat{q}_{\tau
}\right)  -f_{Y}\left(  Q_{\tau}[Y]\right)  }{\hat{f}_{Y}\left(  \hat{q}%
_{\tau}\right)  f_{Y}\left(  Q_{\tau}[Y]\right)  }\\
&  =\frac{n^{-1}\sum_{i=1}^{n}\left[  g_{\dot{\phi}}(Z_{i};\hat{\theta}_{\tau
})\hat{\alpha}_{\tau}\tilde{X}_{i}\right]  -E\left[  g_{\dot{\phi}}%
(Z_{i};\theta_{\tau})\alpha_{\tau}\tilde{X}_{i}\right]  }{\hat{f}_{Y}\left(
\hat{q}_{\tau}\right)  }\\
&  -\frac{E\left[  g_{\dot{\phi}}(Z_{i};\theta_{\tau})\alpha_{\tau}\tilde
{X}_{i}\right]  }{f_{Y}\left(  Q_{\tau}[Y]\right)  ^{2}}\left\{  \dot{f}%
_{Y}(Q_{\tau}[Y])\left(  \hat{q}_{\tau}-Q_{\tau}[Y]\right)  +\hat{f}%
_{Y}\left(  Q_{\tau}[Y]\right)  -f_{Y}\left(  Q_{\tau}[Y]\right)  \right\}
+o_{p}(n^{-1/2}h^{-1/2}).
\end{align*}

Taking a mean-value expansion (element-by-element), we have
\begin{align*}
\frac{1}{n}\sum_{i=1}^{n}g_{\dot{\phi}}(Z_{i};\hat{\theta}_{\tau})\hat{\alpha
}_{\tau}\tilde{X}_{i}  &  =\frac{1}{n}\sum_{i=1}^{n}g_{\dot{\phi}}%
(Z_{i};\theta_{\tau})\alpha_{\tau}\tilde{X}_{i}\\
&  +\left(  \frac{1}{n}\sum_{i=1}^{n}\dot{g}(Z_{i}^{\prime}\hat{\theta}_{\tau
})\dot{\phi}_{\mathrm{x}}\left(  X_{i}\right)  ^{\prime}\tilde{\alpha}_{\tau
}\tilde{X}_{i}Z_{i}^{\prime}\right)  (\hat{\theta}_{\tau}-\theta_{\tau
})+\left(  \frac{1}{n}\sum_{i=1}^{n}g(Z_{i}^{\prime}\tilde{\theta}_{\tau
})\tilde{X}_{i}\dot{\phi}(X_{i})^{\prime}\right)  (\hat{\alpha}_{\tau}%
-\alpha_{\tau}).
\end{align*}
Using the uniform law of large numbers in Assumption
\ref{assumption_logit_probit}(iv), we have
\[
\frac{1}{n}\sum_{i=1}^{n}[\dot{g}(Z_{i}^{\prime}\tilde{\theta}_{\tau}%
)\dot{\phi}_{\mathrm{x}}\left(  X_{i}\right)  ^{\prime}\tilde{\alpha}_{\tau
}]\tilde{X}_{i}Z_{i}^{\prime}\overset{p}{\rightarrow}M_{1}%
:=\underbrace{E\left\{  [\dot{g}(Z_{i}^{\prime}\theta_{\tau})\dot{\phi
}_{\mathrm{x}}\left(  X_{i}\right)  ^{\prime}\alpha_{\tau}]\tilde{X}_{i}%
Z_{i}^{\prime}\right\}  }_{2\times d_{Z}}%
\]
and
\[
\frac{1}{n}\sum_{i=1}^{n}g(Z_{i}^{\prime}\tilde{\theta}_{\tau})\tilde{X}%
_{i}\dot{\phi}_{\mathrm{x}}(X_{i})^{\prime}\overset{p}{\rightarrow}%
M_{2}:=\underbrace{E\left[  g(Z_{i}^{\prime}\theta_{\tau})\tilde{X}_{i}%
\dot{\phi}_{\mathrm{x}}(X_{i})^{\prime}\right]  }_{2\times d_{\phi
_{\mathrm{x}}}}.
\]
Therefore,
\begin{align}
&  \sqrt{n}\left(  \frac{1}{n}\sum_{i=1}^{n}g_{\dot{\phi}}(Z_{i};\hat{\theta
}_{\tau})\hat{\alpha}_{\tau}\tilde{X}_{i}-E\left[  g_{\dot{\phi}}(Z_{i}%
;\theta_{\tau})\alpha_{\tau}\tilde{X}_{i}\right]  \right) \\
&  =\sqrt{n}\left(  \frac{1}{n}\sum_{i=1}^{n}g_{\dot{\phi}}(Z_{i};\theta
_{\tau})\alpha_{\tau}\tilde{X}_{i}-E\left[  g(Z_{i}^{\prime}\theta_{\tau
})\alpha_{\tau}\tilde{X}_{i}\right]  \right)  +M_{1}\sqrt{n}(\hat{\theta
}_{\tau}-\theta_{\tau})+M_{2}\sqrt{n}(\hat{\alpha}_{\tau}-\alpha_{\tau}%
)+o_{p}(1).\nonumber
\end{align}
The first term captures the uncertainty in estimating the expected value, and
the second and third terms capture the uncertainty in estimating the
logit/probit model, and it has already incorporated the contribution of the
preliminary estimator $\hat{q}_{\tau}$ of $Q_{\tau}\left[  Y\right]  $. To
ease notation, define $M:=M_{1}+\left(  M_{2},\,O\right)  $ where $O$ is a
$2\times d_{\phi_{\mathrm{w}}}$ matrix of zeros. Thus, we can write:
\begin{align}
&  \sqrt{n}\left(  \frac{1}{n}\sum_{i=1}^{n}g_{\dot{\phi}}(Z_{i};\hat{\theta
}_{\tau})\hat{\alpha}_{\tau}\tilde{X}_{i}-E\left[  g_{\dot{\phi}}(Z_{i}%
;\theta_{\tau})\alpha_{\tau}\tilde{X}_{i}\right]  \right) \nonumber\\
&  =\sqrt{n}\left(  \frac{1}{n}\sum_{i=1}^{n}g_{\dot{\phi}}(Z_{i};\theta
_{\tau})\alpha_{\tau}\tilde{X}_{i}-E\left[  g_{\dot{\phi}}(Z_{i}^{\prime
}\theta_{\tau})\alpha_{\tau}\tilde{X}_{i}\right]  \right)  +M\sqrt{n}\left(
\hat{\theta}_{\tau}-\theta_{\tau}\right)  +o_{p}(1). \label{eq_decomp_g_2}%
\end{align}

It then follows that
\begin{align*}%
\begin{pmatrix}
\hat{\Pi}_{\tau,L}\\
\hat{\Pi}_{\tau,S}%
\end{pmatrix}
-%
\begin{pmatrix}
\Pi_{\tau,L}\\
\Pi_{\tau,S}%
\end{pmatrix}
&  =\frac{1}{f_{Y}\left(  Q_{\tau}[Y]\right)  }D_{\mu}\left[  \frac{1}{n}%
\sum_{i=1}^{n}g_{\dot{\phi}}(Z_{i};\theta_{\tau})\alpha_{\tau}\tilde{X}%
_{i}-E\left[  g_{\dot{\phi}}(Z_{i};\theta_{\tau})\alpha_{\tau}\tilde{X}%
_{i}\right]  \right] \\
&  +\frac{1}{f_{Y}\left(  Q_{\tau}[Y]\right)  }D_{\mu}M\left(  \hat{\theta
}_{\tau}-\theta_{\tau}\right)  -%
\begin{pmatrix}
\Pi_{\tau,L}\\
\Pi_{\tau,S}%
\end{pmatrix}
\frac{\dot{f}_{Y}(Q_{\tau}[Y])}{f_{Y}\left(  Q_{\tau}[Y]\right)  }\left(
\hat{q}_{\tau}-Q_{\tau}[Y]\right) \\
&  -%
\begin{pmatrix}
\Pi_{\tau,L}\\
\Pi_{\tau,S}%
\end{pmatrix}
\frac{\left(  \hat{f}_{Y}\left(  Q_{\tau}[Y]\right)  -f_{Y}\left(  Q_{\tau
}[Y]\right)  \right)  }{f_{Y}\left(  Q_{\tau}[Y]\right)  }+o_{p}%
(n^{-1/2})+o_{p}(n^{-1/2}h^{-1/2}).
\end{align*}

Plugging the asymptotic representation of $\sqrt{n}\left(  \hat{\theta}_{\tau
}-\theta_{\tau}\right)  $ in \eqref{eq:probit_coef}, we obtain:
\begin{align*}%
\begin{pmatrix}
\hat{\Pi}_{\tau,L}\\
\hat{\Pi}_{\tau,S}%
\end{pmatrix}
-%
\begin{pmatrix}
\Pi_{\tau,L}\\
\Pi_{\tau,S}%
\end{pmatrix}
&  =\frac{1}{f_{Y}\left(  Q_{\tau}[Y]\right)  }D_{\mu}\left[  \frac{1}{n}%
\sum_{i=1}^{n}g_{\dot{\phi}}(Z_{i};\theta_{\tau})\alpha_{\tau}\tilde{X}%
_{i}-E\left[  g_{\dot{\phi}}(Z_{i};\theta_{\tau})\alpha_{\tau}\tilde{X}%
_{i}\right]  \right] \\
&  -\frac{1}{f_{Y}\left(  Q_{\tau}[Y]\right)  }D_{\mu}MH^{-1}\frac{1}{n}%
\sum_{i=1}^{n}s_{i}(\theta_{\tau};Q_{\tau}[Y])\\
&  -\left[
\begin{pmatrix}
\Pi_{\tau,L}\\
\Pi_{\tau,S}%
\end{pmatrix}
\frac{\dot{f}_{Y}(Q_{\tau}[Y])}{f_{Y}\left(  Q_{\tau}[Y]\right)  }+\frac
{1}{f_{Y}\left(  Q_{\tau}[Y]\right)  }D_{\mu}MH^{-1}H_{Q}\right]  \frac{1}%
{n}\sum_{i=1}^{n}\psi(Y_{i},\tau,F_{Y})\\
&  -%
\begin{pmatrix}
\Pi_{\tau,L}\\
\Pi_{\tau,S}%
\end{pmatrix}
\frac{\hat{f}_{Y}\left(  Q_{\tau}[Y]\right)  -f_{Y}\left(  Q_{\tau}[Y]\right)
}{f_{Y}\left(  Q_{\tau}[Y]\right)  }+o_{p}(n^{-1/2})+o_{p}(n^{-1/2}h^{-1/2}).
\end{align*}
Plugging the representation of $\hat{f}_{Y}\left(  Q_{\tau}[Y]\right)
-f_{Y}\left(  Q_{\tau}[Y]\right)  $ in (\ref{density_error}) completes the proof.

\subsection{Proof of Corollary \ref{corollary_est_pi}}

The result has been proved in the main text. Here we give the expressions for
${\hat{M}}$, ${\hat{H},}$ and $\hat{H}_{Q}$. \ For $\hat{M}$ and $\hat{H},$ we
have%
\begin{align*}
{\hat{M}}  &  =\frac{1}{n}\sum_{i=1}^{n}[\dot{g}(Z_{i}^{\prime}\hat{\theta
}_{\tau})\dot{\phi}_{\mathrm{x}}\left(  X_{i}\right)  ^{\prime}\hat{\alpha
}_{\tau}]\left(
\begin{array}
[c]{c}%
1\\
X_{i}%
\end{array}
\right)  \left[  \phi_{\mathrm{x}}\left(  X_{i}\right)  ^{\prime}%
,\phi_{\mathrm{w}}\left(  W_{i}\right)  ^{\prime}\right] \\
&  +\frac{1}{n}\sum_{i=1}^{n}g(Z_{i}^{\prime}\hat{\theta}_{\tau})\left(
\begin{array}
[c]{c}%
1\\
X_{i}%
\end{array}
\right)  \left(
\begin{array}
[c]{cc}%
\dot{\phi}_{\mathrm{x}}\left(  X_{i}\right)  ^{\prime}, & O_{1\times
d_{\phi_{\mathrm{w}}}}%
\end{array}
\right)
\end{align*}
and
\[
\hat{H}=\frac{1}{n}\sum_{i=1}^{n}\frac{g(Z_{i}^{\prime}\hat{\theta}_{\tau
})^{2}}{G(Z_{i}^{\prime}\hat{\theta}_{\tau})(1-G(Z_{i}^{\prime}\hat{\theta
}_{\tau}))}Z_{i}Z_{i}^{\prime}.
\]
For $\hat{H}_{Q}$, we note that
\[
H_{Q}=\frac{\partial E\left[  s_{i}(\theta_{\tau};q)\right]  }{\partial
q}\bigg\vert_{q=Q_{\tau}[Y]}=E\left[  \dfrac{g(Z_{i}^{\prime}\theta_{\tau
})Z_{i}}{G(Z_{i}^{\prime}\theta_{\tau})\left[  1-G(Z_{i}^{\prime}\theta_{\tau
})\right]  }\cdot f_{Y|X,W}(Q_{\tau}[Y]|X_{i},W_{i})\right]  .
\]
Let
\[
\Lambda(Z_{i},\theta_{\tau}):=\dfrac{g(Z_{i}^{\prime}\theta_{\tau})Z_{i}%
}{G(Z_{i}^{\prime}\theta_{\tau})\left[  1-G(Z_{i}^{\prime}\theta_{\tau
})\right]  }.
\]
Then
\begin{align*}
H_{Q}  &  =E[\Lambda(Z_{i},\theta_{\tau})f_{Y|X,W}(Q_{\tau}[Y]|X_{i},W_{i})\\
&  =\int_{\mathcal{W}}\int_{\mathcal{X}}\Lambda(\phi_{\mathrm{x}}\left(
x\right)  ,\phi_{\mathrm{w}}\left(  w\right)  ,\theta_{\tau})f_{Y|X,W}%
(Q_{\tau}[Y]|x,w)f_{XW}(x,w)dxdw\\
&  =f_{Y}(Q_{\tau}[Y])\int_{\mathcal{W}}\int_{\mathcal{X}}\Lambda
(\phi_{\mathrm{x}}\left(  x\right)  ,\phi_{\mathrm{w}}\left(  w\right)
,\theta_{\tau})\frac{f_{Y,X,W}(Q_{\tau}[Y],x,w)}{f_{Y}(Q_{\tau}[Y])f_{XW}%
(x,w)}f_{XW}(x,w)dxdw\\
&  =f_{Y}(Q_{\tau}[Y])\int_{\mathcal{W}}\int_{\mathcal{X}}\Lambda
(\phi_{\mathrm{x}}\left(  x\right)  ,\phi_{\mathrm{w}}\left(  w\right)
,\theta_{\tau})f_{X,W|Y}(x,w|Q_{\tau}[Y])dxw\\
&  =f_{Y}(Q_{\tau}[Y])E[\Lambda(Z,\theta_{\tau})|Y=Q_{\tau}[Y])].
\end{align*}
To estimate the conditional expectation, we may use a vector version of the
Nadaraya-Watson estimator:%
\[
\hat{E}[\Lambda(Z,\hat{\theta}_{\tau})|Y=\hat{q}_{\tau}]=\frac{\sum_{i=1}%
^{n}K_{h}(Y_{i}-\hat{q}_{\tau})\Lambda(Z_{i},\hat{\theta}_{\tau})}{\sum
_{i=1}^{n}K_{h}(Y_{i}-\hat{q}_{\tau})},
\]
where $K_{h}$ is the rescaled kernel $K_{h}(Y_{i}-y)=h^{-1}K\left(
(Y_{i}-y)/h\right)  $ for a kernel function $K\left(  \cdot\right)  .$ We can
then estimate $H_{Q}$ by
\begin{align}
\hat{H}_{Q}  &  =\hat{f}_{Y}(\hat{q}_{\tau})\hat{E}[\Lambda(Z,\hat{\theta
}_{\tau})|Y=\hat{q}_{\tau}]\nonumber\label{eq:H_Q}\\
&  =\left[  \frac{1}{n}\sum_{i=1}^{n}K_{h}(Y_{i}-\hat{q}_{\tau})\right]
\cdot\frac{\sum_{i=1}^{n}K_{h}(Y_{i}-\hat{q}_{\tau})\Lambda(Z_{i},\hat{\theta
}_{\tau})}{\sum_{i=1}^{n}K_{h}(Y_{i}-\hat{q}_{\tau})}\nonumber\\
&  =\frac{1}{n}\sum_{i=1}^{n}K_{h}(Y_{i}-\hat{q}_{\tau})\Lambda(Z_{i}%
,\hat{\theta}_{\tau}).
\end{align}
It is worth pointing out that, in the logistic case, $G(v)=(1+\exp\left(
-v\right)  )^{-1}$ and we have the convenient identity $g(v)=G(v)(1-G(v))$.
Thus, $\Lambda(Z_{i},\hat{\theta}_{\tau})=Z_{i}$ and the estimation of $H$ and
$H_{Q}$ becomes simpler.

\newpage

\section*{Supplementary Appendix}

\renewcommand{\thesubsection}{S.\arabic{subsection}}

\setcounter{assumption}{0}\setcounter{theorem}{0}\setcounter{corollary}{0}\setcounter{equation}{0}\setcounter{subsection}{0}\setcounter{page}{1}
\renewcommand\theequation{S.\arabic{equation}}

\renewcommand\thetheorem{S.\arabic{theorem}}\renewcommand\theassumption{S.\arabic{assumption}}\renewcommand\thecorollary{S.\arabic{corollary}}

\subsection{Simultaneous Policy Changes}

\label{comp_appendix}

Our results focus on the case of a counterfactual policy applied to a
univariate target policy. In this section, we consider the case where a
location shift in one covariate is compensated or amplified by a location
shift in another covariate. In a model $Y=h(X_{1},X_{2},W,U)$ where both
$X_{1}$ and $X_{2}$ are univariate, we consider the limiting effect of the
simultaneous location shift $X_{1\delta}=X_{1}+\ell_{1}\left(  \delta\right)
$ and $X_{2\delta}=X_{2}+\ell_{2}\left(  \delta\right)  $ for some smooth
functions $\ell_{1}\left(  \delta\right)  $ and $\ell_{2}\left(
\delta\right)  $ satisfying $\ell_{1}\left(  0\right)  =\ell_{2}\left(
0\right)  =0.$ Here, $\ell_{1}\left(  \delta\right)  $ and $\ell_{2}\left(
\delta\right)  $ can have the same sign or opposite signs. As a simple
example, we may have $\ell_{1}\left(  \delta\right)  =\delta$ and $\ell
_{2}\left(  \delta\right)  =-p\delta$ for some $p\geq0.$ Here, $p$ can be
interpreted as the \textquotedblleft relative price\textquotedblright\ of
$X_{1}$ in terms of $X_{2}$. A potential application is the following: a
policy targeted towards increasing the level of education can, at the same
time, reduce the experience of workers. As with the case of the scale shift,
neglecting this possible side effect of the policy might lead to an
inconsistent estimator of its effect.

With the above motivation, we now consider a more general setting that allows
for simultaneous changes in $X_{1}$ and $X_{2}$. We induce a change in
$X=\left(  X_{1},X_{2}\right)  ^{\prime}$ so that it becomes $X_{\delta
}=\left(  X_{1\delta},X_{2\delta}\right)  ^{\prime}.$ We do not specify the
exact form of the change, but we use the simultaneous location shift as a
working example. We assume that%
\[
X_{\delta}=\mathcal{G}\left(  x;\delta\right)  =\left(  \mathcal{G}_{1}\left(
X;\delta\right)  ,\mathcal{G}_{2}\left(  X;\delta\right)  \right)  ^{\prime}%
\]
for a smooth and invertible bivariate function $\mathcal{G}=\left(
\mathcal{G}_{1},\mathcal{G}_{2}\right)  ^{\prime}$. We allow $X_{1\delta}$ and
$X_{2\delta}$ to depend on both $X_{1}$ and $X_{2}.$ A special case is that
$\mathcal{G}_{1}\left(  X;\delta\right)  $ is a function of $X_{1}$ only and
$\mathcal{G}_{2}\left(  X;\delta\right)  $ is a function of $X_{2}$ only.

In this general setting, the original outcome is given by%
\[
Y=h\left(  X_{1},X_{2},W,U\right)  =h\left(  X,W,U\right)  ,
\]
and the counterfactual outcome is given by
\begin{equation}
Y_{\delta}=h(X_{1\delta},X_{2\delta},W,U)=h(\mathcal{G}_{1}\left(
X;\delta\right)  ,\mathcal{G}_{2}\left(  X;\delta\right)  ,W,U).
\label{eq_comp_model}%
\end{equation}
The distribution of $\left(  X,W,U\right)  $ is kept the same in the above two
equations. We want to identify the following quantity
\begin{equation}
\Pi_{\tau,C}:=\lim_{\delta\rightarrow0}\frac{Q_{\tau}[Y_{\delta}]-Q_{\tau}%
[Y]}{\delta},
\end{equation}
whenever this limit exists. We refer to $\Pi_{\tau,C}$ as the
\textit{compensated marginal effect for the }$\tau$\textit{-quantile.}

Let $x=\left(  x_{1},x_{2}\right)  ^{\prime}.$ As before, we define
$x^{\delta}=\left(  x_{1}^{\delta},x_{2}^{\delta}\right)  ^{\prime}$ such that
$\mathcal{G}\left(  x^{\delta};\delta\right)  =x.$ By construction,
$X_{\delta}=x$ if and only if $X=x^{\delta}.$ Define the Jacobian matrix as%
\[
J\left(  x^{\delta};\delta\right)  :=\frac{\partial x^{\delta}}{\partial
x^{\prime}}=\left(
\begin{array}
[c]{cc}%
\frac{\partial x_{1}^{\delta}}{\partial x_{1}} & \frac{\partial x_{1}^{\delta
}}{\partial x_{2}}\\
\frac{\partial x_{2}^{\delta}}{\partial x_{1}} & \frac{\partial x_{2}^{\delta
}}{\partial x_{2}}%
\end{array}
\right)  =\left(  \frac{\partial\mathcal{G}\left(  x;\delta\right)  }{\partial
x^{\prime}}\right)  ^{-1}\bigg|_{x=x^{\delta}},
\]
where the second equality follows from differentiating $\mathcal{G}\left(
x^{\delta},\delta\right)  =x$ with respect to $x$ and then solving for
$\partial x^{\delta}/\partial x^{\prime}.$

\begin{assumption}
\label{Assumption:main_C}(i) For some $\varepsilon>0,$ each component function
of $\mathcal{G}\left(  x;\delta\right)  $ is continuously differentiable on
$\mathcal{X\otimes N}_{\varepsilon}.$

(i.b) $\mathcal{G}\left(  x;\delta\right)  $ is an invertible function of $x$
each $\delta\in\mathcal{N}_{\varepsilon}.$

(i.c) $\mathcal{G}\left(  x;0\right)  =x$ for all $x\in\mathcal{X}$.

(ii) for $\delta\in\mathcal{N}_{\varepsilon}$, the conditional density of $U$
satisfies $f_{U|X_{\delta},W}(u|x,w)=f_{U|X,W}(u|x^{\delta},w)$ and the
support $\mathcal{U}$ of $U$ conditional on $X$ and $W$ does not depend on
$\left(  X,W\right)  .$

(iii) Assumption \ref{Assumption:main} (iii.a) holds with $J\left(  x^{\delta
};\delta\right)  $ replaced by $\det\left[  J\left(  x^{\delta};\delta\right)
\right]  $ and Assumption \ref{Assumption:main} (iii.b) holds.

(iv) $f_{X,W}(x,w)$ is equal to $0$ on the boundary of the support of $X_{1}$
given $W=w$ and $X_{2}=x_{2}$ for all $w\in\mathcal{W}$ and $x_{2}%
\in\mathcal{X}_{2},$ the support of $X_{2}$, and symmetrically, $f_{X,W}(x,w)$
is equal to $0$ on the boundary of the support of $X_{2}$ given $W=w$ and
$X_{1}=x_{1}$ for all $w\in\mathcal{W}$ and $x_{1}\in\mathcal{X}_{1},$ the
support of $X_{1}.$

(v) $f_{Y}(Q_{\tau}[Y])>0$.
\end{assumption}

Assumption \ref{Assumption:main_C} is a modified version of Assumption
\ref{Assumption:main} adapted to the case with two target covariates. Under
Assumption \ref{Assumption:main_C}(i.c), we have $J\left(  x;0\right)
=I_{2},$ the $2\times2$ identity matrix. Since $\det\left[  J\left(
x,0\right)  \right]  =1,$ by continuity, $\det\left[  J\left(  x^{\delta
};\delta\right)  \right]  >0$ when $\delta$ is small enough. Hence, there is
no need to take the absolute value of $\det\left[  J\left(  x^{\delta}%
;\delta\right)  \right]  $ when converting the pdf of $\left(  X,W\right)  $
into that of $\left(  X_{\delta},W\right)  .$

Define the local change function as
\[
\kappa\left(  x\right)  =\frac{\partial\mathcal{G}\left(  x;\delta\right)
}{\partial\delta}\bigg|_{\delta=0}.
\]

\begin{theorem}
\label{th:comp_change}Let Assumption \ref{Assumption:main_C} hold. Then
\begin{equation}
\Pi_{\tau,C}=E\left[  \frac{\partial E\left[  \psi\left(  Y,\tau,F_{Y}\right)
|X,W\right]  }{\partial X^{\prime}}\kappa\left(  X\right)  \right]  -E\left[
\psi\left(  Y,\tau,F_{Y}\right)  \frac{\partial\ln f_{U|X,W}(U|X,W)}{\partial
X^{\prime}}\kappa\left(  X\right)  \right]  ,
\end{equation}
where, as before,
\[
\psi\left(  y,\tau,F_{Y}\right)  =\frac{\tau-1\left(  y<Q_{\tau}[Y]\right)
}{f_{Y}(Q_{\tau}[Y])}.
\]

\end{theorem}

The theorem takes the same form as Theorem \ref{Assumption:main}. Under the
assumption that $X_{j\delta}$ is a function of $X_{j}$ only for $j=1$ and $2$,
$\kappa_{j}\left(  x\right)  $ depends on $x_{j}$ only, and the effect from
changing $X_{1}$ into $X_{1\delta}$ and that from changing $X_{2}$ into
$X_{2\delta}$ are additively separable.

\begin{corollary}
\label{corollary:comp_change}Let Assumptions \ref{Assumption ID} and
\ref{Assumption:main_C} hold. Then
\[
\Pi_{\tau,C}=E\left[  \frac{\partial E\left[  \psi\left(  Y,\tau,F_{Y}\right)
|X,W\right]  }{\partial X^{\prime}}\kappa\left(  X\right)  \right]  .
\]
For the case of a simultaneous location shift $X_{1\delta}=X_{1}+\ell
_{1}\left(  \delta\right)  $ and $X_{2\delta}=X_{2}+\ell_{2}\left(
\delta\right)  $, we have%
\[
\kappa\left(  x\right)  =\left(  \dot{\ell}_{1}\left(  0\right)  ,\dot{\ell
}_{2}\left(  0\right)  \right)  ^{\prime}\text{,}%
\]
and so
\begin{align}
\Pi_{\tau,C}  &  =\frac{\dot{\ell}_{1}\left(  0\right)  }{f_{Y}(Q_{\tau}%
[Y])}\int_{\mathcal{W}}\int_{\mathcal{X}}\frac{\partial\mathcal{S}%
_{Y|X,W}\left(  Q_{\tau}[Y]|x,w\right)  }{\partial x_{1}}f_{X,W}%
(x,w)dxdw\nonumber\label{eq:pi_comp}\\
&  +\frac{\dot{\ell}_{2}\left(  0\right)  }{f_{Y}(Q_{\tau}[Y])}\int%
_{\mathcal{W}}\int_{\mathcal{X}}\frac{\partial\mathcal{S}_{Y|X,W}\left(
Q_{\tau}[Y]|x,w\right)  }{\partial x_{2}}f_{X,W}(x,w)dxdw.
\end{align}

\end{corollary}

Corollary \ref{corollary:comp_change} shows that the compensated effect from
the simultaneous location shift is a linear combination of two location
effects: one where the target variable is $X_{1}$ and the other where the
target variable is $X_{2}$. Thus, we can write: $\Pi_{\tau,C}=\Pi_{\tau
,L,1}+\Pi_{\tau,L,2}$. This additive result follows because we have two
unrelated location shifts whose effects are, in essence, captured by the sum
of two partial derivatives. This is convenient since it immediately allows us
to obtain the bias if we omit the possible simultaneous change in a covariate
different from the target variable.

Corollary 1 in \cite{FirpoFortinLemieux09} considers the case of a
simultaneous location shift in $k$ covariates, and delivers a $k\times1$
vector of marginal effects. Theorem \ref{th:comp_change} and Corollary
\ref{corollary:comp_change} complement such a result by showing how to
interpret a linear combination of the entries of the vector of marginal
effects.\ Furthermore, Theorem \ref{th:comp_change} and Corollary
\ref{corollary:comp_change} allow for the intervention of a target covariate
to depend on another target covariate. Here we consider only two target
covariates for ease of exposition. Our results can be easily extended to the
case with more than two target covariates.

Our framework can accommodate more complicated policy interventions, such as
simultaneous location-scale shifts in two target variables. In a potential
application, a compensated change may substitute the mean of one target
variable with the variance of another target variable. Given the generality of
$\mathcal{G}\left(  x;\delta\right)  $, Corollary \ref{corollary:comp_change}
is general enough to accommodate various compensating policies.

\subsection{Estimation of Simultaneous Effects}

In this section, we focus on the estimation of $\Pi_{\tau,C}$ given in
\eqref{eq:pi_comp}. We use the same estimators of the quantile, the density of
$Y,$ and the parameters in the probit/logit model. We only need to make some
minor notational changes. As before $\theta_{\tau}=\left(  \alpha_{\tau
}^{\prime},\beta_{\tau}^{\prime}\right)  ^{\prime},$ $\hat{\theta}_{\tau
}=(\hat{\alpha}_{\tau}^{\prime},\hat{\beta}_{\tau}^{\prime})^{\prime}$ and
$Z_{i}=(\phi_{\mathrm{x}}(X_{i})^{\prime},\phi_{\mathrm{w}}(W_{i})^{\prime
})^{\prime}$ but now $\alpha_{\tau}=\left(  \alpha_{\tau,1}^{\prime}%
,\alpha_{\tau,2}^{\prime}\right)  ^{\prime},\hat{\alpha}_{\tau}=\left(
\hat{\alpha}_{\tau,1}^{\prime},\hat{\alpha}_{\tau,2}^{\prime}\right)
^{\prime}$ and $X_{i}=(X_{1i}^{\prime},X_{2i}^{\prime})^{\prime}.$
%
As in the case with the location-scale effect, we estimate $\Pi_{\tau,C}$ by
\[
\hat{\Pi}_{\tau,C}=\hat{\Pi}_{\tau,L,1}+\hat{\Pi}_{\tau,L,2}%
\]
where
\begin{align}
\hat{\Pi}_{\tau,L,1}  &  =-\frac{\dot{\ell}_{1}\left(  0\right)  }{\hat{f}%
_{Y}\left(  \hat{q}_{\tau}\right)  }\frac{1}{n}\sum_{i=1}^{n}g(Z_{i}^{\prime
}\hat{\theta}_{\tau})\frac{\partial\phi_{\mathrm{x}}(X_{i})^{\prime}}{\partial
X_{1i}}\hat{\alpha}_{\tau,1},\label{eq:est_pi_c_1}\\
\hat{\Pi}_{\tau,L,2}  &  =-\frac{\dot{\ell}_{2}\left(  0\right)  }{\hat{f}%
_{Y}\left(  \hat{q}_{\tau}\right)  }\frac{1}{n}\sum_{i=1}^{n}g(Z_{i}^{\prime
}\hat{\theta}_{\tau})\frac{\partial\phi_{\mathrm{x}}(X_{i})^{\prime}}{\partial
X_{2i}}\hat{\alpha}_{\tau,2}. \label{eq:est_pi_c_2}%
\end{align}

For the next theorem, we define the diagonal matrix:
\[
\dot{\phi}_{\mathrm{x}}\left(  X_{i}\right)  =\left(
\begin{array}
[c]{cc}%
\frac{\partial\phi_{\mathrm{x}}(X_{i})}{\partial X_{1i}} & O\\
O & \frac{\partial\phi_{\mathrm{x}}(X_{i})}{\partial X_{2i}}%
\end{array}
\right)  .
\]
We need the following modification of Assumption \ref{assumption_logit_probit}.

\begin{assumption}
\label{assumption_logit_probit_2}\textbf{Logit/Probit II.} Assumption
\ref{assumption_logit_probit} holds with (iv) replaced by the following:%
\begin{align*}
M_{1L}\left(  \theta\right)   &  =E\left\{  \left[  \dot{g}(Z_{i}^{\prime
}\theta)\dot{\phi}_{\mathrm{x}}\left(  X_{i}\right)  ^{\prime}\alpha\right]
Z_{i}^{\prime}\right\} \\
M_{2L}\left(  \theta\right)   &  =E\left[  g(Z_{i}^{\prime}\theta)\dot{\phi
}_{\mathrm{x}}\left(  X_{i}\right)  ^{\prime}\right]
\end{align*}
are well defined for any $\theta\in\mathcal{N}_{\theta_{\tau}}$ and
\begin{align*}
&  \sup_{\theta\in\mathcal{N}_{\theta_{\tau}}}\bigg\|\frac{1}{n}\sum_{i=1}%
^{n}\left[  \dot{g}(Z_{i}^{\prime}\theta)\dot{\phi}_{\mathrm{x}}\left(
X_{i}\right)  ^{\prime}\alpha\right]  Z_{i}^{\prime}-M_{1L}\left(
\theta\right)  \bigg\|\overset{p}{\rightarrow}0,\\
&  \sup_{\theta\in\mathcal{N}_{\theta_{\tau}}}\bigg\|\frac{1}{n}\sum_{i=1}%
^{n}g(Z_{i}^{\prime}\theta)\dot{\phi}_{\mathrm{x}}\left(  X_{i}\right)
^{\prime}-M_{2L}\left(  \theta\right)  \bigg\|\overset{p}{\rightarrow}0.
\end{align*}

\end{assumption}

\begin{theorem}
\label{theorem_est_pi_c} Under Assumptions \ref{assumption_quantile},
\ref{assumption_density}, and \ref{assumption_logit_probit_2}, the estimators
given in \eqref{eq:est_pi_c_1} and \eqref{eq:est_pi_c_2} satisfy%
\[%
\begin{pmatrix}
\hat{\Pi}_{\tau,L,1}\\
\hat{\Pi}_{\tau,L,2}%
\end{pmatrix}
-%
\begin{pmatrix}
\Pi_{\tau,L,1}\\
\Pi_{\tau,L,2}%
\end{pmatrix}
=\frac{1}{n}\sum_{i=1}^{n}\Phi_{i,\tau}^{L}+O_{P}\left(  h^{2}\right)
+o_{p}(n^{-1/2})+o_{p}(n^{-1/2}h^{-1/2}),
\]
where
\begin{align*}
\Phi_{i,\tau}^{L}  &  =\frac{1}{f_{Y}\left(  Q_{\tau}[Y]\right)  }%
D_{L}\left\{  g(Z_{i}^{\prime}\theta_{\tau})\dot{\phi}_{\mathrm{x}}\left(
X_{i}\right)  ^{\prime}\alpha_{\tau}-E\left[  g(Z_{i}^{\prime}\theta_{\tau
})\dot{\phi}_{\mathrm{x}}\left(  X_{i}\right)  ^{\prime}\alpha_{\tau}\right]
\right\} \\
&  -\frac{1}{f_{Y}\left(  Q_{\tau}[Y]\right)  }D_{L}M_{L}H^{-1}s_{i}%
(\theta_{\tau};Q_{\tau}[Y])\\
&  -\left[
\begin{pmatrix}
\Pi_{\tau,L,1}\\
\Pi_{\tau,L,2}%
\end{pmatrix}
\frac{\dot{f}_{Y}(Q_{\tau}[Y])}{f_{Y}\left(  Q_{\tau}[Y]\right)  }+\frac
{1}{f_{Y}\left(  Q_{\tau}[Y]\right)  }D_{L}M_{L}H^{-1}H_{Q}\right]  \psi
(Y_{i},\tau,F_{Y})\\
&  -%
\begin{pmatrix}
\Pi_{\tau,L,1}\\
\Pi_{\tau,L,2}%
\end{pmatrix}
\frac{1}{f_{Y}\left(  Q_{\tau}[Y]\right)  }\left\{  \mathcal{K}_{h}\left(
Y_{i}-Q_{\tau}[Y]\right)  -E\mathcal{K}_{h}\left(  Y_{i}-Q_{\tau}[Y]\right)
\right\}  ,
\end{align*}%
\[
D_{L}=%
\begin{pmatrix}
-\dot{\ell}_{1}(0) & 0\\
0 & -\dot{\ell}_{2}(0)
\end{pmatrix}
\]
and
\[
M_{L}=M_{1L}\left(  \theta_{\tau}\right)  +%
\begin{pmatrix}
M_{2L}\left(  \theta_{\tau}\right)  , & O_{2\times d_{\phi_{\mathrm{w}}}}%
\end{pmatrix}
.
\]

\end{theorem}

For the asymptotic normality, the discussions after Theorem
\ref{theorem_est_pi} are still applicable.

In the special case that $\ell_{1}\left(  \delta\right)  =\delta$ and
$\ell_{2}\left(  \delta\right)  =-p\delta$, it suffices to change $D_{L}$ to
$\mathrm{diag}(1,-p).$ It is possible that $p$, the relative price $X_{1}$ in
terms of $X_{2}$, has to be estimated by $\hat{p}$ based on an independent
sample. In that case, the estimator of the compensated effect would be
\[
\hat{\Pi}_{\tau,L}=\hat{\Pi}_{\tau,L,1}-{\hat{p}}\hat{\Pi}_{\tau,L,2}.
\]
If the sample size $\tilde{n}$ of the independent sample for estimating $p$ is
much larger than $n$ (i.e., $\tilde{n}/n\rightarrow\infty),$ then the
expansion in Theorem \ref{theorem_est_pi_c} still holds.

\subsection{Proof of Theorem \ref{th:comp_change}}

The proof of this Theorem is very similar to the proof of Theorem
\ref{th:uqpe_scale}. The following decomposition still holds
\[
\frac{F_{Y_{\delta}}(y)-F_{Y}(y)}{\delta}:=G_{1,\delta}\left(  y\right)
+G_{2,\delta}\left(  y\right)  ,
\]
where
\begin{align*}
G_{1,\delta}\left(  y\right)   &  =\int_{\mathcal{W}}\int_{\mathcal{X}%
}F_{Y|X,W}\left(  y|x,w\right)  \frac{\left[  f_{X_{\delta},W}(x,w)-f_{X,W}%
(x,w)\right]  }{\delta}dxdw,\\
G_{2,\delta}\left(  y\right)   &  =\int_{\mathcal{W}}\int_{\mathcal{X}}%
\int_{\mathcal{U}}\mathds1\left\{  h(x,w,u)\leq y\right\}  \frac{\left[
f_{U|X,W}(u|x^{\delta},w)-f_{U|X,W}(u|x,w)\right]  }{\delta}f_{X_{\delta}%
,W}(x,w)dudxdw.
\end{align*}

We first consider the term $G_{1,\delta}\left(  y\right)  .$ Under the
assumptions given, we have
\[
f_{X_{\delta},W}(x,w)=\det\left[  J(x^{\delta};\delta)\right]  f_{X,W}%
(x^{\delta},w).
\]
Evaluated at $\delta=0,$ $f_{X_{\delta},W}(x,w)$ is $f_{X,W}(x,w)$. Given
this, we expand $f_{X_{\delta},W}(x,w)-f_{X,W}(x,w)$ around $\delta=0$, which
is possible under Assumptions \ref{Assumption:main_C}(i) and (iii). We have%
\begin{align}
&  f_{X_{\delta},W}(x,w)-f_{X,W}(x,w)\nonumber\\
&  =\delta\frac{\partial\left[  \det\left[  J\left(  x^{\delta};\delta\right)
\right]  f_{X,W}(x^{\delta},w)\right]  }{\partial\delta}\bigg|_{\delta
=0}+\delta R_{1}(x,w,\delta)\nonumber\\
&  =\delta\frac{\partial\det\left[  J\left(  x^{\delta};\delta\right)
\right]  }{\partial\delta}\bigg|_{\delta=0}f_{X,W}(x,w)+\delta\det\left[
J(x^{\delta};\delta)\right]  \left(  \frac{\partial x^{\delta}}{\partial
\delta}\right)  ^{\prime}\frac{\partial f_{X,W}(x^{\delta},w)}{\partial
x^{\delta}}\bigg|_{\delta=0}+\delta R_{1}(x,w,\delta)\nonumber\\
&  =\delta\frac{\partial\det\left[  J\left(  x^{\delta};\delta\right)
\right]  }{\partial\delta}\bigg|_{\delta=0}f_{X,W}(x,w)+\left(  \frac{\partial
x^{\delta}}{\partial\delta}\right)  \bigg|_{\delta=0}\frac{\partial
f_{X,W}(x,w)}{\partial x}+\delta R_{1}(x,w,\delta),
\end{align}
where, for $\tilde{\delta}(x,w)$ between $0$ and $\delta$,
\[
R_{1}(x,w,\delta)=\left\{  \frac{\partial\left[  \det\left[  J\left(
x^{\delta};\delta\right)  \right]  f_{X,W}(x^{\delta},w)\right]  }%
{\partial\delta}\bigg|_{\delta=\tilde{\delta}(x,w)}-\frac{\partial\left[
\det\left[  J\left(  x^{\delta};\delta\right)  \right]  f_{X,W}(x^{\delta
},w)\right]  }{\partial\delta}\bigg|_{\delta=0}\right\}  .
\]
Using the arguments similar to those in the proof of Theorem
\ref{th:uqpe_scale}, we can show that $G_{1,\delta}\left(  y\right)  $
converges to
\begin{align*}
G_{1,0}\left(  y\right)   &  :=\int_{\mathcal{W}}\int_{\mathcal{X}}%
\frac{\partial\det\left[  J\left(  x^{\delta};\delta\right)  \right]
}{\partial\delta}\bigg|_{\delta=0}F_{Y|X,W}(y|x,w)f_{X,W}(x,w)dxdw\\
&  +\int_{\mathcal{W}}\int_{\mathcal{X}}F_{Y|X,W}(y|x,w)\left(  \frac{\partial
x^{\delta}}{\partial\delta}\right)  ^{\prime}\bigg|_{\delta=0}\frac{\partial
f_{X,W}(x,w)}{\partial x}dxdw\\
&  :=G_{1,0}^{\left(  1\right)  }\left(  y\right)  +G_{1,0}^{\left(  2\right)
}\left(  y\right)
\end{align*}
uniformly in $y\in\mathcal{Y}$, as $\delta\rightarrow0$.

Using Assumption \ref{Assumption:main_C} and the fact that $\frac{\partial
x_{i}^{\delta}}{\partial x_{j}}|_{\delta=0}=\mathds1\left\{  i=j\right\}  $,
we have%
\begin{align*}
\frac{\partial\det\left[  J\left(  x^{\delta};\delta\right)  \right]
}{\partial\delta}\bigg|_{\delta=0}  &  =\frac{\partial}{\partial\delta}\left(
\frac{\partial x_{1}^{\delta}}{\partial x_{1}}\frac{\partial x_{2}^{\delta}%
}{\partial x_{2}}-\frac{\partial x_{1}^{\delta}}{\partial x_{2}}\frac{\partial
x_{2}^{\delta}}{\partial x_{1}}\right)  \bigg|_{\delta=0}\\
&  =-\left(  \frac{\partial\kappa_{1}\left(  x\right)  }{\partial x_{1}}%
\frac{\partial x_{2}^{\delta}}{\partial x_{2}}+\frac{\partial x_{1}^{\delta}%
}{\partial x_{1}}\frac{\partial\kappa_{2}\left(  x\right)  }{\partial x_{2}%
}-\frac{\partial\kappa_{1}\left(  x\right)  }{\partial x_{2}}\frac{\partial
x_{2}^{\delta}}{\partial x_{1}}-\frac{\partial x_{1}^{\delta}}{\partial x_{2}%
}\frac{\partial\kappa_{2}\left(  x\right)  }{\partial x_{1}}\right)
\bigg|_{\delta=0}\\
&  =-\left(  \frac{\partial\kappa_{1}\left(  x\right)  }{\partial x_{1}}%
+\frac{\partial\kappa_{2}\left(  x\right)  }{\partial x_{2}}\right)  .
\end{align*}
So%
\[
G_{1,0}^{\left(  1\right)  }\left(  y\right)  =-\int_{\mathcal{W}}%
\int_{\mathcal{X}}\left(  \frac{\partial\kappa_{1}\left(  x\right)  }{\partial
x_{1}}+\frac{\partial\kappa_{2}\left(  x\right)  }{\partial x_{2}}\right)
F_{Y|X,W}(y|x,w)f_{X,W}(x,w)dxdw.
\]

Next, note that
\[
\left(  \frac{\partial x^{\delta}}{\partial\delta}\right)  ^{\prime
}\bigg|_{\delta=0}\frac{\partial f_{X,W}(x,w)}{\partial x}=-\kappa\left(
x\right)  ^{\prime}\frac{\partial f_{X,W}(x,w)}{\partial x}.
\]
Using integration by parts, we can show that for $j=1$ and $2,$
\begin{align*}
&  -\int_{\mathcal{W}}\int_{\mathcal{X}}F_{Y|X,W}(y|x,w)\left[  \kappa
_{j}\left(  x\right)  \frac{\partial f_{X,W}(x,w)}{\partial x_{j}}\right]
dxdw\\
&  =\int_{\mathcal{W}}\int_{\mathcal{X}}f_{X,W}(x,w)\frac{\partial\left[
F_{Y|X,W}(y|x,w)\kappa_{j}\left(  x\right)  \right]  }{\partial x_{j}}dxdw.
\end{align*}
So
\begin{align*}
G_{1,0}^{\left(  2\right)  }\left(  y\right)   &  =\int_{\mathcal{W}}%
\int_{\mathcal{X}}f_{X,W}(x,w)\left(  \frac{\partial\left[  F_{Y|X,W}%
(y|x,w)\kappa_{1}\left(  x\right)  \right]  }{\partial x_{1}}+\frac
{\partial\left[  F_{Y|X,W}(y|x,w)\kappa_{2}\left(  x\right)  \right]
}{\partial x_{2}}\right)  dxdw\\
&  =\int_{\mathcal{W}}\int_{\mathcal{X}}f_{X,W}(x,w)\left(  \frac
{\partial\left[  F_{Y|X,W}(y|x,w)\right]  }{\partial x_{1}}\kappa_{1}\left(
x\right)  +\frac{\partial\left[  F_{Y|X,W}(y|x,w)\right]  }{\partial x_{2}%
}\kappa_{2}\left(  x\right)  \right)  dxdw\\
&  +\int_{\mathcal{W}}\int_{\mathcal{X}}f_{X,W}(x,w)F_{Y|X,W}(y|x,w)\left(
\frac{\partial\kappa_{1}\left(  x\right)  }{\partial x_{1}}+\frac
{\partial\kappa_{2}\left(  x\right)  }{\partial x_{2}}\right)  dxdw.
\end{align*}
Therefore,
\begin{align*}
&  G_{1,0}\left(  y\right) \\
&  =\int_{\mathcal{W}}\int_{\mathcal{X}}\left[  \frac{\partial F_{Y|X,W}%
(y|x,w)}{\partial x_{1}}\kappa_{1}\left(  x\right)  +\frac{\partial
F_{Y|X,W}(y|x,w)}{\partial x_{2}}\kappa_{2}\left(  x\right)  \right]
f_{X,W}(x,w)dxdw\\
&  =\int_{\mathcal{W}}\int_{\mathcal{X}}\left[  \frac{\partial\left[
F_{Y|X,W}(y|x,w)\right]  }{\partial x^{\prime}}\kappa\left(  x\right)
\right]  f_{X,W}(x,w)dxdw\\
&  =E\left[  \frac{\partial F_{Y|X,W}(y|X,W)}{\partial X^{\prime}}%
\kappa\left(  X\right)  \right]  .
\end{align*}

For $G_{2,\delta}\left(  y\right)  ,$ the proof of Theorem \ref{th:uqpe_scale}
remains valid, and we have that $G_{2,\delta}\left(  y\right)  $ converges to
\[
G_{2,0}\left(  y\right)  :=-E\left[  \mathds1\left\{  h(X,W,U)\leq y\right\}
\frac{\partial\ln f_{U|X,W}(U|X,W)}{\partial X^{\prime}}\kappa\left(
X\right)  \right]
\]
uniformly in $y\in\mathcal{Y}$, as $\delta\rightarrow0$.

Invoking the same argument as that in the proof of Theorem \ref{th:uqpe_scale}%
, we obtain the desired result.

\subsection{Proof of Theorem \ref{theorem_est_pi_c}}

The proof of this theorem is similar to that of Theorem \ref{theorem_est_pi}.
We outline the main steps and omit the details here. We have
\[%
\begin{pmatrix}
\hat{\Pi}_{\tau,L,1}\\
\hat{\Pi}_{\tau,L,2}%
\end{pmatrix}
-%
\begin{pmatrix}
\Pi_{\tau,L,1}\\
\Pi_{\tau,L,2}%
\end{pmatrix}
=D_{L}\left[  \frac{n^{-1}\sum_{i=1}^{n}g(Z_{i}^{\prime}\hat{\theta}_{\tau
})\dot{\phi}_{\mathrm{x}}\left(  X_{i}\right)  ^{\prime}\hat{\alpha}_{\tau}%
}{\hat{f}_{Y}\left(  \hat{q}_{\tau}\right)  }-\frac{E\left[  g(Z_{i}^{\prime
}\theta_{\tau})\dot{\phi}_{\mathrm{x}}\left(  X_{i}\right)  ^{\prime}%
\alpha_{\tau}\right]  }{f_{Y}\left(  Q_{\tau}[Y]\right)  }\right]  .
\]
But%
\begin{align*}
&  \frac{n^{-1}\sum_{i=1}^{n}g(Z_{i}^{\prime}\hat{\theta}_{\tau})\dot{\phi
}_{\mathrm{x}}\left(  X_{i}\right)  ^{\prime}\hat{\alpha}_{\tau}}{\hat{f}%
_{Y}\left(  \hat{q}_{\tau}\right)  }-\frac{E\left[  g(Z_{i}^{\prime}%
\theta_{\tau})\dot{\phi}_{\mathrm{x}}\left(  X_{i}\right)  ^{\prime}%
\alpha_{\tau}\right]  }{f_{Y}\left(  Q_{\tau}[Y]\right)  }\\
&  =\frac{n^{-1}\sum_{i=1}^{n}\left[  g(Z_{i}^{\prime}\hat{\theta}_{\tau}%
)\dot{\phi}_{\mathrm{x}}\left(  X_{i}\right)  ^{\prime}\hat{\alpha}_{\tau
}\right]  -E\left[  g(Z_{i}^{\prime}\theta_{\tau})\dot{\phi}_{\mathrm{x}%
}\left(  X_{i}\right)  ^{\prime}\alpha_{\tau}\right]  }{f_{Y}\left(  Q_{\tau
}[Y]\right)  }\\
&  -\frac{E\left[  g(Z_{i}^{\prime}\theta_{\tau})\dot{\phi}_{\mathrm{x}%
}\left(  X_{i}\right)  ^{\prime}\alpha_{\tau}\right]  }{f_{Y}\left(  Q_{\tau
}[Y]\right)  ^{2}}\left\{  \dot{f}_{Y}(Q_{\tau}[Y])\left(  \hat{q}_{\tau
}-Q_{\tau}[Y]\right)  +\hat{f}_{Y}\left(  Q_{\tau}[Y]\right)  -f_{Y}\left(
Q_{\tau}[Y]\right)  \right\}  +o_{p}(n^{-1/2}h^{-1/2}).
\end{align*}
Now
\begin{align*}
&  \frac{1}{n}\sum_{i=1}^{n}g(Z_{i}^{\prime}\hat{\theta}_{\tau})\dot{\phi
}_{\mathrm{x}}\left(  X_{i}\right)  ^{\prime}\hat{\alpha}_{\tau}\\
&  =\frac{1}{n}\sum_{i=1}^{n}g(Z_{i}^{\prime}\theta_{\tau})\dot{\phi
}_{\mathrm{x}}\left(  X_{i}\right)  ^{\prime}\alpha_{\tau}\\
&  +\left(  \frac{1}{n}\sum_{i=1}^{n}\dot{g}(Z_{i}^{\prime}\tilde{\theta
}_{\tau})\dot{\phi}_{\mathrm{x}}\left(  X_{i}\right)  ^{\prime}\tilde{\alpha
}_{\tau}Z_{i}^{\prime}\right)  (\hat{\theta}_{\tau}-\theta_{\tau})+\left(
\frac{1}{n}\sum_{i=1}^{n}g(Z_{i}^{\prime}\tilde{\theta}_{\tau})\dot{\phi
}_{\mathrm{x}}\left(  X_{i}\right)  ^{\prime}\right)  (\hat{\alpha}_{\tau
}-\alpha_{\tau})\\
&  =\frac{1}{n}\sum_{i=1}^{n}g(Z_{i}^{\prime}\theta_{\tau})\dot{\phi
}_{\mathrm{x}}\left(  X_{i}\right)  ^{\prime}\alpha_{\tau}+M_{L}(\hat{\theta
}_{\tau}-\theta_{\tau})+o_{p}\left(  n^{-1/2}\right)  .
\end{align*}
Therefore,
\begin{align*}
&
\begin{pmatrix}
\hat{\Pi}_{\tau,L,1}\\
\hat{\Pi}_{\tau,L,2}%
\end{pmatrix}
-%
\begin{pmatrix}
\Pi_{\tau,L,1}\\
\Pi_{\tau,L,2}%
\end{pmatrix}
\\
&  =\frac{1}{f_{Y}\left(  Q_{\tau}[Y]\right)  }D_{L}\left[  \frac{1}{n}%
\sum_{i=1}^{n}g(Z_{i}^{\prime}\theta_{\tau})\dot{\phi}_{\mathrm{x}}\left(
X_{i}\right)  ^{\prime}\alpha_{\tau}-E\left[  g(Z_{i}^{\prime}\theta_{\tau
})\dot{\phi}_{\mathrm{x}}\left(  X_{i}\right)  ^{\prime}\alpha_{\tau}\right]
\right] \\
&  -\frac{1}{f_{Y}\left(  Q_{\tau}[Y]\right)  }D_{L}M_{L}H^{-1}\frac{1}{n}%
\sum_{i=1}^{n}s_{i}(\theta_{\tau};Q_{\tau}[Y])\\
&  -\left[
\begin{pmatrix}
\Pi_{\tau,L,1}\\
\Pi_{\tau,L,2}%
\end{pmatrix}
\frac{\dot{f}_{Y}(Q_{\tau}[Y])}{f_{Y}\left(  Q_{\tau}[Y]\right)  }+\frac
{1}{f_{Y}\left(  Q_{\tau}[Y]\right)  }D_{L}M_{L}H^{-1}H_{Q}\right]  \frac
{1}{n}\sum_{i=1}^{n}\psi(Y_{i},\tau,F_{Y})\\
&  -%
\begin{pmatrix}
\Pi_{\tau,L,1}\\
\Pi_{\tau,L,2}%
\end{pmatrix}
\frac{\hat{f}_{Y}\left(  Q_{\tau}[Y]\right)  -f_{Y}\left(  Q_{\tau}[Y]\right)
}{f_{Y}\left(  Q_{\tau}[Y]\right)  }+o_{p}(n^{-1/2})+o_{p}(n^{-1/2}h^{-1/2}).
\end{align*}
Combining this with (\ref{density_error}) leads to the desired result.

\subsection*{Details of Example \ref{example_normal_location_model}
\label{Sec: Details Example Normal Location Model}}

Before the location-scale shift,
\[
Y=\lambda+X\gamma+U=\lambda+\mu_{X}\gamma+U^{\circ}%
\]
and the unconditional $\tau$-quantile $Q_{\tau}[Y]$ of $Y$ is $\lambda+\mu
_{X}\gamma+Q_{\tau}[U^{\circ}]$. After the location-scale shift with $\mu
=\mu_{X}$, $\ $we have
\[
X_{\delta}=\left(  X-\mu_{X}\right)  s\left(  \delta\right)  +\mu_{X}%
+\ell\left(  \delta\right)  ,
\]
and so%
\[
Y_{\delta}=\lambda+X_{\delta}\gamma+U=\lambda+\left[  \mu_{X}+\ell\left(
\delta\right)  \right]  \gamma+U_{\delta}^{\circ}%
\]
where $U_{\delta}^{\circ}=U+\left(  X-\mu_{X}\right)  \gamma s\left(
\delta\right)  .$ The unconditional $\tau$-quantile $Q_{\tau}[Y_{\delta}]$ of
$Y_{\delta}$ is $\lambda+\left[  \mu_{X}+\ell\left(  \delta\right)  \right]
\gamma+Q_{\tau}[U_{\delta}^{\circ}].$ Hence%
\begin{align*}
\Pi_{\tau}  &  =\lim_{\delta\rightarrow0}\frac{\ell\left(  \delta\right)
-\ell\left(  0\right)  }{\delta}\gamma+\lim_{\delta\rightarrow0}\frac{Q_{\tau
}[U_{\delta}^{\circ}]-Q_{\tau}[U^{\circ}]}{\delta}\\
&  =\dot{\ell}\left(  0\right)  \gamma+\lim_{\delta\rightarrow0}\frac{Q_{\tau
}[U_{\delta}^{\circ}]-Q_{\tau}[U^{\circ}]}{\delta}.
\end{align*}
The first term is the location effect, and the second term is the scale effect.

Now, we write
\[
Q_{\tau}\left(  U_{\delta}^{\circ}\right)  =\sqrt{\sigma_{U}^{2}+\sigma
_{X}^{2}\gamma^{2}s\left(  \delta\right)  }Q_{\tau}\left[  \varepsilon
_{\delta}^{\circ}\right]  ,
\]
where $\varepsilon_{\delta}^{\circ}:=U_{\delta}^{\circ}/\sqrt{\sigma_{U}%
^{2}+\sigma_{X}^{2}\gamma^{2}s\left(  \delta\right)  }$ is a random variable
with zero mean and unit variance. So
\begin{align*}
&  \lim_{\delta\rightarrow0}\frac{Q_{\tau}[U_{\delta}^{\circ}]-Q_{\tau
}[U^{\circ}]}{\delta}\\
&  =\lim_{\delta\rightarrow0}\frac{\sqrt{\sigma_{U}^{2}+\sigma_{X}^{2}%
\gamma^{2}s\left(  \delta\right)  }-\sqrt{\sigma_{U}^{2}+\sigma_{X}^{2}%
\gamma^{2}}}{\delta}Q_{\tau}\left[  \varepsilon_{\delta}^{\circ}\right]
+\sqrt{\sigma_{U}^{2}+\sigma_{X}^{2}\gamma^{2}}\lim_{\delta\rightarrow0}%
\frac{Q_{\tau}\left[  \varepsilon_{\delta}^{\circ}\right]  -Q_{\tau}\left[
\varepsilon_{0}^{\circ}\right]  }{\delta}\\
&  =\frac{\sigma_{X}^{2}\gamma^{2}}{\sqrt{\sigma_{U}^{2}+\sigma_{X}^{2}%
\gamma^{2}}}\dot{s}\left(  0\right)  Q_{\tau}\left[  \varepsilon_{0}^{\circ
}\right]  +\sqrt{\sigma_{U}^{2}+\sigma_{X}^{2}\gamma^{2}}\lim_{\delta
\rightarrow0}\frac{Q_{\tau}\left[  \varepsilon_{\delta}^{\circ}\right]
-Q_{\tau}\left[  \varepsilon_{0}^{\circ}\right]  }{\delta}.
\end{align*}
If $Q_{\tau}\left[  \varepsilon_{\delta}^{\circ}\right]  =Q_{\tau}\left[
\varepsilon_{0}^{\circ}\right]  $ for any $\delta\geq0,$ then the second term
is zero, and we obtain
\begin{align*}
\Pi_{\tau,S}  &  :=\lim_{\delta\rightarrow0}\frac{Q_{\tau}[U_{\delta}^{\circ
}]-Q_{\tau}[U^{\circ}]}{\delta}=\frac{\sigma_{X}^{2}\gamma^{2}}{\sqrt
{\sigma_{U}^{2}+\sigma_{X}^{2}\gamma^{2}}}\dot{s}\left(  0\right)  \cdot
Q_{\tau}\left[  \varepsilon_{0}^{\circ}\right] \\
&  =\frac{\sigma_{X}^{2}\gamma^{2}}{\sigma_{U}^{2}+\sigma_{X}^{2}\gamma^{2}%
}\dot{s}\left(  0\right)  \cdot Q_{\tau}\left[  U^{\circ}\right]  .
\end{align*}
Furthermore, if $Q_{\tau}\left[  X_{\gamma}^{\circ}/\sqrt{\sigma_{X}^{2}%
\gamma^{2}}\right]  =Q_{\tau}\left[  \varepsilon_{0}^{\circ}\right]  ,$ then
\[
\Pi_{\tau,S}=\sqrt{\frac{\sigma_{X}^{2}\gamma^{2}}{\sigma_{U}^{2}+\sigma
_{X}^{2}\gamma^{2}}}\dot{s}\left(  0\right)  Q_{\tau}\left[  X_{\gamma}%
^{\circ}\right]  =\dot{s}\left(  0\right)  \sqrt{R_{YX}^{2}}Q_{\tau}\left[
X_{\gamma}^{\circ}\right]  .
\]
Note that if $X$ and $U$ are normals, then $\varepsilon_{\delta}^{\circ
},\varepsilon_{0}^{\circ}$ and $X_{\gamma}^{\circ}/\sqrt{\sigma_{X}^{2}%
\gamma^{2}}$ are all standard normals, and hence they have the same quantiles.
Therefore, $\Pi_{\tau,S}=\dot{s}\left(  0\right)  \sqrt{R_{YX}^{2}}Q_{\tau
}\left[  X_{\gamma}^{\circ}\right]  $, as given in Example
\ref{example_normal_location_model}.
\end{document}